\documentclass{aa}
\graphicspath{images/}  
\usepackage{graphicx}
\usepackage{txfonts}
\usepackage{environ}
\usepackage{booktabs}
\usepackage{floatrow}
\usepackage[colorlinks=true,linkcolor=blue,citecolor=blue]{hyperref}

\def\smbh{SMBH}
\def\ism{ISM}
\def\sf{SF}

\def\cii{[CII]}
\def\sfr{SFR}
\def\mum{$\mu$m}
\def\kms{km s$^{-1}$}
\def\lbol{$L_{\rm AGN}$}

\def\lcii{$L_{\rm [CII]}$}
\def\ergs{erg s$^{-1}$}
\def\fwhm{FWHM$_{\rm [CII]}$}
\def\lbroad{$L_{\rm [CII]}^{\rm broad}$}
\def\lfir{$L_{\rm FIR}$}
\def\msun{M$_\odot$}
\def\msunyr{M$_\odot$ yr$^{-1}$}
\def\lsun{L$_\odot$}
\def\zcii{$z_{\rm [CII]}$}

\begin{document}

   \title{Widespread QSO-driven outflows in the early Universe}

   \author{M. Bischetti \inst{1,2,3}
   	\and R. Maiolino \inst{2,3}
   	\and S. Carniani \inst{2,3}
   	\and F. Fiore \inst{4,1}
   	\and E. Piconcelli \inst{1}
   	\and A. Fluetsch \inst{2,3}
   	       }

   \institute{INAF - Osservatorio Astronomico di Roma, Via Frascati 33, I--00078 Monte Porzio Catone (Roma), Italy
   	\and Cavendish Laboratory, University of Cambridge, 19 J. J. Thomson Avenue, Cambridge CB3 0HE, UK
   	\and Kavli Institute for Cosmology, University of Cambridge, Madingley Road, Cambridge CB3 0HA, UK
   	\and INAF - Osservatorio Astronomico di Trieste, via G.B. Tiepolo 11, I--34143 Trieste, Italy 
             }

   \date{}

 
  \abstract{We present the stacking analysis of a sample of 48 QSOs at $4.5<z<7.1$ detected by ALMA in the $\rm [CII]$ $\lambda158$ $\mu$m emission line to investigate the presence and the properties of massive, cold outflows associated with broad wings in the \cii\ profile. 
  The high sensitivity reached through this analysis allows us to reveal very broad \cii\ wings tracing the presence of outflows with velocities in excess of 1000 \kms. We find that the luminosity of the broad \cii\ emission increases with $L_{\rm AGN}$, while it does not 
  significantly depend on the SFR of the host galaxy, indicating that the central AGN is the main driving mechanism of the \cii\ outflows in these powerful, distant QSOs. 
  From the stack of the ALMA cubes, we derive an average outflow spatial extent of $\sim3.5$ kpc.
  The average atomic neutral mass outflow rate inferred from the stack of the whole sample is
  $\dot{M}_{\rm out}\sim100$ \msunyr, while for the most luminous systems it increases to $\sim200$ \msunyr. The associated outflow kinetic power is about 0.1\% of $L_{\rm AGN}$, while the outflow momentum rate is $\sim L_{\rm AGN}/c$ or lower, suggesting that these outflows are either driven by radiation pressure onto
  dusty clouds or, alternatively, are driven by the nuclear wind and energy conserving but with low coupling with the ISM.
  We discuss the implications of the resulting feedback effect on galaxy
  evolution in the early Universe.}

  \keywords{galaxies:~active -- galaxies:~high-redshift -- galaxies:~nuclei -- quasars:~emission lines --quasars:~general -- techniques:~interferometric}

  \maketitle


\section{Introduction}\label{sect:intro}
The growth of super-massive black holes (\smbh) at the centres of galaxies and
the properties and evolution of the interstellar medium (\ism) in their hosts are expected to be connected \citep[e.g.][]{DiMatteo05,Hopkins08}. There are in fact well established correlations observed between the black hole mass and the physical properties of the host galaxy \citep{Kormendy&Ho13} such as the bulge mass or velocity dispersion, suggesting that the energy output of the accretion onto SMBH may be communicated to the surrounding \ism\ and affect star formation (\sf).
Indeed, active galactic nuclei (AGN) feedback onto their host galaxies is expected to proceed via kpc-scale, wide-angle outflows \citep{Menci08,Faucher-Giguere12}, capable of heating and removing gas, therefore suppressing \sf. AGN feedback is one of the main mechanisms invoked in cosmological simulations to prevent an excessive growth of massive galaxies and make gas-rich starburst galaxies quickly evolve to quiescence.

Growing observational evidence of massive AGN-driven outflows has been collected, involving different gas phases (ionised, atomic and molecular) extending from sub-pc to kpc scales.
While recent works, based on local AGN, use a multi-phase study of outflows to quantify their impact on the host galaxy \citep[e.g.][]{Feruglio15,Tombesi15,Veilleux17,Longinotti18}, at high redshift ($z\sim1-3$) studies of outflows are still mostly limited to the ionised phase \citep[see][and references therein]{Fiore17}. There are only few detections of fast molecular gas observed in CO high-\textit{J} rotational transitions \citep{Carniani17,Feruglio17,Vayner17,Brusa18}.
However massive, quiescent systems and old (aged 2$-$3 Gyr) galaxies have been observed already at $z\sim2-3$
\citep{Cimatti04,Whitaker13,Straatman14}, indicating that a feedback mechanism must have been in place even at very early epochs, around $z\sim5-6$.

Observations of AGN-driven outflows at high redshift have targeted the \cii\ fine-structure emission line at 158 \mum,  which is
generally the strongest emission line in galaxies at far infrared (FIR) wavelengths. Typically \cii\ is a tracer of both the neutral atomic gas, primarily in
Photo-Dissociated Regions (PDRs), but can be in part emitted also from the (partly) ionised medium \citep[e.g.][]{Carilli13}. Since PDR are produced by
the UV radiation emitted by young stars, [CII] has also been used as a tracer of SF
\citep{Maiolino05,DeLooze11,Carniani13,Carniani18a}.

Recently, \cii\ has also been exploited to trace cold gas in galactic outlows.
Indeed, broad wings \cii\ emission has been observed in the hyper-luminous quasi-stellar object (QSO) J1148$+$5251 at z $\sim$ 6.4
by \cite{Maiolino12} and \cite{Cicone15}, revealing outflowing gas extended up to $\sim30$ kpc and escaping with velocities in excess
of 1000 \kms. The Herschel Space Observatory has enabled the detection of cold outflows through broad wings of the [CII] line also
in local active galaxies \citep{Janssen16}.

The exploitation of the bright [CII] line at high redshift has been increasing in the last few years with the
advent of ALMA. In particular,
the population of high-$z$ luminous QSOs with detected \cii\ emission has been rapidly growing. Previous works have exploited
the [CII] emission to investigate the properties of their host galaxies, such as the \sfr, the dynamical mass, and the presence of
merging companions \citep[e.g.][]{Wang13,Wang16,Venemans16,Venemans17,Willott15,Willott17,Trakhtenbrot17,Decarli17,Decarli18}. In none
of these high-$z$ QSOs evidence of \cii\ outflows has been reported. However, most of the \cii\ observations in distant QSOs are still
rather short (10$-$20 minutes of on-source time), with a sensitivity generally not yet adequate to individually detect weak \cii\ broad wings.

We have collected a sample of 48 QSOs with ALMA \cii\ detection to investigate the presence of outflows, as traced by weak \cii\ broad wings, by performing a stacking analysis. We will show that the stacked data achieve a sensitivity that is more than an order of magnitude deeper than that reached in the previous \cii\ outflow detection by \cite{Maiolino12,Cicone15} and enable us to reveal very broad wings tracing cold outflows associated with distant QSOs.

\section{Sample and data reduction}\label{sect:sample}

\begin{table*}
	\caption{Main information about the ALMA observations and source properties of the QSOs in our sample.
	Columns give the following information: (1) source ID, (2) \cii-based redshift, (3) beamsize, (4) continuum
	rms sensitivity, (5) representative rms of the \cii\ spectral region for a channel width of 30 \kms, (6)
	continuum flux, (7) \cii\ luminosity, (8) FWHM of the \cii\ line, (9) FIR luminosity in the range 8-1000 \mum\ derived from the ALMA continuum flux, (10) AGN bolometric luminosity.}
	\centering
	\small
	\makebox[1\textwidth]{
	\setlength{\tabcolsep}{3 pt}	
\begin{tabular}{lcccccccccc}
	\toprule
Source ID   & $z_{\rm [CII]}^*$ & Beam &   Cont rms  &  \cii\ rms  &  $f_{\rm cont}$    & $L_{\rm [CII]}^{\rm core}$   & FWHM$_{\rm [CII]}^{\rm core}$ & Log(\lfir/\lsun) &  Log(\lbol/\ergs)$^{**}$ & Stack\\
			& 				  & [arcsec] & [mJy/beam]  & [mJy/beam]  & [mJy/beam]	& [$10^9$ \lsun]	& [\kms] 	& & \\
(1) & (2) & (3) & (4) & (5) & (6) & (7) & (8) & (9) & (10) & (11) \\			
\midrule
PJ007+04    &    6.001   &  0.47$\times$0.69 &  0.05   &    0.58   &    1.87$\pm$0.04  &     1.30$\pm$0.15    &    365$\pm$45   &   12.89   &   46.77 & BF\\
PJ009-10    &    6.003   &  0.45$\times$0.66 & 0.05   &    0.49   &    2.43$\pm$0.12  &    2.28$\pm$0.15      &    290$\pm$35   &   12.98   &   46.73 & AF\\
J0055+0146  &    6.005   &  0.60$\times$0.72 & 0.03   &    0.30   &    0.22$\pm$0.02  &   0.60$\pm$0.07      &    330$\pm$40   &   11.80   &   46.04 & AE\\
J0109-3047  &    6.790   &  0.51$\times$0.80 & 0.05   &    0.84   &    0.58$\pm$0.04  &     1.64$\pm$0.18    &    310$\pm$40   &   12.30   &   46.37 & AE\\
J0129-0035  &    5.779   &  0.36$\times$0.45 &  0.02   &    0.24   &    3.04$\pm$0.05  &     1.73$\pm$0.04    &    200$\pm$30   &   12.91   &   45.67& AF\\
J0142-3327  &    6.337   &  0.75$\times$0.87 &  0.04   &    0.61   &    1.70$\pm$0.06  &     2.94$\pm$0.16    &    300$\pm$30   &   12.71   &   47.24& BF\\
J0210-0456  &    6.433   &  0.61$\times$0.90 &  0.03   &    0.29   &    0.16$\pm$0.03  &     0.37$\pm$0.04     &   185$\pm$35   &   11.70   &   45.93& AE\\
J0305-3150  &    6.615   &  0.51$\times$0.72 &  0.03   &    0.37   &    3.20$\pm$0.05  &     2.34$\pm$0.09     &   215$\pm$30   &   13.02   &   46.59& AF\\
J0331-0741  &    4.737   &  0.31$\times$0.40 &  0.05   &    0.45   &    3.75$\pm$0.07  &     2.84$\pm$0.11     &   475$\pm$35   &   12.84   &   47.39& DF\\
PJ065-26    &    6.187   &  0.87$\times$1.11 &  0.05   &    0.76   &    1.05$\pm$0.07  &     1.94$\pm$0.19     &   405$\pm$40   &   12.48   &   47.01& DE\\
PJ065-19    &    6.125   &  0.75$\times1.09$ & 0.04   &    1.32   &    0.42$\pm$0.04  &     1.80$\pm$0.40     &   315$\pm$60   &   12.11   &   46.76& BE\\
J0454-4448  &    6.058   &  0.80$\times$1.18 &  0.04   &    0.63   &    0.68$\pm$0.05  &     0.62$\pm$0.10     &   360$\pm$70   &   12.30   &   46.68& AE\\
J0807+1328  &    4.879   &  0.25$\times$0.40 &  0.03   &    0.67   &    6.64$\pm$0.13  &     2.44$\pm$0.19     &   435$\pm$38   &   13.14   &   47.07 & DF\\
J0842+1218  &    6.076   &  1.14$\times$1.27 &  0.05   &    0.77   &    0.57$\pm$0.04  &     1.62$\pm$0.22     &   480$\pm$55   &   12.24   &   46.88 & DE\\
J0923+0247  &    4.655   &  0.29$\times$0.51 &  0.04   &    0.30   &    2.94$\pm$0.08  &    2.55$\pm$0.09      &   325$\pm$30   &   12.76   &   46.96 & BF\\
J0935+0801  &    4.682   &  0.29$\times$0.55 &  0.04   &    0.29   &    1.39$\pm$0.05  &     0.70$\pm$0.07     &   385$\pm$40   &   12.48   &   47.25 & BE\\
J1017+0327  &    4.949   &  0.30$\times$0.38 &  0.03   &    0.32   &    1.23$\pm$0.07  &     1.02$\pm$0.05     &   270$\pm$30   &   12.42   &   46.27& AE\\
PJ159-02    &    6.381   &  0.99$\times$1.27 &  0.04   &    0.55   &    0.60$\pm$0.05  &      1.24$\pm$0.15    &    385$\pm$45  &    12.27  &   46.83 & BE\\
J1044-0125  &    5.785   &  0.66$\times$0.72 &  0.02   &    0.29   &    3.07$\pm$0.03  &      1.62$\pm$0.08    &    470$\pm$35  &    12.92  &   47.07 & DF\\
J1048-0109  &    6.676   &  1.00$\times$1.43 &  0.03   &    0.51   &    2.57$\pm$0.03  &     2.42$\pm$0.13     &   350$\pm$35   &   12.94   &   46.51 & AF\\
PJ167-13    &    6.515   &  0.98$\times$1.27 &  0.04   &    0.51   &    0.69$\pm$0.04  &     3.15$\pm$0.19     &    480$\pm$35  &    12.35  &   46.36 & CE\\
J1120+0641  &    7.086   &  0.29$\times$0.32 &  0.01   &    0.13   &    0.40$\pm$0.02  &     0.69$\pm$0.05     &    540$\pm$40  &    12.19  &   46.77 & DE\\
J1152+0055  &    6.365   &  1.02$\times$1.36 &  0.04   &    0.70   &    0.50$\pm$0.06  &     0.51$\pm$0.10     &    140$\pm$50  &    12.23  &   46.17 & AE\\
J1207+0630  &    6.037   &  0.89$\times$1.63 &  0.06   &    0.90   &    0.56$\pm$0.04  &     1.16$\pm$0.18     &   490$\pm$95   &   12.20   &   46.77 & DE\\
PJ183+05    &    6.439   &  1.06$\times1.24$ &  0.04   &    0.59   &    4.62$\pm$0.05  &     6.02$\pm$0.19     &   370$\pm$30   &   13.17   &   46.93 & BF\\
J1306+0356  &    6.033   &  0.98$\times$1.17 &  0.05   &    0.74   &    1.20$\pm$0.05  &     1.87$\pm$0.17     &   265$\pm$35   &   12.53   &   46.84 & BE\\
J1319+0950  &    6.132   &  1.10$\times$1.26 &  0.03   &    0.43   &    5.00$\pm$0.05  &     3.85$\pm$0.18     &   520$\pm$35   &   13.17   &   46.93 & DF\\
J1321+0038  &    4.722   &  0.34$\times$0.39 &  0.02   &    0.19   &    1.49$\pm$0.04  &     1.19$\pm$0.06     &   560$\pm$35   &   12.47   &   46.70 & CE\\
J1328-0224  &    4.646   &  0.31$\times$0.48 &  0.04   &    0.37   &    1.58$\pm$0.04  &     1.56$\pm$0.06     &   300$\pm$30   &   12.49   &   47.05 & BE\\
J1341+0141  &    4.700   &  0.30$\times$0.38 &  0.06   &    0.45   &   17.74$\pm$0.33  &     3.06$\pm$0.15     &   435$\pm$35   &   13.55   &  47.50 & DF\\
J1404+0314  &    4.924   &  0.34$\times$0.40 &  0.05   &    0.58   &   10.98$\pm$0.20  &     3.14$\pm$0.15     &   515$\pm$35   &   13.37   &  47.02 & DF\\
PJ217-16    &    6.149   &  0.92$\times$1.19 &  0.05   &    0.71   &    0.40$\pm$0.02  &      0.89$\pm$0.17    &    510$\pm$70  &    12.14   &  46.89& DE\\
J1433+0227  &    4.727   &  0.34$\times$0.44 &  0.05   &    0.43   &    7.69$\pm$0.13  &     2.52$\pm$0.08     &   415$\pm$30   &   13.19   &   47.37& DF\\
J1509-1749  &    6.122   &  0.92$\times$1.43 &  0.04   &    0.67   &    1.34$\pm$0.04  &     1.72$\pm$0.20     &   615$\pm$75   &   12.59   &   46.9& DE\\
J1511+0408  &    4.679   &  0.31$\times$0.53 &  0.06   &    0.46   &   10.08$\pm$0.19  &     2.78$\pm$0.18     &   580$\pm$45   &   13.30   &  47.25& DF\\
PJ231-20    &    6.587   &  0.94$\times$1.29 &  0.04   &    0.66   &    3.80$\pm$0.10  &     2.97$\pm$0.21     &   410$\pm$35   &   13.09   &   46.99& DF\\
J1554+1937  &    4.627   &  0.74$\times$1.26 &  0.16   &    1.67   &   11.98$\pm$0.42  &     6.86$\pm$0.43     &   800$\pm$45   &   13.37   &  47.70& DF\\
PJ308-21    &    6.234   &  0.68$\times$0.89 &  0.03   &    0.54   &    1.12$\pm$0.08  &     2.17$\pm$0.18     &   575$\pm$45   &   12.53   &   46.65& CE\\
J2054-0005  &    6.039   &  0.73$\times$0.76 &  0.02   &    0.40   &    2.89$\pm$0.04  &     2.46$\pm$0.07     &   230$\pm$30   &   12.92   &   46.60& AF\\
J2100-1715  &    6.082   &  0.66$\times$0.78 &  0.05   &    0.60   &    0.46$\pm$0.02  &     1.27$\pm$0.17     &   390$\pm$60   &   12.22   &   46.33& AE\\
J2229+1457  &    6.151   &  0.70$\times$0.80 &  0.03   &    0.42   &    0.14$\pm$0.02  &     0.34$\pm$0.06     &   240$\pm$50   &   11.62   &   46.03& AE\\
J2244+1346  &    4.661   &  0.33$\times$0.40 &  0.03   &    0.31   &    3.26$\pm$0.05  &     1.74$\pm$0.04     &   270$\pm$30   &   12.80   &   46.58& AF\\
W2246-0526  &    4.601   &  0.35$\times$0.37 &  0.05   &    0.52   &    7.18$\pm$0.12  &     6.12$\pm$0.19     &   740$\pm$35   &   13.14   &   48.12& DF \\
J2310+1855  &    6.002   &  0.79$\times$1.18 &  0.04   &    0.41   &    7.62$\pm$0.12  &     5.74$\pm$0.15     &   405$\pm$30   &   13.34   &   47.23& DF\\
J2318-3029  &    6.148   &  0.75$\times$0.87 &  0.05   &    0.82   &    2.87$\pm$0.05  &     1.73$\pm$0.18     &   275$\pm$35   &   12.93   &   46.60& AF\\
J2318-3113  &    6.444   &  0.79$\times$0.89 &  0.06   &    0.92   &    0.32$\pm$0.04  &     1.26$\pm$0.20     &   305$\pm$55   &   12.00   &   46.56& AE\\
J2348-3054  &    6.902   &  0.62$\times$0.82 &  0.05   &    0.79   &    1.90$\pm$0.05  &     1.52$\pm$0.23     &   455$\pm$65   &   12.82   &   46.43& CF\\
PJ359-06    &    6.172   &  0.64$\times$1.14 &  0.07   &    0.77   &    0.76$\pm$0.05  &     1.66$\pm$0.17     &   305$\pm$40   &   12.46   &   46.83& BE\\

	\bottomrule
	
\end{tabular}
}
\flushleft
$^*$ Given the statistical error on the centroid of the best-fit Gaussian modelling the [CII] line and the systematics associated with the 30 \kms\ channel width of our ALMA spectra, the typical error on redshift is $\Delta z_{\rm [CII]}=0.001$.\\
$^{**}$ We consider as error on \lbol\ the 0.1 dex scatter associated with the bolometric correction by \citet{Runnoe12}. 
\label{tab:sample}
\end{table*}	

\begin{figure}
	\centering
	\includegraphics[width=1\columnwidth]{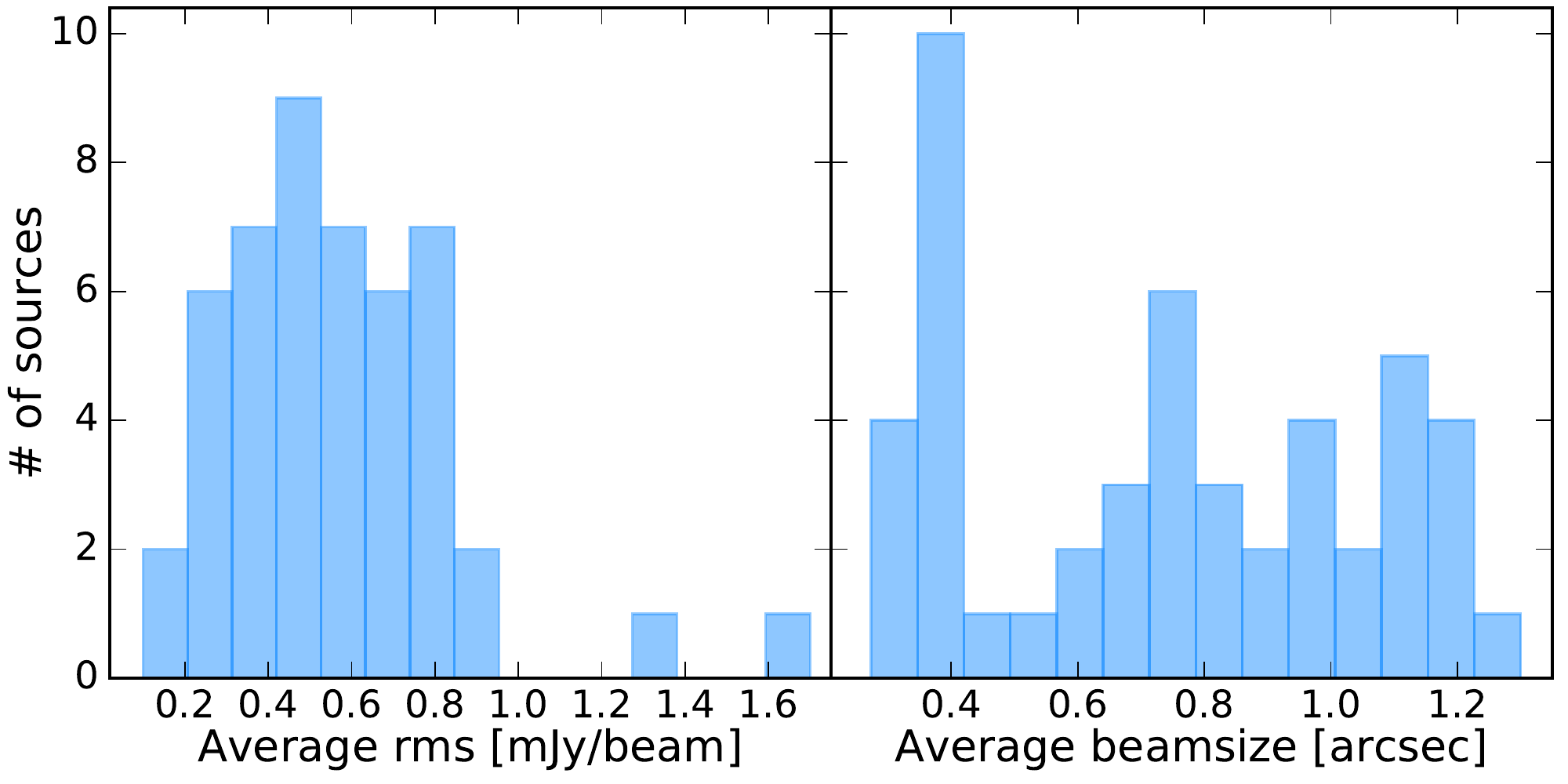}
	\caption{Sensitivity and beam size distributions of the ALMA \cii\ observations for the high-$z$ QSOs in our sample. \textit{Left panel:} number of sources as a function of the mean (averaged over the spectral range covered by the stack) sensitivity for a 30 \kms\ channel. \textit{Right panel:} histogram of the mean beam axis size.}
	\label{fig:rmshisto}

\end{figure}

\begin{figure}
	\centering
	\includegraphics[width=0.55\columnwidth]{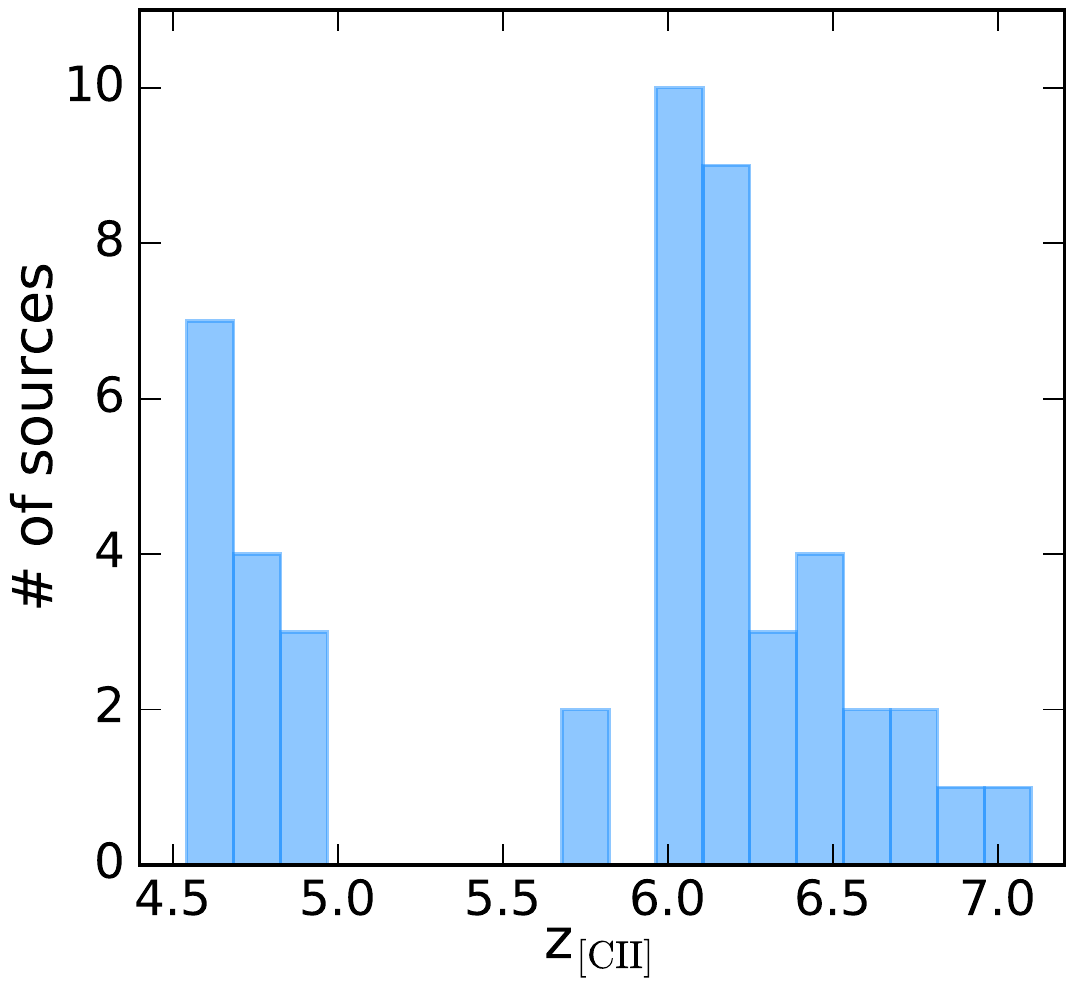}
	\includegraphics[width=1\columnwidth]{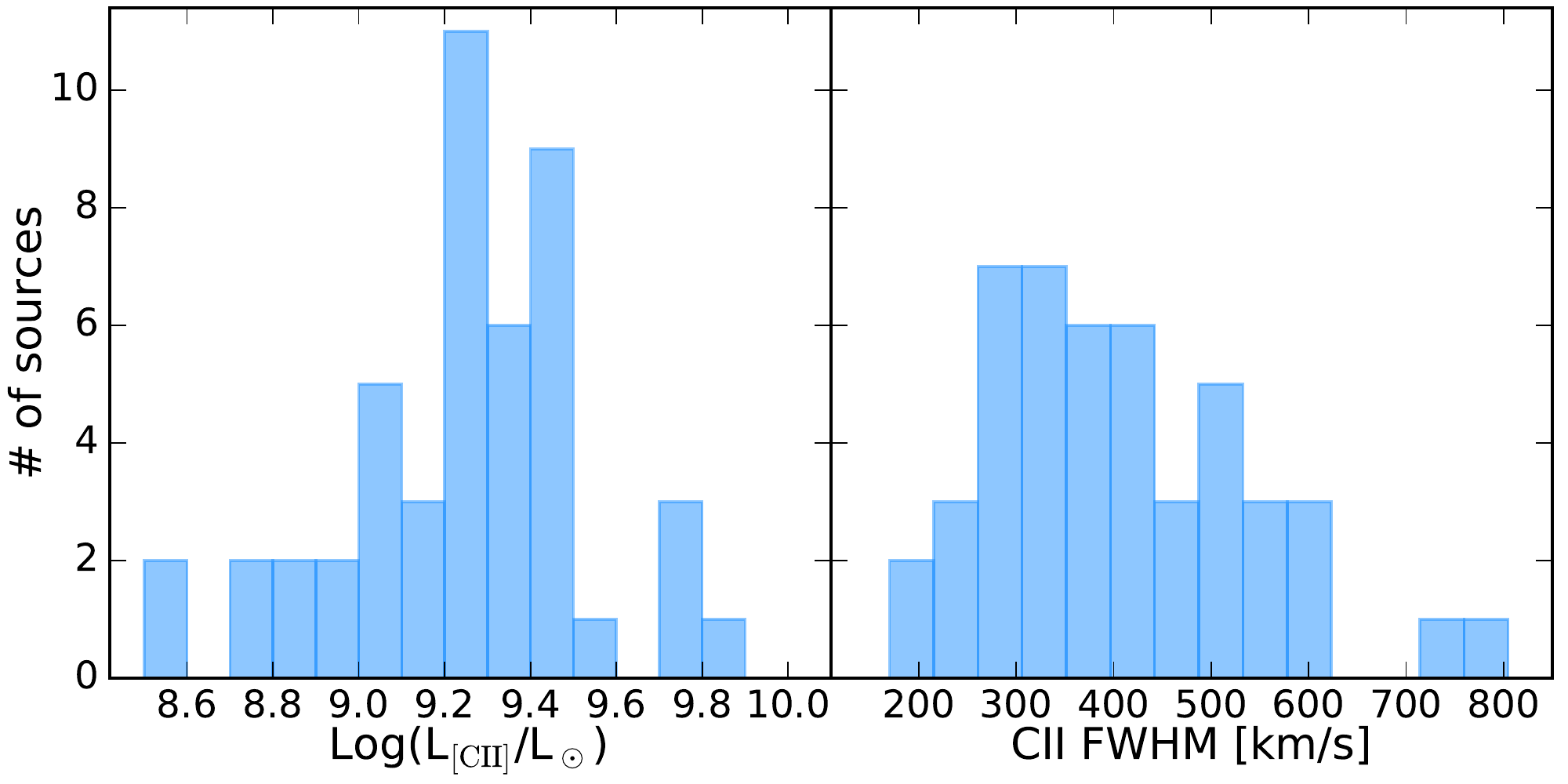}
	\includegraphics[width=1\columnwidth]{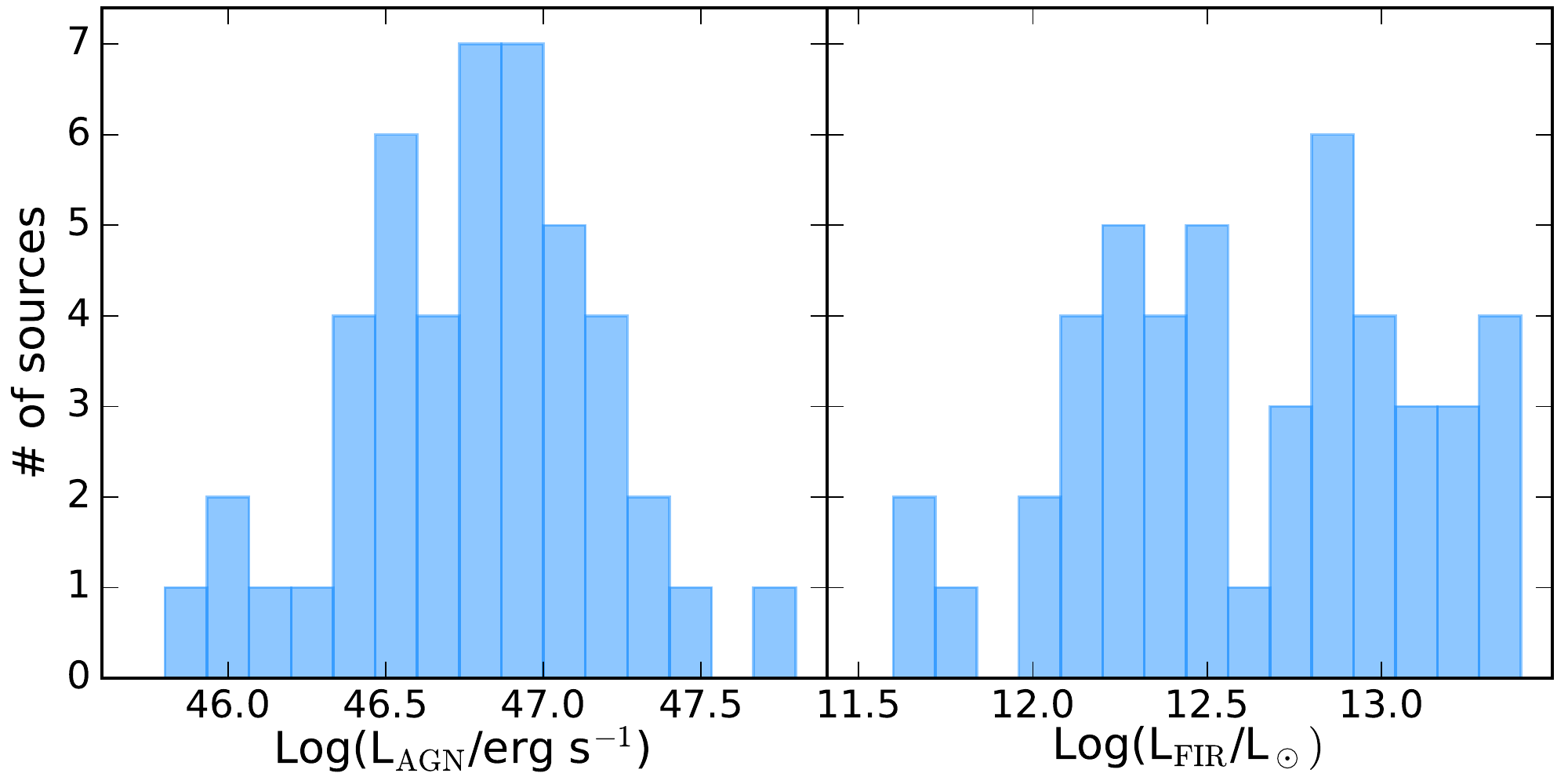}
	\caption{Properties of the high-$z$ QSOs sample considered in this work. \textit{Top panel:} redshift distribution. \textit{Middle panel:} luminosity and FWHM of the \cii\ core emission. \textit{Bottom panel:} AGN luminosity and FIR luminosity.}
	\label{fig:sample-distrib}
\end{figure}

We collected all \cii\ observations of $z>4.5$ QSOs on the ALMA archive public
as of March 2018 and selected the sources with a \cii\ detection significant at $\gtrsim 5\sigma$. 
Specifically, we used data from ALMA projects 2011.0.00243.S (P.I. C. Willott), 2012.1.00604.S (P.I. A. Kimball),
2012.1.00676.S (P.I. C. Willott), 2012.1.00882.S (P.I. B. Venemans), 2013.1.01153.S (P.I. P. Lira), 2015.1.01115.S (P.I. F. Walter) and
2016.1.01515.S (P.I. P. Lira). Details about individual QSOs in our sample, for those
which have been published, can be found in
\cite{Wang13}, \cite{Willott13}, \cite{Willott15}, \cite{Willott17},
\cite{Kimball15}, \cite{Diaz-Santos16}, \cite{Venemans16},
\cite{Venemans17}, \cite{Decarli17}, \cite{Decarli18}, and  \cite{Trakhtenbrot17};
however, we also included some ALMA not yet published archival data from project 2015.1.00997.S (P.I. R. Maiolino, Carniani et al. in prep.). The assembled sample consists of the most luminous QSOs with rest-frame absolute UV magnitude $-28.5 \lesssim M_{1450\AA} \lesssim -23.9$ mag and black hole masses $10^8 \lesssim M_{\rm BH} \lesssim 10^{10}$ \msun.
As mentioned, in total we combined ALMA data for 48 QSOs, and equivalent to a total of $\sim34$ hours of
on-source observing time.

Observations involve ALMA bands 6 or 7, depending on the redshift of
the individual source. The distribution of the average rms sensitivity, representative of the \cii\ spectral region, and that of the size of the ALMA beam are shown in Fig. \ref{fig:rmshisto}. Individual values are listed in Table \ref{tab:sample}. Except for few outliers, the bulk of the observations have similar rms sensitivities from $\sim 0.3$ to $\sim 0.8$ mJy/beam for a 30 \kms channel. The angular resolutions, computed as average beam axis, range from $\sim 0.3$ to 1.2 arcsec.
Data were calibrated using the CASA 4.7.2 software \citep{McMullin07} in manual or pipeline mode. The default phase, bandpass and flux calibrators were used unless differently indicated by the ALMA observatory. Where necessary, extra flagging and improvement in the flux calibration was done.
Data cubes were produced by using the CASA task clean by using the Hogbom algorithm with no cleaning mask and a number of iterations $N_{\rm iter}=500-1000$ according to the significance of the detection, together with a threshold of three times the sensitivity limit given by the rms. We chose a natural weighting to maximise the sensitivity of the individual observations, a common pixel size of 0.05$''$ and a common spectral bin of 30 \kms. 

Continuum maps were obtained by averaging over all the four spectral windows and excluding the spectral range covered by the \cii\ emission and possible \cii\ broad wings. Continuum flux densities were derived by fitting a 2D Gaussian model to the ALMA maps.
Furthermore, to model the continuum emission we combined the two adjacent spectral windows of the sideband containing the \cii\ line to increase the available
spectral range, for a total of $\sim3.7$ GHz. We did not consider the two additional spectral windows in the sideband not including
\cii because of the large spectral separation ($\sim15$ GHz in the observed frame). The expected intrinsic differences in the QSO continuum
flux ($\sim15-20$\%) among this large spectral range and, mainly, the systematics in the relative calibration of distant spectral windows may affect the detection of broad wings.
We thus fitted a zeroth order continuum model in the UV plane to all the
available channels (of the spectral windows adjacent to \cii\, where the QSO continuum variation is expected to be $<1$\%) with a velocity $|v| > 1500$ \kms with respect to the
centroid of the (core) [CII] emission. This choice represents a trade-off between maximising the number of channels ($\sim1/4$ of each spectral window) available to the fit and avoiding spectral regions where broad \cii\ wings might be present. Moreover, spectral regions corresponding to
an atmospheric transmission $<0.5$ for a 1 mm precipitable water vapour were excluded from the fit. We verified that modelling the continuum emission with a first order polynomial did not significantly affect our results, given the limited frequency range covered by our stack.

To determine the properties of the host galaxy emission, we extracted the continuum-subtracted \cii\ spectra from a region with an area of four beams (see Sect. \ref{sect:methods}). The line parameters describing the \cii\ core emission were derived by fitting each spectrum with one Gaussian component model. Specifically, redshifts (\zcii) were derived from the centroid of the best-fit \cii\ model (see Table \ref{tab:sample}). 
 
The main properties of our high-$z$ QSOs sample are shown in Fig.
\ref{fig:sample-distrib} and listed in Table \ref{tab:sample}. The QSOs in the sample are distributed
in two main redshift bins, i.e. a first group at $4.5<z<5$ and a second,
higher-$z$ group at $z\gtrsim6$. The bulk of the sample is characterised by a luminosity of the \cii\ core emission in the range Log(\lcii/\lsun)
$\sim9.0-9.5$ and \cii\ line profiles with a Full Width Half Maximum
(FWHM$_{\rm [CII]}^{\rm core}$) in the range between 300 and 500 \kms. We computed the FIR luminosity by using an
Mrk231-like template \citep{Polletta07} normalised to the observed continuum flux density at (rest frame) 158 \mum. The resulting \lfir\ (see Fig. \ref{fig:sample-distrib})
span almost two orders of magnitude, i.e. Log(\lfir/\lsun)$=11.6-13.4$. The AGN bolometric luminosity (\lbol) was derived from the monochromatic luminosity at 1450 \AA\ by applying the bolometric correction from \cite{Runnoe12}. All sources in our sample are luminous and hyper-luminous QSOs with \lbol\ $\gtrsim10^{46}$ \ergs, with an average \lbol\ of $6.3\times10^{46}$ \ergs.

\section{Methods} \label{sect:methods}

In order to investigate the presence of high velocity wings of the \cii\
emission, we performed a stacking analysis of the distant QSOs in our sample.
The stacking technique has the potential to greatly increase the sensitivity of
the stacked spectrum or stacked cube and, therefore, favours the detection of even modest outflows traced by weak \cii\ wings.

As a first step, the cubes were aligned at the \cii\ rest frequency (1900.5369 GHz) according to \zcii\ and spatially centred on the peak of the QSO continuum emission. 
We did not include in the stack spectral regions corresponding to an atmospheric transmission $<0.5$ for a 1 mm precipitable water vapour.
Then, we combined the data from the 48 sources in our sample according to the relation below, defining the weighted intensity $I^\prime_k$ of a generic spatial pixel ($x^\prime, y^\prime$) in the stacked cube for each spectral channel $k$, and the relative weight $W^\prime_{\rm k}$ as follows \citep{Fruchter&Hook02}:

\begin{equation}\label{eq:stack_1}
W^\prime_{\rm k} = \sum_{j=1}^n w_{\rm j,k}  = \sum_{j=1}^n \frac{1}{\sigma_{\rm j,k}^2} = \frac{1}{\sigma^{\prime 2}_k}
\end{equation}

\begin{equation}\label{eq:stack_2}
I^\prime_k = \frac{\sum_{j=1}^n \left(i_{\rm j,k}\cdot w_{\rm j,k}\right)}{W^\prime_k}
\end{equation}

\noindent where $i_{\rm j,k}$ is the intensity at the same spatial pixel ($x_j,
y_j$) and same spectral channel $k$ of source $j$, and $n=48$.  We applied a standard variance-weighted stacking, i.e. we used a weighting factor $w_{\rm j,k}=1/\sigma_{\rm j,k}^2$, where $\sigma_{\rm j,k}$ is the rms noise estimated channel by channel from cube $j$. Furthermore, with this method we accounted for the noise variation with frequency in the spectral range covered by the ALMA \cii\ spectra, i.e. $\sim 3.7$ GHz, and considered only the contribution of sources with available spectral coverage in our weighted mean.
We performed the stacking in two alternative, complementary ways: by stacking the 1D spectra extracted from the individual cubes and by stacking
the 3D cubes into a single stacked cube.

\begin{table*}
	\centering
	\caption{Variance-weighted properties of the stacked QSO samples and the corresponding \cii\ emission properties. Specifically, rows give the following information: (1) rms sensitivity representative of the
		\cii\ spectral region for a channel width of 30 \kms, (2) average \lbol\ in the (sub-)sample,
		(3) average FIR-based SFR, (4) FWHM of the [CII] core, (5) average luminosity of the broad \cii\ wings, (6) their significance, (7) FWHM, and (8) velocity shift. (9) peak and (10) integrated flux density ratios of the broad \cii\ with respect to the core \cii\ emission.}
	\setlength{\tabcolsep}{3 pt}
	\begin{tabular}{llccccccc}
		\toprule
		  & &  \textbf{Whole} & \multicolumn{6}{c}{Subsamples} \\\cline{4-9}
		  &  & \textbf{sample} & A & B & C & D & E & F \\
		 \midrule
		 (1) & rms  [mJy beam$^{-1}$] & \textbf{0.06} & 0.11 & 0.16 & 0.16 & 0.09 & 0.09 & 0.10 \\
		 (2) & \lbol\  [\ergs] & \textbf{47.0} & 46.3 & 47.1 & 46.7 & 47.2 & 46.7 & 47.1 \\
		 (3) & SFR$_{\rm FIR}$  [\msunyr] & \textbf{790} & 570 & 540 & 360 & 1270 & 260 & 1750 \\
		 (4) & FWHM$_{\rm [CII]}^{\rm core}$  [\kms] & \textbf{390$\pm$30} & 210$\pm30$ & 330$\pm$30 & 600$\pm$40 & 510$\pm40$ & 390$\pm$30 & 360$\pm30$\\
		 (5) & \lbroad\  [10$^8$ \lsun] & \textbf{4.1$\pm$0.7} & $2.5\pm0.8$ & 4.6$\pm$1.9 & 3.7$\pm$1.2 & 6.9$\pm$1.5 & 4.2$\pm$1.1 & 3.8$\pm$0.8 \\
		 (6) & SNR$_{\rm [CII]}^{\rm broad}$  & \textbf{5.6$^{\ast}$, 10.2$^{\ast\ast}$, 7.2$^{\ast\ast\ast}$} & 3.0, 5.6, 3.5 & 2.4, 7.5, 3.3 & 3.0, 2.9, 2.4 & 4.6, 7.0, 9.8 & 3.7, 6.5, 5.1 & 4.8, 7.8, 5.4 \\
		 (7) & FWHM$_{\rm [CII]}^{\rm broad}$  [\kms] & \textbf{1730 $\pm$ 210} & 850 $\pm$ 160 & 710 $\pm$ 130 & 2360 $\pm$ 640 & 1920 $\pm$ 250 & 2210 $\pm$ 430 & 1380 $\pm$ 200 \\
		 (8) & $\Delta v$  [\kms] & \textbf{-90 $\pm$ 40} & -110 $\pm$ 70 & -70 $\pm$ 50 & -10 $\pm$ 100 & -180 $\pm$ 70 & 130 $\pm$ 100 & -240 $\pm$ 90 \\
		 (9) & $p_{\rm\cii}$  & \textbf{0.05 $\pm$ 0.01} & 0.05 $\pm$ 0.01 & 0.1 $\pm$ 0.03 &  0.05 $\pm$ 0.02 & 0.07 $\pm$ 0.01 & 0.07 $\pm$ 0.01 & 0.04 $\pm$ 0.01\\
		 (10) & $f_{\rm\cii}$  & \textbf{0.22 $\pm$ 0.04} & 0.18 $\pm$ 0.06 & 0.23 $\pm$ 0.08 & 0.22 $\pm$ 0.07 & 0.31 $\pm$ 0.06 &0.39 $\pm$ 0.09 & 0.14 $\pm$ 0.02 \\
		\bottomrule
	\end{tabular}
	\flushleft
	$^{\ast}$ Computed from the fit parameters errors, accounts for the uncertainty in modelling the narrow component.\\
	$^{\ast\ast}$ Computed from the pure statistical uncertainty. \\
	$^{\ast\ast\ast}$ Computed as in $^{\ast\ast}$, but excluding the central channels affected by the \cii\ core.\\
	\label{tab:outflow}	
\end{table*}

In the first case the continuum-subtracted spectrum of each target was
extracted from an elliptical aperture with same position angle of the beam, but over an area four times larger. This approach allows us to collect most of the flux from the QSO (for a point source, $\sim$ 94\% of the flux lies within two beam axes) and limits the contamination of possible companions. The angular size of the systemic [CII] emission is comparable to the ALMA beam for most of the QSOs in our sample \citep[e.g.][]{Venemans16, Venemans17, Decarli18}. The chosen extraction areas therefore maximise the significance of possible high-velocity [CII] wings if outflowing and systemic gas are distributed over similar scales \citep{Cicone14}. 
As a drawback of our approach, emission from different physical scales may contribute to the stacked spectra. The individual spectra were stacked according to Eq. 1 and  Eq. 2. 

In the second approach, the continuum-subtracted {\it cubes} of the single sources were stacked by applying Eqs. 1,2 to each spaxel. This resulted in a stacked data cube, containing the contribution of each source to the different channels and spatial positions.

Table \ref{tab:outflow} reports the statistical uncertainties of the stacked spectrum and cube in spectral channel of 30 \kms, which have been estimated excluding those spaxels contaminated by the QSO emission. We note that the statistical uncertainty of a spectrum extracted from an area of N beams was computed as $\sqrt{N}$ times the rms. Similarly, the sensitivity of a map integrated over K channels was derived as $\left(\sum_{i=1}^K \sigma _i^{-2}\right)^{-1/2}$, where $\sigma _i$ is the rms within each channel slice.
	
To ensure that the presence of broad \cii\ wings in the stacked spectrum is not an artefact of the stacking procedure, we extracted individual integrated spectra from 100 "empty" positions randomly-selected within the ALMA field of view and stacked them as described above. An upper limit on the significance of the integrated flux associated with a spurious broad component can be derived by fitting the stacked noise spectra with one broad Gaussian profile with a FWHM $>500$ \kms\ and centroid in the velocity range $v\in[-500,+500]$ \kms. This resulted into an average spurious signal-to-noise ratio of $\sim0.4$ for the stack of the total sample.

We also verified that the presence of broad \cii\ wings was not associated with a few QSOs but instead a general property of our sample. For this purpose, we recomputed 1000 times the stack of the integrated spectrum on different subgroups, excluding each time a combination of five randomly selected sources (i.e. $\sim10\%$ of the sample). 
The resulting rms variation of the \cii\ wings in the velocity range $400<|v|<1500$ \kms\ corresponds to $\sim$20\% of the peak flux density of the broad \cii\ wings presented in Sect. \ref{sect:totstack}. The average luminosity variation of the \cii\ wings is $\sim11$\%, with a maximum variation of 40\%.

The uncertainty on the continuum fitting of the individual spectra would result into a simple pedestal, as we modelled the continuum emission with a zero order. However, the fitting of the total stacked spectrum does include a continuum component which is fully consistent with zero, confirming on average a proper continuum subtraction in the individual spectra.

\begin{figure}
	\centering
	\includegraphics[width=1\columnwidth]{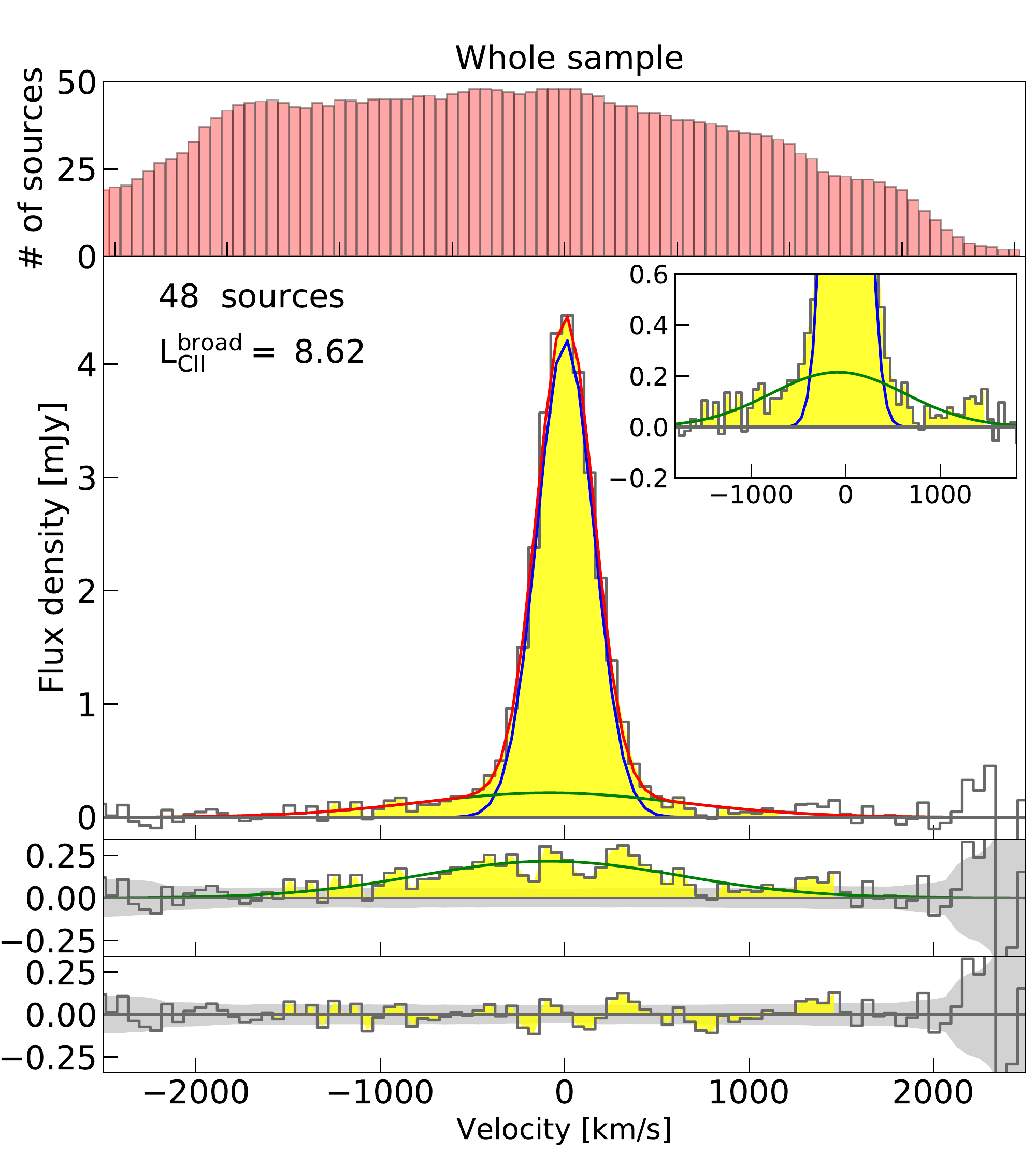}
	\caption{Whole sample stacked integrated spectrum. \textit{First panel from top:} number of sources contributing to
		the stack at different velocities. \textit{Second panel from top:} \cii\ flux density as a function of velocity, in spectral
		bins of 60 \kms. The red curve represents the best-fit 2 Gaussian components model: the combination of a
		core component (blue) and a broad component (green) is needed to properly reproduce the data. Labels
		indicate the number of stacked sources and the luminosity of the broad \cii\ wings. The inset shows a zoom
		on the broad component. \textit{Third panel from top:} residuals from the subtraction of the core component (blue line in the second panel). The green curve shows the best fit broad component. \textit{Fourth panel:} residuals from the two Gaussian components fitting. The 1$\sigma$ rms of the spectrum is also indicated by the shaded region.}
	\label{fig:stackedspec}
\end{figure}

\section{Results}

\subsection{Stacked spectrum} \label{sect:totstack}

The integrated spectrum resulting from the stack of all 48 QSO individual (1D) spectra
in our sample is shown in Fig. \ref{fig:stackedspec}. The stacked spectrum reveals very broad wings beneath the narrow line core, tracing fast outflows of cold gas.

We modelled the spectrum with a single Gaussian component. The $\chi^2$ minimisation of the fit was performed by using for each channel the weight $W^\prime_{\rm k}$ (see Eq. 1) and all the model parameters were free to vary in the fit with no constrains. However, a single Gaussian could not account for the emission at $v>500$ \kms. This can be seen in the first bottom panel of Fig. \ref{fig:stackedspec} as also indicated by the resulting large $\chi^{2}_{\nu\rm,1G} = 8.6$.  The addition of a second unconstrained Gaussian component gives a $\chi^2_\nu =
3.7$ (see Fig. \ref{fig:stackedspec}), i.e. a factor of $\gtrsim2$ smaller, which indicates that the second, broad Gaussian component is required with very high confidence level ($>99.9$\%). The reduced $\chi^2$ is yet larger than unity, hence suggesting that the line profile might be more complex than two simple Gaussian components. Details on the fitting procedure, uncertainties and confidence ellipses associated with the parameters of the broad Gaussian modelling the \cii\ wings, are reported in Appendix A. In Appendix A we also show the results from a single-Gaussian model fitting, indicating the reliability of the broad component.

The significance estimated through the fitting analysis is $5.6\sigma$. The significance based on the simple integration of the flux associated with the broad component (and the statistical uncertainty calculated in the same spectral region) gives a significance of $\sim10\sigma$. This is a pure statistical significance of the broad signal, higher than the confidence obtained from the fit, as it does not take into account the uncertainties associated with the subtraction of the narrow component. However, even ignoring the central channels affected by the core, the statistical significance of the wings alone is $\sim7\sigma$ (see Table \ref{tab:outflow}). 

The median rms of the stacked spectrum is $\sim0.06$ mJy/beam (see Table \ref{tab:outflow}), which is consistent with the noise expected by stacking the original spectra if the noise is Gaussian.
To give an idea of the significant improvement in sensitivity obtained with the stack, the sensitivity level reached in this work is a factor of $\sim14$ lower than that of the J1148$+$5251 observations of \cite{Cicone15}, where a massive \cii\ outflow was found.

\begin{figure*}[]
	\centering
	\includegraphics[width=0.367\textwidth]{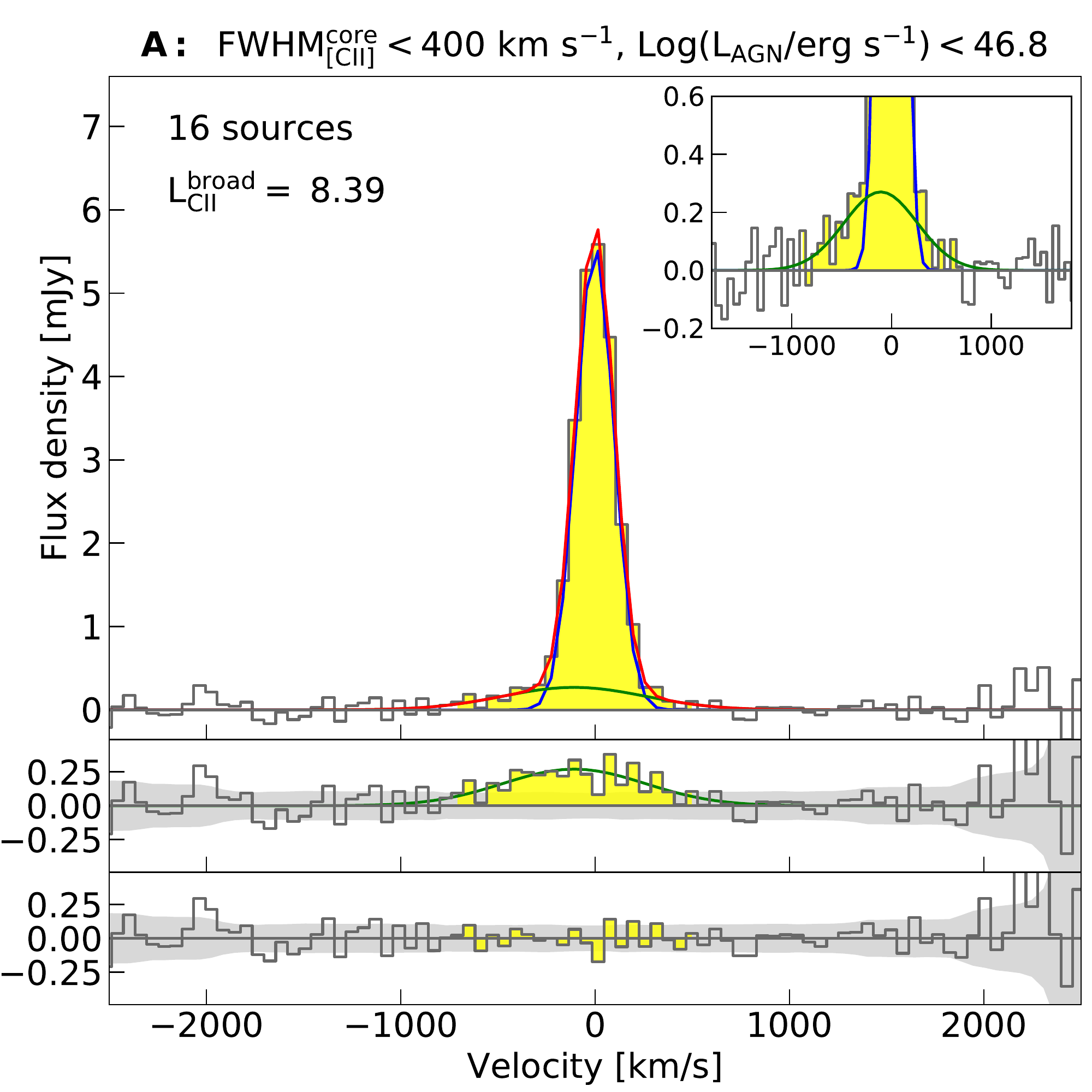}
	\includegraphics[width=0.367\textwidth]{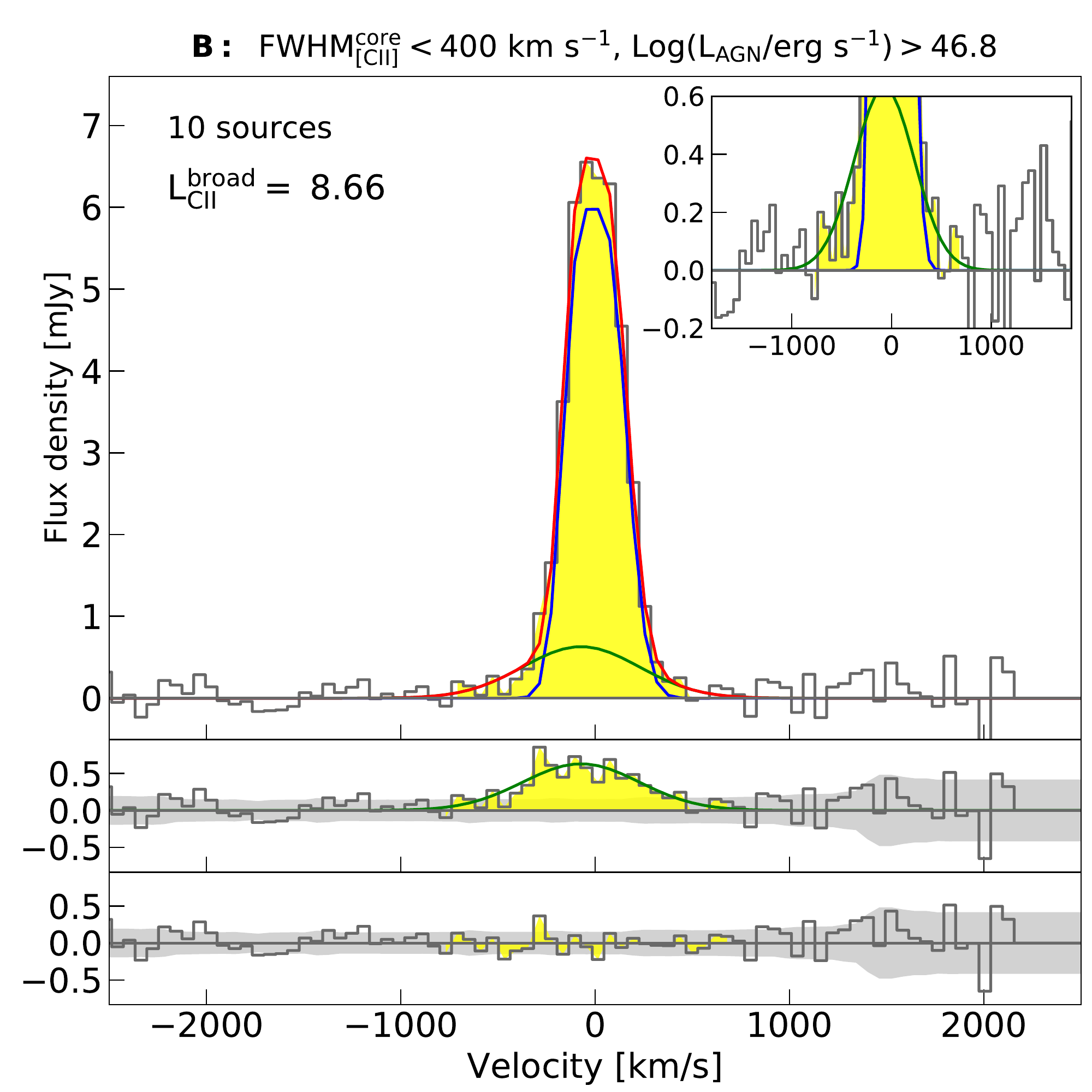}
	\includegraphics[width=0.367\textwidth]{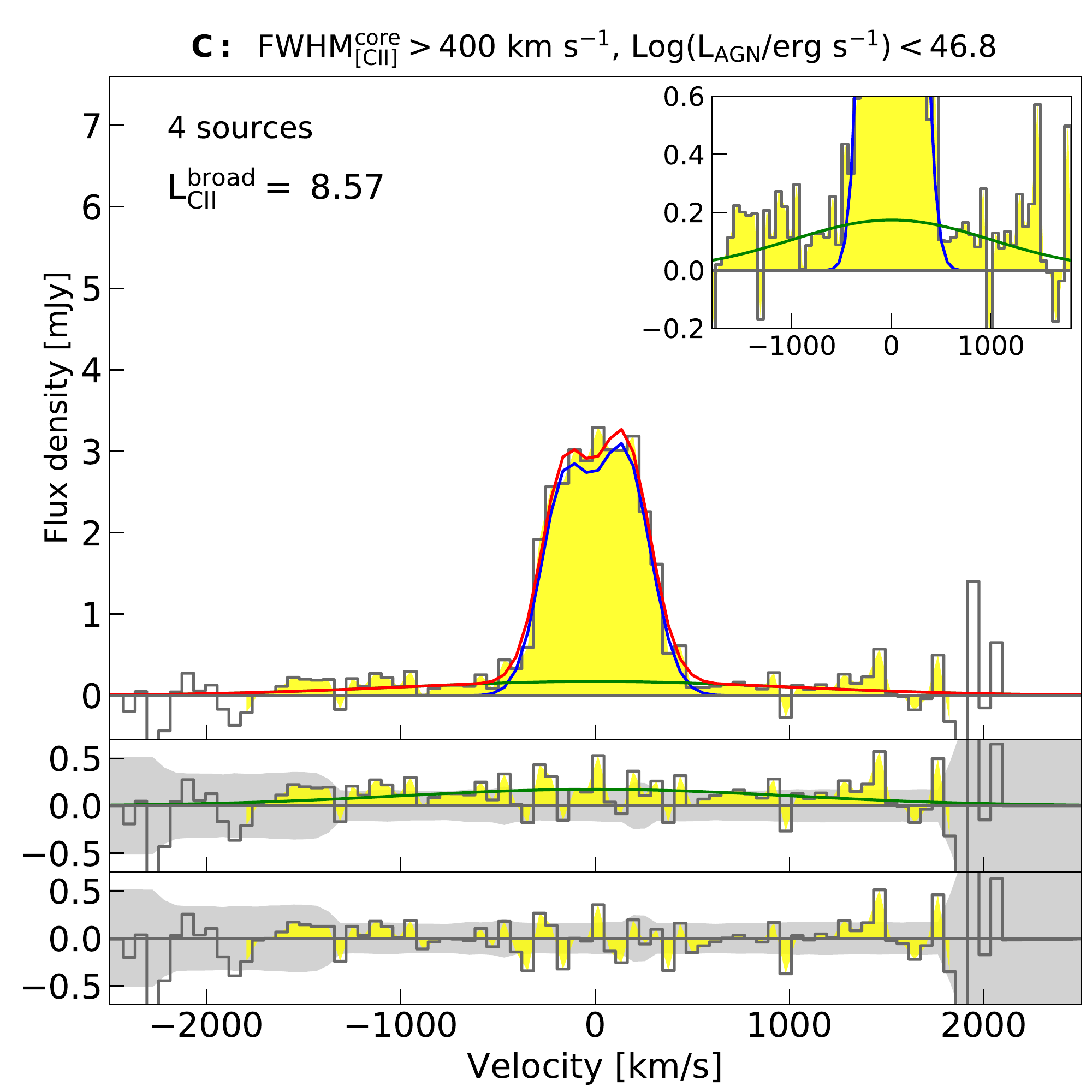}
	\includegraphics[width=0.367\textwidth]{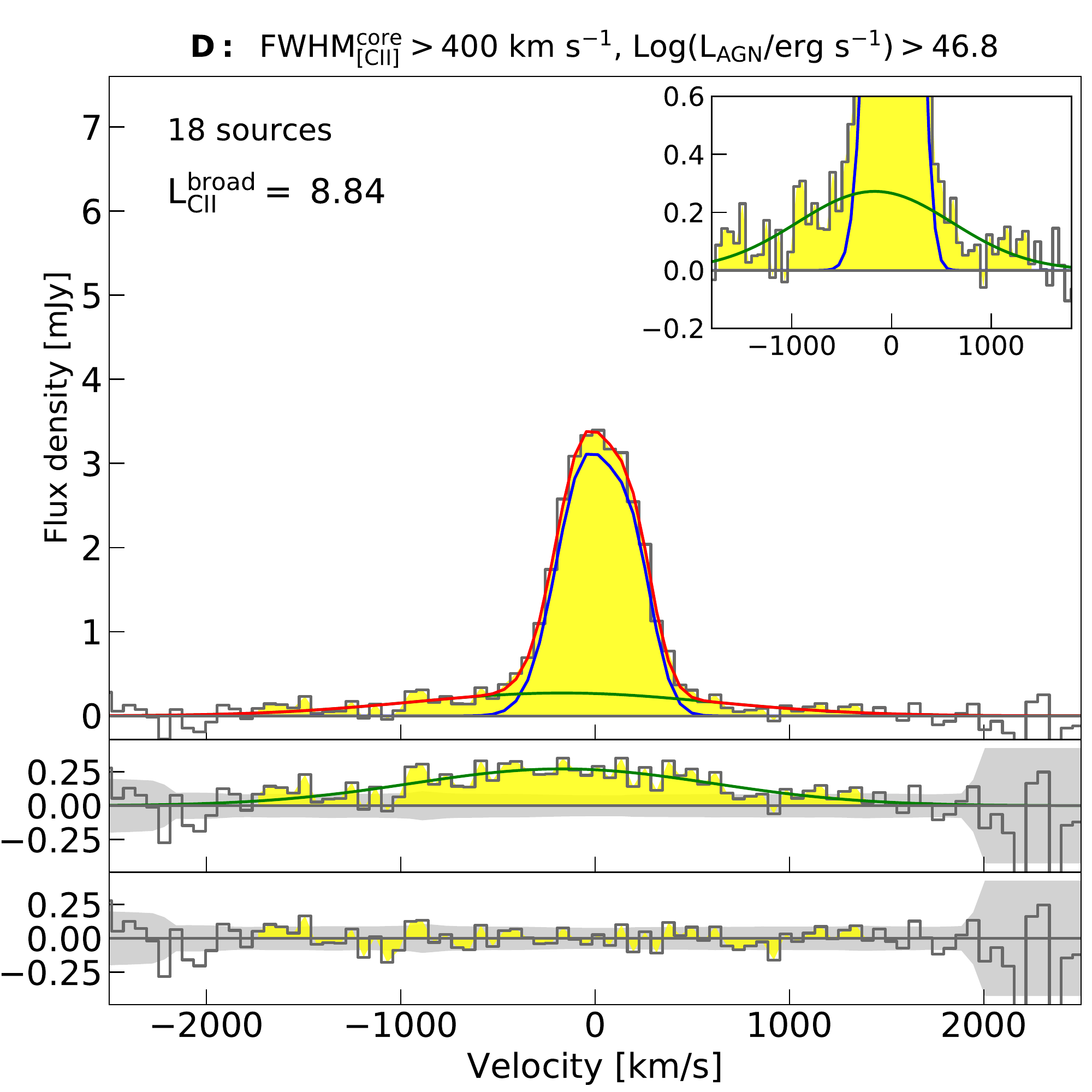}
	\caption{Stacked integrated spectra for the different QSOs subgroups \textit{A, B, C, D} (properties of the
		individual samples are indicated in the top labels). For each plot, the \textit{first panel from top} shows the \cii\
		flux density as a function of velocity, in bins of 60 \kms. The red curve represents the best-fit 2
		Gaussian components model; the two individual components are shown with blue and green curves.
		Labels indicate the number of stacked sources and the luminosity of the broad \cii\ wings. The inset
		shows a zoom on the broad component. \textit{Second panel from top:} residuals from the subtraction of the core component (blue line in first panel). The green curve shows the best fit broad component. \textit{Third panel:}  residuals from the two Gaussian components fitting. The 1$\sigma$ rms of the spectrum is also indicated by the shaded region.}
	\label{fig:s-fwhm-l-lbol-stack-fit}
\end{figure*}

In the stacked spectrum the core emission component has a width of FWHM$^{\rm core}_{\rm\cii}$ = 390 $\pm30$ \kms, while the underlying very broad component has a width of
FWHM$^{\rm broad}_{\rm\cii}$ = 1730 $\pm$ 210 \kms (see Table \ref{tab:outflow}).
The broad wings are not symmetric, the blue side being much more prominent
than the red side, resulting in the overall broad Gaussian used to fit
the broad component being slightly blueshifted (by $\sim$ 90 \kms, see Table \ref{tab:outflow}) with respect to
the systemic \cii\ emission. This might be an artefact resulting from the
asymmetric distribution of the data, with the red wing being contributed by fewer
spectra than the blue wing (top panel of Fig. \ref{fig:stackedspec}).
Alternatively, at such early epochs, the host galaxies
of these hyper-luminous QSOs may be similar to extreme ultra-luminous infrared
galaxies (ULIRGs), which have been found to be optically thick even at far-IR and sub-millimetre
wavelengths \citep{Papadopoulos10,Neri14,Gullberg15}, which
may result into absorption of the receding (redshifted) component of the outflow, even at the wavelength of [CII].
The latter interpretation is supported by the fact that, as we will show in Sect. \ref{sect:subsamples}, when we produce stacks
by splitting the
sample between galaxies with high and low SFR (hence high/low gas and dust content), the stack associated with
low SFR (hence low dust content) does not show a blueshift of the broad component, while the
sample with high SFR (hence high dust content) exhibits a large blueshift of the [CII] broad component.

The peak flux density of the broad \cii\ component is about 5\%  of that of the core, while the integrated broad-to-narrow \cii\
flux density ratio is $f_{\rm [CII]}\sim0.22$ (see Table \ref{tab:outflow}).
In order to estimate the luminosity of the [CII] broad component representative of our sample,
we computed a weighted $L_{\rm [CII]}^{\rm stack}$ by
applying Eqs. (1) and (2) to the individual narrow \cii\ luminosities of our targets. We therefore derived the luminosity associated with the broad \cii\ wings as $L_{\rm [CII]}^{\rm broad}=L_{\rm [CII]}^{\rm stack}\times f_{\rm\cii}$.

The individual \cii\ spectra contributing to  the stacked spectrum are characterised by FWHM of the line core in the range $\sim150-800$ \kms\ (Table \ref{tab:sample}). The combination of line profiles with different widths may potentially result into a stacked profile similar to the combination of a narrow and a broad Gaussian curve. Although the latter could not be as broad as the wings observed in the stacked spectrum of Fig. \ref{fig:stackedspec}
, this could still contribute to \lbroad. We quantified this contribution by stacking Gaussian curves with the same FWHM and \lcii\ distribution of the QSOs in our sample, according to Eq. (1) and Eq (2). In the stacked profile, we computed the \cii\ emission in excess of a single Gaussian curve to be $\sim17$~\% of \lbroad, indicating that the effect mentioned above can only have a marginal contribution to the flux of the broad component.

\begin{figure}[]
	\centering
	\includegraphics[width=1\columnwidth]{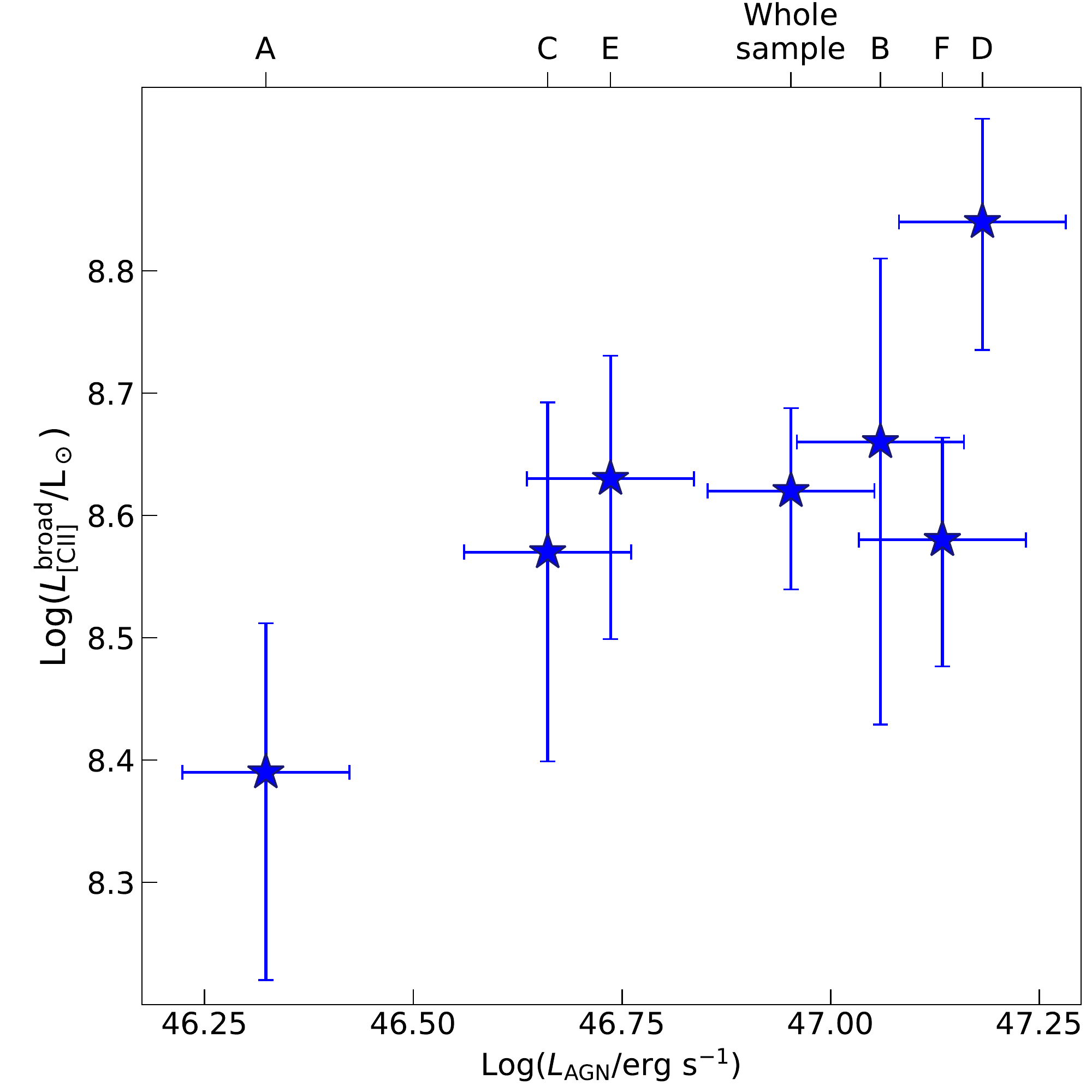}
	\caption{Broad \cii\ wings luminosity  as a function of the AGN bolometric luminosity for the different stacks performed (indicated by the top labels). Error bars on \lbol\ correspond to the 0.1 dex associated with the UV-based bolometric correction by \cite{Runnoe12}. }
	\label{fig:summary}
\end{figure} 

\begin{figure}[]
	\centering
	\includegraphics[width=0.8\columnwidth]{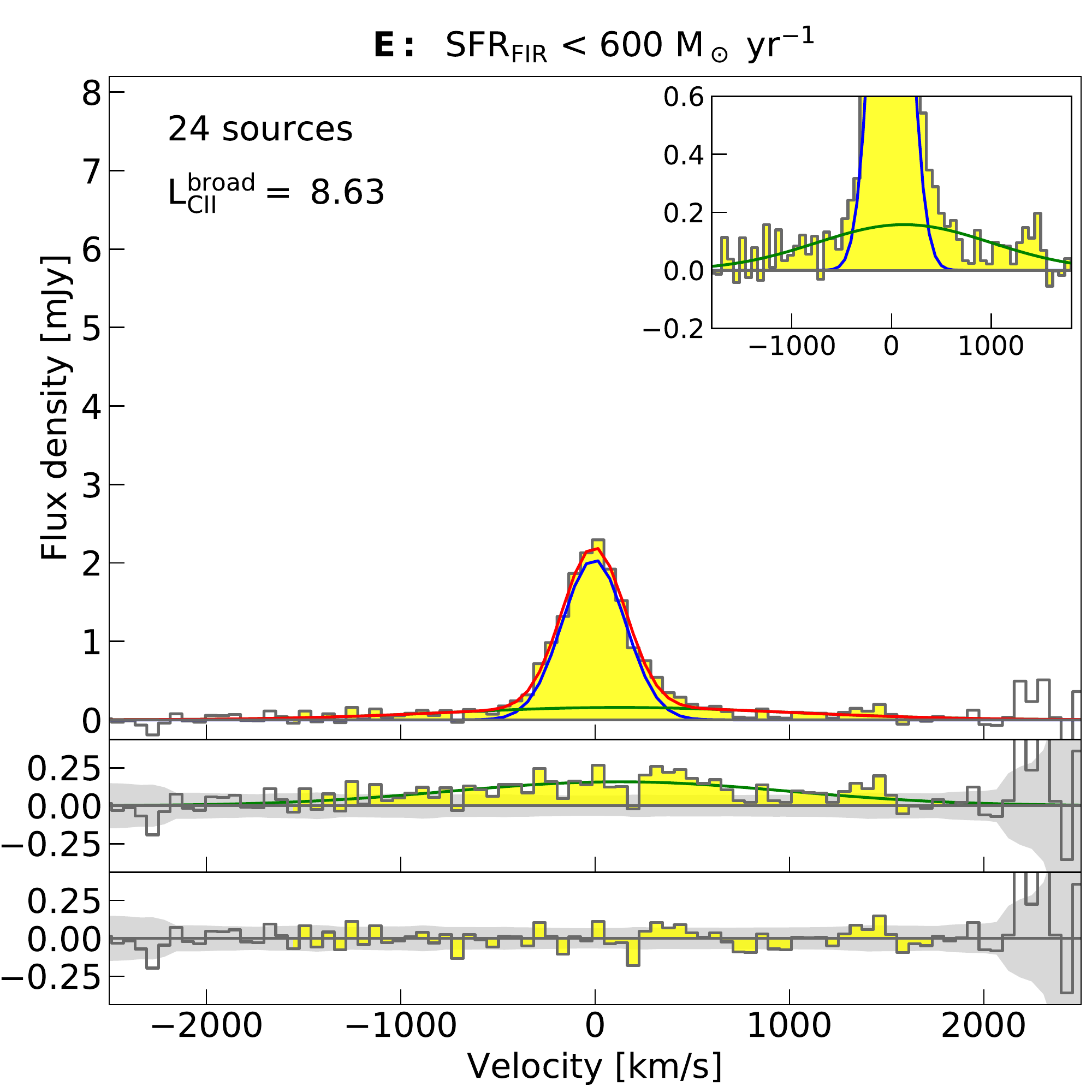}
	\includegraphics[width=0.8\columnwidth]{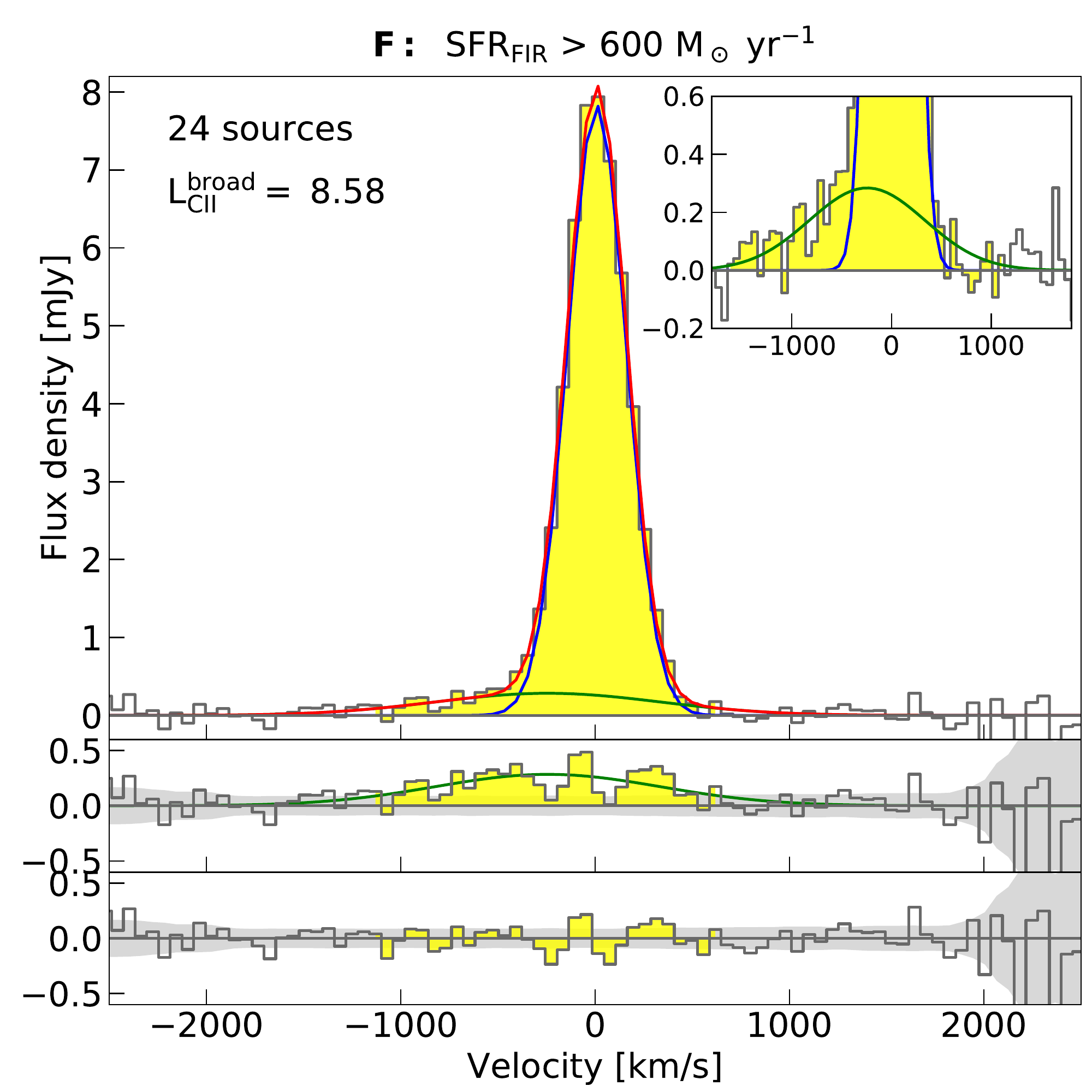}
	\caption{Stacked integrated spectra for the low SFR ($\textit{E}$) and high SFR ($\textit{F}$) subgroups
		(properties of the individual samples are indicated in the top labels). For each plot, the \textit{first panel from top} shows the \cii\ flux density as a function of velocity, in bins of 60 \kms, and the associated
		2-Gaussians best-fit model; the two individual components are shown with blue and green curves. Labels indicate the number of stacked sources and the luminosity of the broad \cii\ wings. The inset
		shows a zoom on the broad component. \textit{Second panel from top:} residuals from the subtraction of the core component (blue line in first panel). The green curve shows the best fit broad component. \textit{Third panel:} residuals from the two Gaussian components fitting. The 1$\sigma$ rms of the spectrum is also indicated by the shaded region.}
	\label{fig:sfr-stack}	
\end{figure}

\subsection{Outflow relation with QSO-galaxy properties}\label{sect:subsamples}

In this section, we study the relation between the presence of cold outflows traced by the \cii\ wings and the properties of the QSO-host galaxy system, such as AGN luminosity and SFR. 
Furthermore, to investigate the presence of broad \cii\ wings without the shortcomings of combining \cii\ profiles with significantly different widths, we
separate the QSOs in two subsamples with \fwhm$<400$ \kms\ (median linewidth of the whole sample) and \fwhm$>400$
\kms, respectively. This roughly corresponds to discriminate between less and more massive systems, given that the \cii\ linewidth is a proxy of the dynamical mass of the galaxy (modulo disc inclination effects). We further separate our sample in two AGN luminosity bins: specifically \lbol$<10^{46.8}$ \ergs\ (median \lbol\ of the whole sample) and \lbol$>10^{46.8}$ \ergs. This allowed us to investigate the relation between the \cii\ outflow strength and \lbol.
For simplicity, hereafter the different subsamples will be referred to as:  
\begin{itemize}
	\item \textit{A}:\hspace{0.1cm} \fwhm$<$ 400 \kms, \lbol$<10^{46.8}$ \ergs
	\item \textit{B}:\hspace{0.1cm} \fwhm$<$ 400 \kms, \lbol$>10^{46.8}$ \ergs 
	\item \textit{C}:\hspace{0.1cm} \fwhm$>$ 400 \kms, \lbol$<10^{46.8}$ \ergs
	\item \textit{D}:\hspace{0.1cm} \fwhm$>$ 400 \kms, \lbol$>10^{46.8}$ \ergs
\end{itemize}

\noindent The stacked spectra for the different subsamples are shown in Fig. \ref{fig:s-fwhm-l-lbol-stack-fit}. Because of less
statistics, the sensitivity improvement is modest compared to the stack of the whole sample
(see Table \ref{tab:outflow})
and the individual source contribution to the stacked spectrum is more evident. This is particularly
evident in Fig. \ref{fig:s-fwhm-l-lbol-stack-fit} for stacks \textit{C} and \textit{D}, where the core of the stacked \cii\ profile is broadened by few sources exhibiting a rotation pattern in their \cii\ spectra. We fit \textit{A} and \textit{B} with a two Gaussian components model, while for stack \textit{D} and \textit{E} we use a combination of two Gaussians to account for the broadening of the \cii\ core, and a third Gaussian to reproduce the \cii\ wings. Similarly to sect. \ref{sect:stacked-cube}, all parameters were let free to very in the fit with no constrains. The best fit models are shown in Fig. \ref{fig:s-fwhm-l-lbol-stack-fit}.

A faint broad \cii\ emission component is still observed in the stacked spectrum of the subgroups, although with lower significance if compared to the whole sample stack presented in Sect. \ref{sect:totstack} and, in few cases, with only marginal significance (see Table \ref{tab:outflow}).
In sources with small FWHM of the [CII] core emission  (stacks \textit{A} and \textit{B}), wings are characterised by
FWHM$_{\rm\cii}^{\rm broad}$ up to $\sim850$ \kms, while broader wings with FWHM$_{\rm\cii}^{\rm broad} \sim 2000$ \kms\
are observed in the subsamples with broader [CII] cores (stacks \textit{C} and \textit{D}).
Similarly to what we found in the whole sample stack, the peak of the broad \cii\ wings is 5\% to 10\% of the core peak flux density, while the integrated flux of the broad component corresponds to $20-30$\% of the core component.
Following the same method presented in Sect. \ref{sect:methods} for the stack of the total sample, the average signal-to-noise ratio of a spurious broad component is $\sim0.4-0.6$.

For each subsample, \lbroad\ and \lbol\ have been computed following the same method of Sect. \ref{sect:totstack} (see
Table \ref{tab:outflow}). We observe an increased \lbroad\ in the high \lbol\ sources (see Fig. \ref{fig:s-fwhm-l-lbol-stack-fit}), despite the limited luminosity range
spanned by the sources considered in our analysis. Fig. \ref{fig:summary} shows indeed that the stacked \lbroad\ follow a trend with \lbol\ similar to what observed by previous works in individual sources at lower redshift \citep[e.g.][]{Cicone14,Fiore17,Flutsch18}, finding that the outflow strength correlates with the AGN luminosity. This result indicates that the observed \cii\ outflows are primarily QSO-driven.
Instead, we see only marginal variations of  \lbroad\ with respect to the width of the line core, indicating that the dynamics of the galactic disc does not significantly affect the detectability  of the broad \cii\ components associated with the outflow.

An alternative driving mechanism of the fast \cii\ emission could be the starburst in the QSO host galaxy through supernovae and radiation pressure.
To investigate this possibility in more detail, we considered two subgroups according to their SFR, as inferred from their
\lfir\ (computed following \cite{Kennicutt12}), assuming that the bulk of the far-IR emission is associated with SF in the
host galaxy:
\begin{itemize}
	\item\textit{E}: SFR$_{\rm FIR}$ $<$ 600 \msunyr
	\item\textit{F}: SFR$_{\rm FIR}$ $>$ 600 \msunyr
\end{itemize}

The corresponding stacked spectra are shown in Fig. \ref{fig:sfr-stack}. It is evident that the \cii\ core emission is
mainly associated with SF activity, confirming that [CII] is a tracer of star formation as previously found by e.g. \cite{DeLooze14,HerreraCamus15}. The \cii\ flux density of the core in
stack \textit{E}, characterised by a variance-weighted SFR of 260 \msunyr, is in fact a factor of $\sim3.5$ lower with respect to
the highly star forming sources stacked in \textit{F} (whose variance-weighted SFR is $\sim$ 1750 \msunyr). However, the broad
\cii\ wings are well present in both stacks with comparable luminosity \lbroad$\sim4\times10^{8}$ \lsun, indicating that SF
does not significantly contribute to the outflows in the hosts of these powerful QSOs. 
Interestingly, as mentioned,
we observe the largest blueshift ($\sim240$ \kms) of the broad \cii\ wings in the high-SFR QSOs, which are those hosted
in dustier galaxies, hence possibly corroborating the interpretation that the blueshift of the [CII] broad component
is associated with heavy obscuration by the host galaxy. For stacks {\it E} and {\it F} we calculate an average signal-to-noise ratio of a spurious broad component of $\sim0.4$ and $\sim0.5$, respectively (see Sect. \ref{sect:methods}). Moreover, we estimate the contribution to \lbroad\ due to the combination of \cii\ profiles with different FWHM to be only $\sim10$\% for stack \textit{E} and $\sim20$\% for stack \textit{F}. 

\subsection{Outflow detectability}\label{sect:detectability}

As mentioned in Sect. \ref{sect:intro}, up to date few tens of high-z QSOs have been targeted in \cii, some of them with deep ALMA observations. Despite this fact, J1148$+$5251 remains the only source where a massive cold outflow was detected by \cite{Maiolino12} and \cite{Cicone15}. 
	
Among the deepest observations \cite{Venemans17} targeted  the \cii\ emission in the $z=7.1$ QSO J1120$+$0641, with no detection of fast \cii\ emission. The sensitivity reached by \cite{Venemans17} is comparable to that of our B and C subgroups, where the broad \cii\ wings are only marginally detected (see Sect. \ref{sect:subsamples}).
A forthcoming work reaching similar depths (Carniani et al., in prep.) but exploiting configurations more sensitive to the
extended, diffuse emission (hence more suitable to detect extended outflows), will present two QSOs were \cii\ fast emission associated with AGN-driven
outflows may be present. Similarly to our work, \cite{Decarli18}  computed the variance-weighted stacked spectrum of a sample of 23 ALMA
\cii-detected QSOs but finding no emission in excess of a Gaussian profile. However, the observations presented by \cite{Decarli18} consist of very
short ($\sim8$ min) integrations and, therefore, sensitivities from $\sim$0.5 to 1.0 mJy beam$^{-1}$ covering the high-rms half of our sample. These observations correspond to $\sim10\%$ of the total on-source time covered by the QSOs in our sample. By applying our stacking procedure (Sect. \ref{sect:methods}) to the \citet{Decarli18} sample alone, we derive a median rms sensitivity of 0.17 mJy/beam, comparable to that reached by subsamples \textit{B} and \textit{C}, in which we find only marginal presence of \cii\ wings. We note that our approach differs from that by \citet{Decarli18} as they extract the individual QSOs spectra from the brightest pixel, while we choose extraction apertures of four ALMA beams recovering emission from extended scales. By fitting a two-Gaussian components model to the resulting stacked spectrum, we find a SNR$_{\rm [CII]}^{\rm broad}\sim2$. Accordingly, we agree with \citet{Decarli18} in finding no clear evidence of outflow signatures when stacking only their sample.

\begin{figure*}[htb]
	\centering		
	\includegraphics[width=1\linewidth]{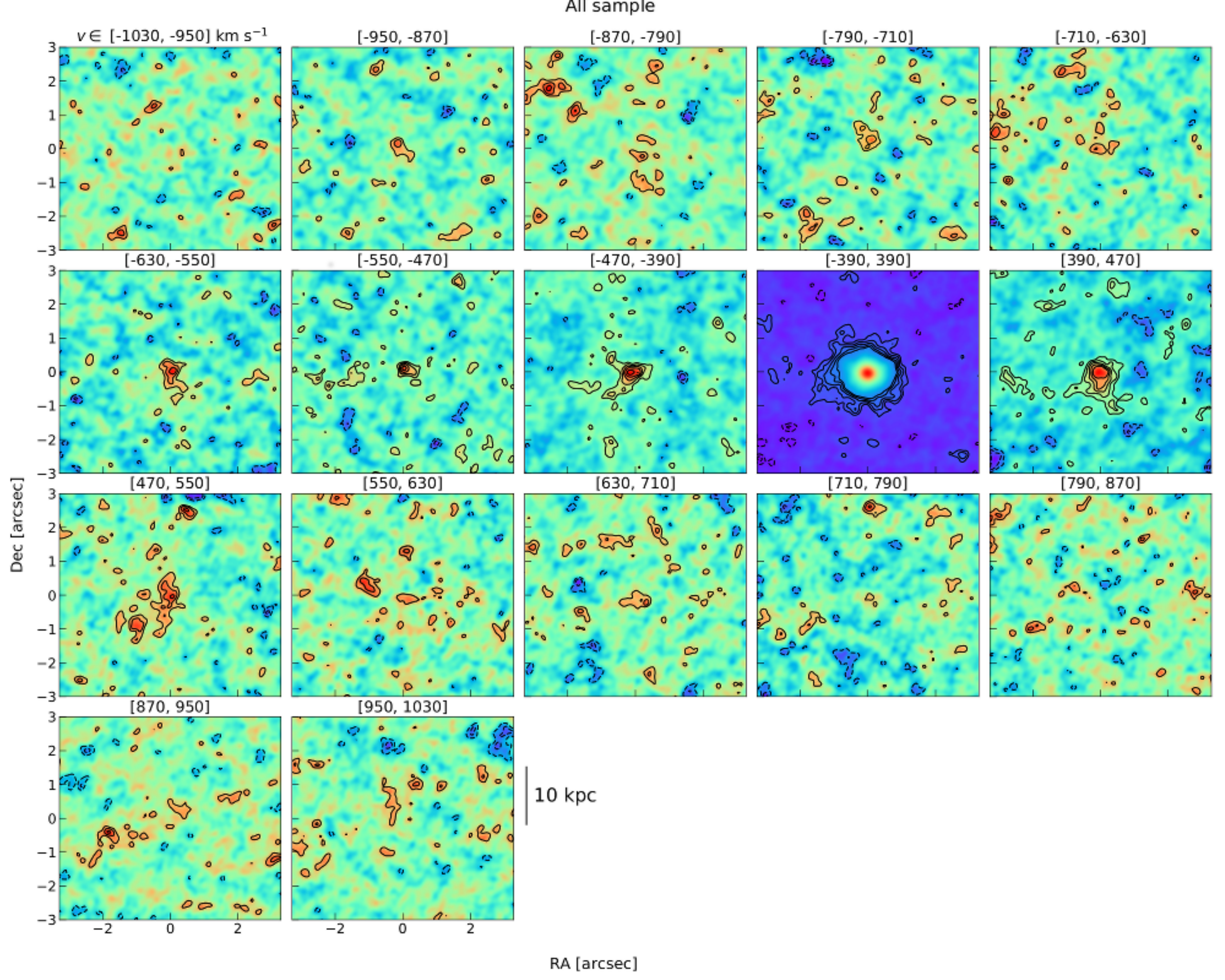}
	\caption{Channel maps of the whole sample stacked cube, corresponding to the central $6''\times6''$ in the
		velocity range $v\in$ [$-$1000, 1000] \kms\ (in bins of 80 \kms, as indicated by the top labels). The bulk
		of the \cii\ core emission is collapsed in the channel $v\in$[-390, 390] \kms. Contours correspond to [-3,
		-2, 2, 3, 4, 5, 6]$\sigma$, where $\sigma$ is the rms sensitivity evaluated for each channel.}
	\label{fig:stackedcube}
\end{figure*}

In our stacked spectra we limited the contamination from companions that are usually observed around a fraction of high-z, high-luminosity QSOs, that can be as high as 50\% \citep[e.g.][]{Trakhtenbrot17, Fan2018}. Companions might in fact mimic a tail in the \cii\ line profile similar to the broad \cii\ wings indicative of outflowing gas. Most of these companions are located at much larger angular separations than the extraction regions of our spectra (see Sect. \ref{sect:totstack}) and, therefore, do not contaminate our spectra. However, a few of the high-z QSOs in our sample are known to have close companions with angular separation of about 1-2 arcsec which, despite the small extraction region used in our work, may partly contaminate the QSO emission. 
Specifically, the QSOs PJ231$-$20 and PJ308$-$21 from \cite{Decarli17} show a close companion galaxy with a [CII] luminosity
comparable to that of the QSO. In both cases the companion is slightly redshifted and may consequently contaminate the red wing of the \cii\ stacked spectrum. The QSO PJ167$-$13 presented by \cite{Willott17} is also likely associated with a companion at 0.9 arcsec separation, whose \cii\ blueshifted ($\sim270$ \kms) emission corresponds to about 20\% of the QSO \cii\ luminosity. As mentioned in Sect. \ref{sect:methods}, individual sources do not significantly affect the luminosity of the whole sample stacked spectrum. As further verification, we also excluded these particular three QSOs from the stack and found no significant variation in the luminosity of the broad \cii\ wings.

In our stacking procedure we assumed no relation between the luminosity of the broad \cii\ wings and the AGN-host galaxy properties, such as the luminosity of the core of the \cii\ emission line. Therefore, as mentioned in Sect. \ref{sect:methods}, we performed a standard variance weighted stack in flux density units. However, as our sample spans a factor of $\sim1.7$ in luminosity distance, we verified that our results persist if performing a stack in luminosity density (see Appendix B). We derived a \lbroad$\sim4.7\times10^8$ \lsun, comparable to that found in the original stacked spectrum (Table \ref{tab:outflow}).

\begin{figure*}[htb]
	\centering		
	\includegraphics[width=0.8\linewidth]{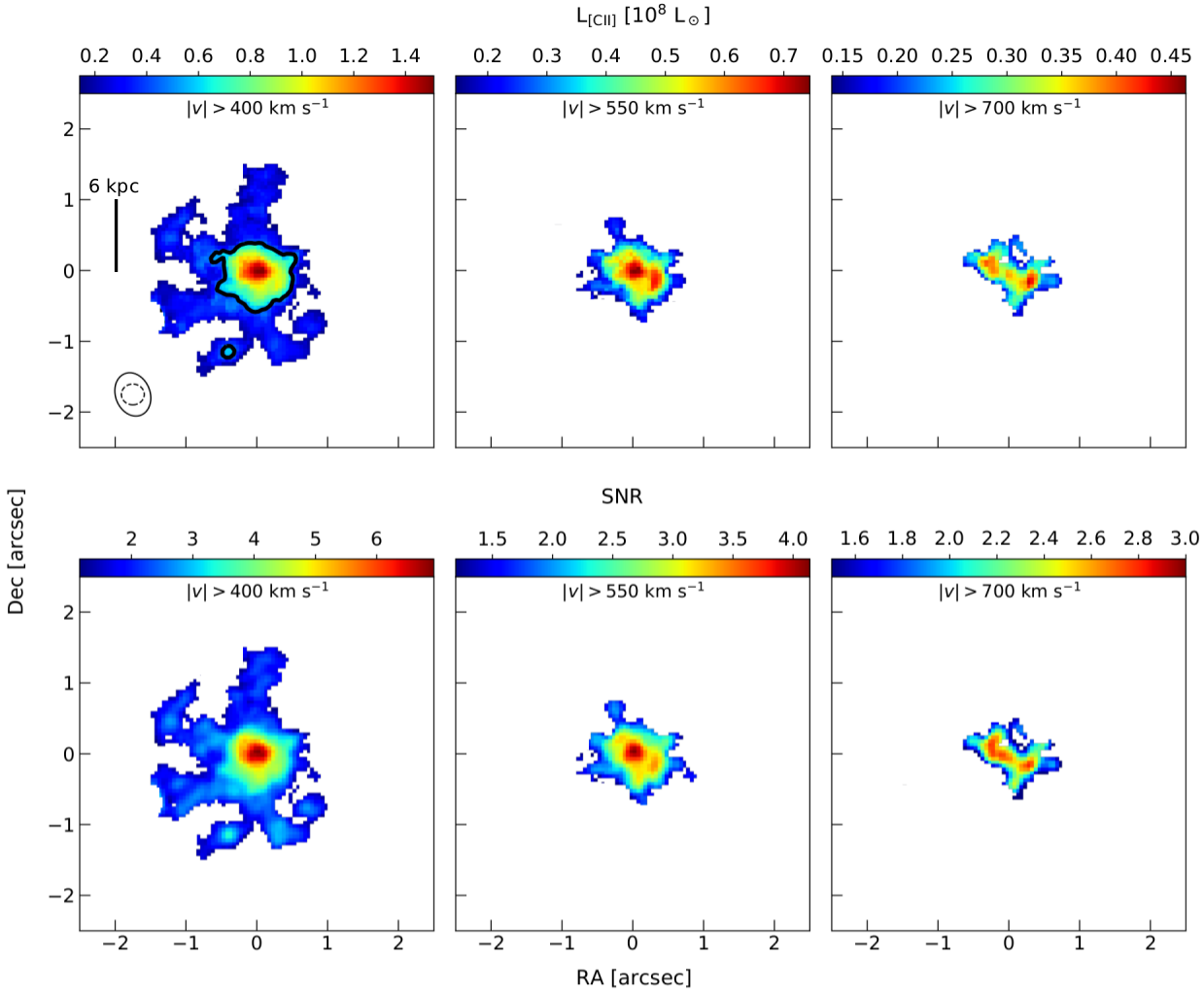}	
	\caption{\textit{Top:} luminosity maps of the high-velocity \cii\ emission derived from the whole sample
		stacked cube. From left to right, panels correspond to emission at increasing absolute velocities,
		specifically $|v|>400$ \kms, $|v|>550$ \kms\ and $|v|>700$ \kms. Maps were obtained by summing the emission
		at $>3\sigma$ in 80 \kms\ channel maps for at least three channels (i.e. $\gtrsim$ 250 \kms). The
		variance-weighted beam of the stacked cube is also indicated in the first map (solid line), together with
		the smallest beam contributing to the stack (dashed line). The thick solid contour encloses the region from
		which 50\% of \lbroad\ arises. \textit{Bottom:} signal-to-noise maps associated with the different velocity
		bins.}
	\label{fig:summaps}
\end{figure*}

\begin{figure*}[htb]
	\centering
	\floatbox[{\capbeside\thisfloatsetup{capbesideposition={left,center},capbesidewidth=5cm}}]{figure}[\FBwidth]
	{\caption{Channel maps of the stacked cube obtained by removing the higher resolution ($<0.6$ arsec) observations (see Sect. \ref{sect:sample}) and tapering the remaining data to a common resolution of 1.2 arcsec. The displayed region corresponds to the central $6''\times6''$ in the velocity range $v\in$ [$-$1000, 1000] \kms\ (in bins of 80 \kms, as indicated by the top labels). The bulk of the \cii\ core emission is collapsed in the channel $v\in$[-390, 390] \kms. Contours correspond to [-3,-2, 2, 3, 4, 5, 6]$\sigma$, where $\sigma$ is the rms sensitivity evaluated for each channel.}\label{fig:tapered_maps}}
	{\includegraphics[width=0.7\textwidth]{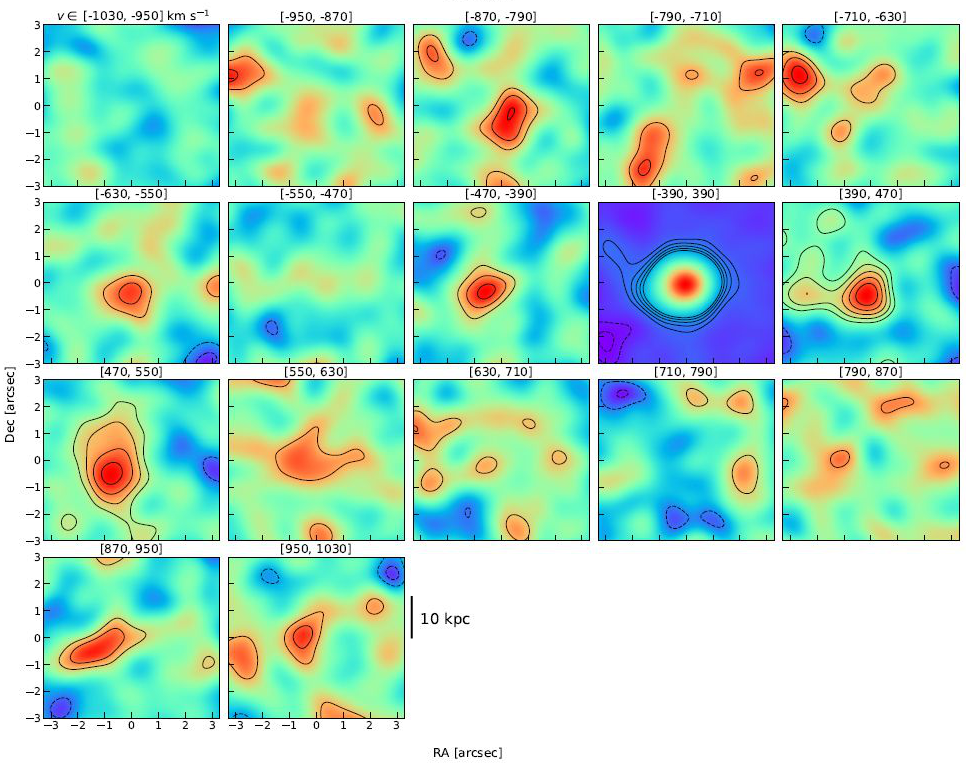}}
	\vspace{-0.5cm}	
\end{figure*}
\begin{figure*}[htb]
	\centering
	\includegraphics[width=0.7\textwidth]{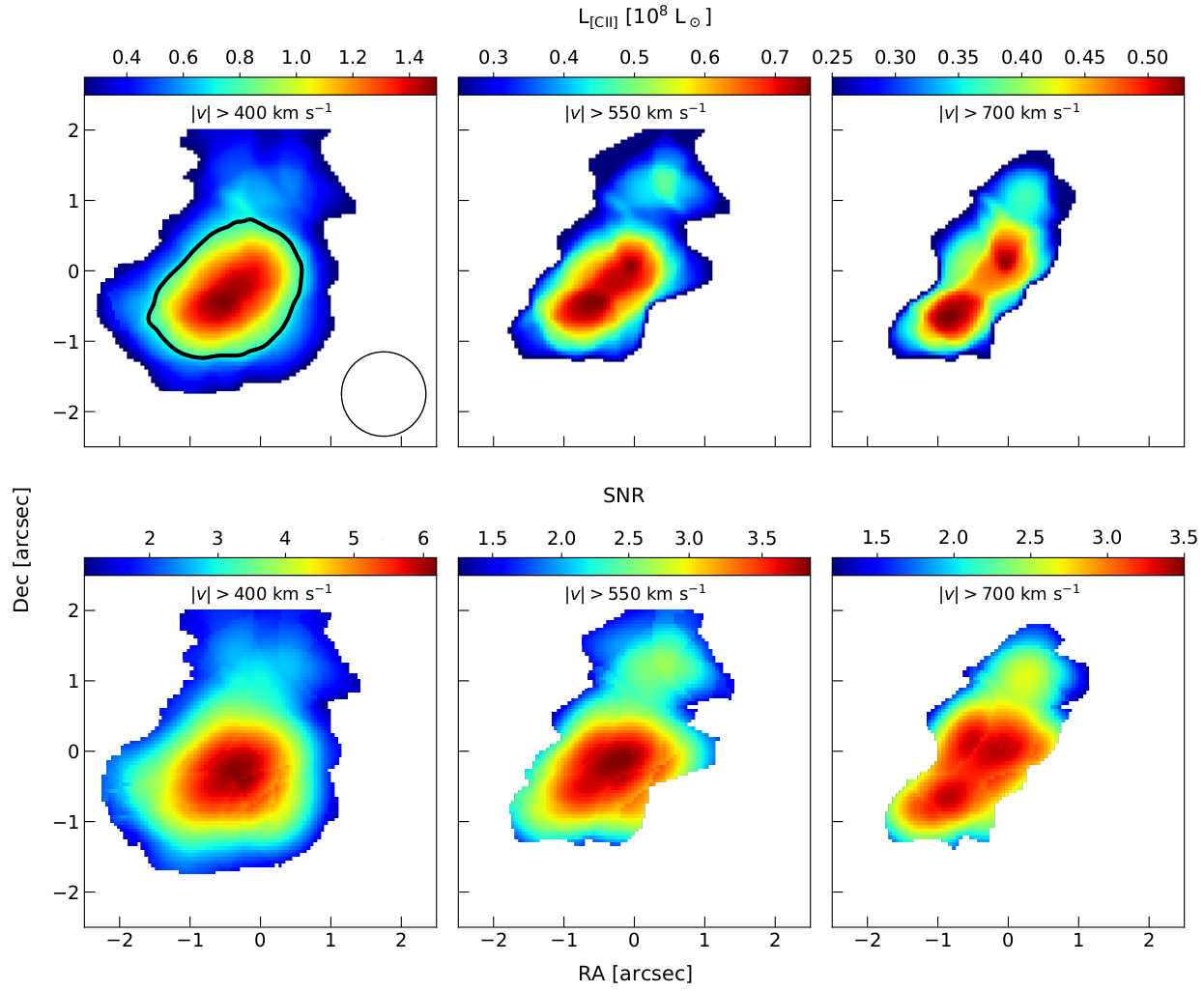}
	\caption{\textit{Top:} luminosity maps of the high-velocity \cii\ emission derived from the stacked cube of the $\gtrsim0.6$ arcsec sample, after having applied a tapering to a common 1.2 arcsec resolution. From left to right, panels correspond to emission at increasing absolute velocities, specifically $|v|>400$ \kms, $|v|>550$ \kms\ and $|v|>700$ \kms. Maps were obtained by summing the emission at $>3\sigma$ in 80 \kms\ channel maps for at least three channels (i.e. $\gtrsim$ 250 \kms). The 1.2 arcsec beam is also indicated in the first map (solid line). The thick solid contour encloses the region from which 50\% of \lbroad\ arises. \textit{Bottom:} signal-to-noise maps associated with the different velocity bins.}
	\label{fig:tapered_total}
\end{figure*}

\subsection{Stacked Cube}\label{sect:stacked-cube}

In this section we present the results from the stacking of the ALMA data cubes for the QSOs in our sample. We produced a stacked cube by applying the stacking technique presented in Sect. \ref{sect:methods}, i.e. we used Eqs. (1) and (2) to compute the variance-weighted stacked flux density of each spaxel in the final cube. 
It is a very simple approach primarily aimed at investigating the spatial scale of the \cii\ outflows in the high-$z$ QSOs host galaxies. Differently from the analysis of the integrated spectra of Sect. \ref{sect:totstack}, here we did not not choose an extraction region but only piled up the emission contributions to each pixel of the map.

However, the application of this stacking method to heterogeneous observations
may lead to a few issues, as discussed in the following:
\begin{itemize}

	\item Combining observations with different angular resolutions (see Fig. \ref{fig:rmshisto}) implies that emission from different physical scales may contribute to the total flux density of a same pixel.
	Degrading the observations to the lowest angular resolution would allow to stack emission arising from similar physical scales (given that the physical-to-angular scale ratio changes only by a factor $\lesssim1.3$ in the redshift range of our sources). However, in interferometric data this would imply tapering the visibilities, i.e. lowering the weight of the extended baselines to the final map, at the expenses of the sensitivity. Therefore, we preferred not to modify the angular resolutions. 
	However, by computing the variance-weighted beamsize of our stacked cube,
	it is possible to have an indication of the angular scale above which the
	most of the emission is resolved. For the all-sample stacked cube we
	computed an average
	angular resolution $\theta_{\rm res} = 0.52'' \times 0.68''$.
	In the case of a point source emission, the flux density contribution to the scale of the beam axes is only $\sim$6\%.
	We may therefore safely assess that emission on larger scales is mainly
	associated with extended, resolved emission.

	\item Lacking {\it a-priori}
	information about the structure and orientation of possible
	\cii\ extended emission, in particular at high velocities, may cause
	outflowing anisotropic or clumpy \cii\ emission to be diluted in the stack. As
	a consequence, the true fraction of \cii\ emission associated with extended structures may be significantly higher.

	\item The different angular resolutions of the interferometric observations are the
	result of different array configurations, which may filter out emission
	on different large
	angular scales. As a rough estimate the largest angular scale ($LAS$) that can be recovered by interferometric observations is
	$LAS\sim(4-6)\times\theta_{\rm res}$, where $\theta_{\rm res}$ is the
	angular resolution. In the case of our stacked cube the flux loss of
	extended emission due to filtering starts to become important
	at $\sim2$ arcsec.
	
\end{itemize}

\noindent Aware of these potential issues,
Fig. \ref{fig:stackedcube} shows the central $6'' \times 6''$ region of the stacked cube obtained by combining all the high-$z$ QSOs in our sample. Specifically, channel maps of the \cii\ emission are shown for a velocity range $v\in$ [$-$1000, 1000] \kms\, in bins of 80 \kms. The bulk of the \cii\ core emission is in the central $v\in$ [-390, 390] \kms. Compact \cii\ emission is observed in almost all channels at $\gtrsim2\sigma$ up to $\sim6\sigma$, in addition to the presence of few offset clumps. At $|v|\sim400-600$ \kms\ there is also some indication of extended \cii\ emission.

The channel maps of Fig. \ref{fig:stackedcube} suggest that we might be
observing  \cii\ emission clumps moving at different velocities and
characterised by a range of velocity dispersions. To build a global picture of
the \cii\ outflows, we created an integrated luminosity map of the high-velocity
\cii\ emission by summing the emission contributions, in the 80 \kms\ channel
maps, detected at $> 3\sigma$ significance for at least three channels (i.e.
$\gtrsim$ 250 \kms) in the whole-sample stacked cube. The result is shown in Fig.
\ref{fig:summaps}, where the maps corresponding to the velocity bins $|v|>400$
\kms, $|v|>550$ \kms\ and $|v|>700$ \kms\ are displayed. We also plotted the
associated signal-to-noise ratio maps.

As expected, most of the fast \cii\ emission arises from the central regions,
where all sources contribute in the stack. At the highest velocities
($>700$ \kms) the nuclear outflow is still present at $\sim3\sigma$ significance. At moderate velocities $|v|\sim400-550$ \kms,
we observe extended emission up to $\sim1.5$ arcsec, corresponding to $\sim9$ kpc at $\langle z_{\rm stack}\rangle = 5.8$,
and fully resolved compared with the average beam of the observations in the stack. We cannot exclude that part of this extended emission is due to contamination from the \cii\ core emission.
Marginally resolved emission is observed also at higher velocities. However, we stress that we might be losing a significant part of the extended emission in our stack.
It is interesting that the apparent size of the outflow appears to decrease as a function of velocity. This is
what is expected in an approximate spherically symmetric outflow as a consequence of projection effects. On the other hand, the stacked cube is a combination of outflows which may have different size, morphology and orientation between sources. The modest significance of the stacked data does not allow to draw conclusions on the outflow geometry.

Nonetheless, we can compute the spatial scale at which the bulk of the observed fast \cii\ is
emitted  as the half light radius of the $|v|>400$ \kms\ map (see Fig.
\ref{fig:summaps}), which has the highest SNR. This radius corresponds to the
average extent of the region enclosing 50\% of the \cii\ emission, indicated by
the black contour in Fig. \ref{fig:summaps}. We derived a beam-deconvolved half light radius $R_{\rm out} = \sqrt{(R_{\rm 50\%}^2 - R_{\rm beam}^2})\sim 0.60$ arcsec, where $R_{\rm beam}$ is the weighted beam radius of the stacked cube.
$R_{\rm out}$ corresponds to $\sim3.5$ kpc at $\langle z_{\rm stack}\rangle$.

To ensure that the presence of spatially extended \cii\ emission in the stacked cube was not an artefact of the combination of different ALMA beam sizes, we performed the whole sample stack of the data cubes after degrading the observations to a common angular resolution.
We produced ALMA cubes at the worse angular resolution in our sample by tapering the visibilities to 1.2 arcsec. We did not consider observations with angular resolution $<0.6''$, as tapering to a worse resolution by a factor of $>2$ would result into a major loss of the original flux. 
The channel maps from the resulting stacked cube are shown in Fig. \ref{fig:tapered_maps}. The corresponding integrated luminosity map of the high-velocity emission is shown in Fig. \ref{fig:tapered_total}. In agreement with Fig. \ref{fig:summaps}, we found that the high-velocity \cii\ emission is mainly located in a bright central component and extends up to $\sim1.5-2$ arcsec. Indeed, combining the observations associated with the shorter baselines in our sample allowed us to recover some extended \cii\ emission also in the high-velocity ($|v|>550$ \kms and $|v|>700$ \kms) bins. 
By following the same procedure presented in Sect. \ref{sect:stacked-cube}, we derived a beam-corrected half light radius $R_{\rm out}^{\rm taper} \sim 4.6$ kpc at the representative redshift of the stack $\langle z_{\rm stack}^{\rm taper}\rangle\simeq6.2$, slightly larger than the radius inferred from the stack of the total sample. 
 


\section{Discussion}

As mentioned in Sect. \ref{sect:intro}, most of \cii\ emission in IR-bright galaxies is expected to arise from PDRs \citep[e.g.][]{Sargsyan12}, accounting for about 70\% of the total \cii\ emission. Under the assumption that \cii\ emission is optically thin, it is possible to link the luminosity of the broad \cii\ wings to the mass of the outflowing atomic gas. In case of optically thick \cii, the true outflowing gas mass would be larger. It is therefore
possible to estimate the typical energetics of \cii\ outflows in high-redshift, high-luminosity QSOs, in the
central $\sim3$ kpc regions (see Sect. \ref{sect:stacked-cube}). Specifically, to compute the outflow mass of
atomic neutral gas we use the relation from \cite{Hailey-Dunsheath10}:

\begingroup\makeatletter\def\f@size{9.2}\check@mathfonts
\begin{equation}
M_{\rm out}/{\rm M_\odot} = 0.77 \left( \frac{0.7L_{\rm [CII]}}{\rm L_\odot} \right)\left( \frac{1.4\times10^{-4}}{X_{\rm C^+}} \right)\times\frac{1 + 2e^{-91K/T}+n_{\rm crit}/n}{2e^{-91K/T}}
\label{eq:m_cii}
\end{equation}
\endgroup

\noindent where $X_{\rm C^+}$ is the C$^+$ fraction per hydrogen atom, $T$ is the gas temperature, $n$ is the gas density and $n_{\rm crit}\sim3\times10^3$ cm$^{-3}$ is the \cii$\lambda$158 $\mu$m critical density. 
We use Eq. \ref{eq:m_cii} in the approximation of $n>>n_{\rm crit}$, thus deriving a lower limit on the outflowing gas mass. This choice is in agreement
with \cite{Maiolino05} who estimated a gas density of $\sim10^5$ cm$^{-3}$ in J1148$+$5251, but also confirmed by the large
densities typically observed in QSO outflows \citep{Aalto12,Aalto15}, and allows us to directly compare with the energetics of the outflow detected in this QSO. Following \cite{Maiolino12} and \cite{Cicone15} we consider a conservative $X_{\rm C^+}\sim10^{-4}$ and a gas temperature of 200 K, both typical of PDRs \citep{Hailey-Dunsheath10}.
We recall that, although the molecular gas phase in the ISM has typically lower temperatures, in the outflow even the molecular gas is expected to have higher temperatures, of a few 100 K \citep{RichingsGiguere18a}.
Assuming a temperature from 100 K to 1000 K would imply a variation of only 20\% in the resulting gas mass.
The 0.7 factor in the first parenthesis of Eq. \ref{eq:m_cii} accounts for the fraction of neutral [CII] typically arising from PDRs, while 30\%
typically comes from the partially ionised phase \citep{Stacey10, Maiolino12, Cicone14}.
By applying Eq. \ref{eq:m_cii} to the stack of the whole sample we infer a mass of the outflowing neutral gas 
$M_{\rm out} = (3.7\pm0.7) \times10^8$ M$_\odot$, see Table \ref{tab:outrate}.

To compute the \cii\ outflows energetics of the high-$z$ QSOs in our sample, we assume the
sceario of time-averaged expelled shells or clumps \citep{Rupke&Veilleux05b}:
\begin{equation}\label{eq:moutrate} 
\dot{M}_{\rm out} = \frac{v_{\rm out} \times M_{\rm out}}{R_{\rm out}}
\end{equation}
where $v_{\rm out}= |\Delta v_{\rm broad}| + \rm FWHM_{[CII]}^{broad}/2$ (see Table \ref{tab:outrate}) and
$\Delta v_{\rm broad}$ is the velocity shift of the centroid of the broad \cii\ wings with respect to the
systemic emission, $R_{\rm out}\sim3.5$ kpc as derived in Sect. \ref{sect:stacked-cube} from the extension
of the [CII] broad wings as inferred from the stacked cube. We calculate the kinetic power associated with the \cii\ outflows as: 
\begin{equation}
\dot{E}_{\rm out} = \frac{1}{2} \dot{M}_{\rm out} \times v_{\rm out}^2
\end{equation}
and the momentum load:
\begin{equation}\label{eq:pout}
\dot{P}_{\rm out}/\dot{P}_{\rm AGN} = \frac{\dot{M}_{\rm out}\times v_{\rm out}}{L_{\rm AGN}/c}
\end{equation}
where $\dot{P}_{\rm AGN}=L_{\rm AGN}/c$ is the AGN radiation momentum rate.  This approach allows us to directly compare our findings to the collection of 30 low redshift AGN by \citep{Flutsch18}, for which the energetics of spatially resolved molecular and (in $\sim$ one third of the same sources) neutral \cii\ and ionised outflows has been homogeneously calculated.

\begin{figure*}[]
	\centering
	\includegraphics[width=0.48\textwidth]{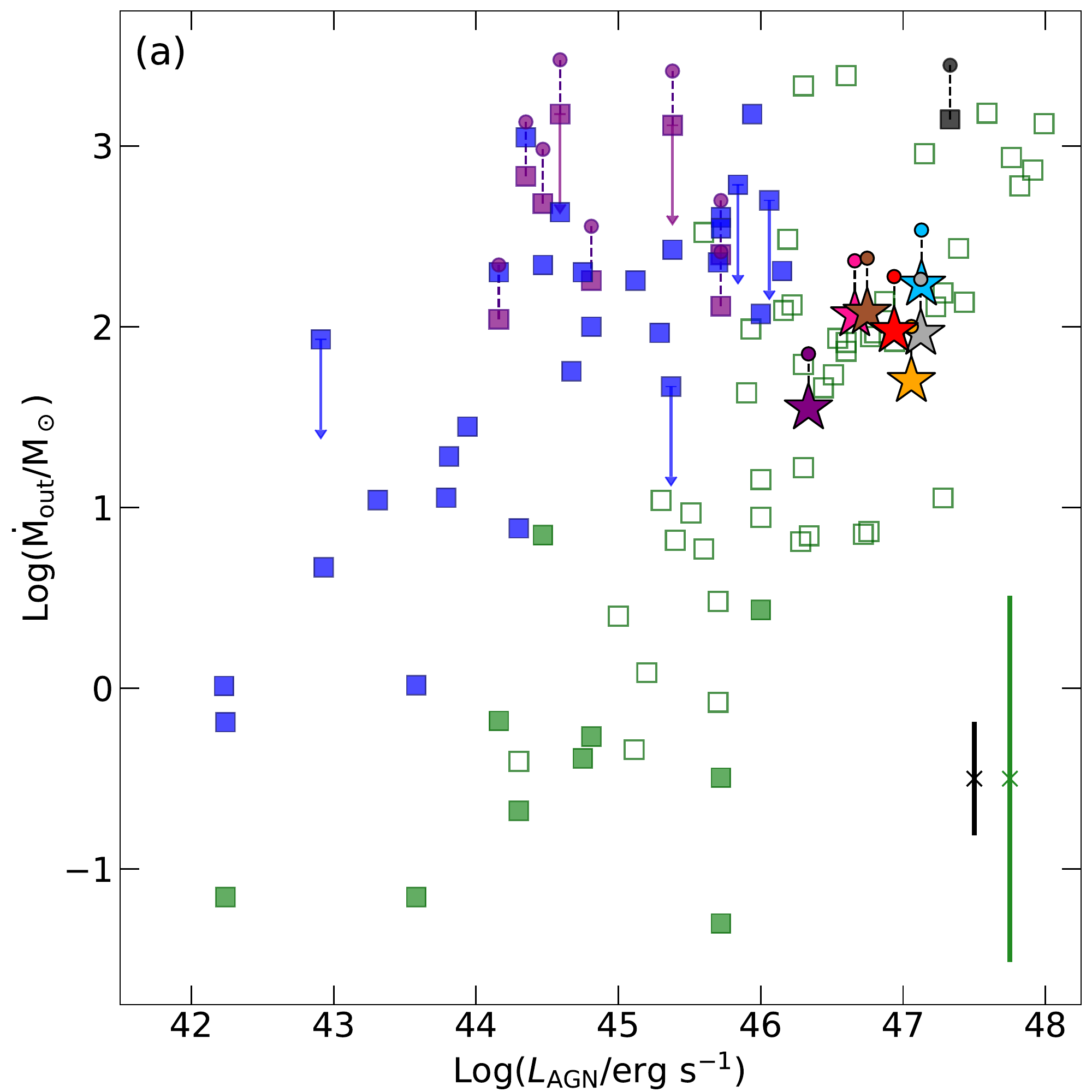}
	\includegraphics[width=0.48\textwidth]{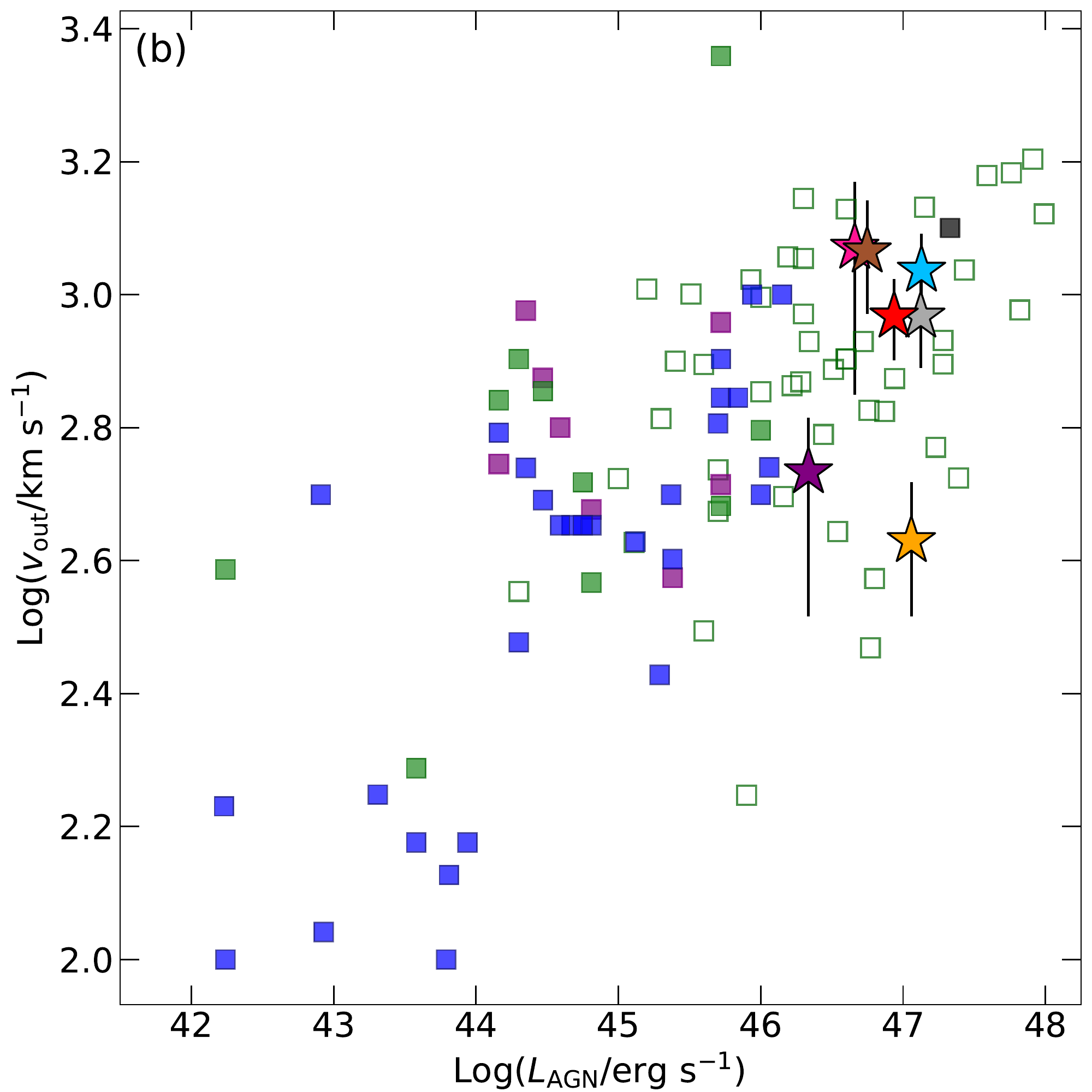}
	\includegraphics[width=0.96\textwidth]{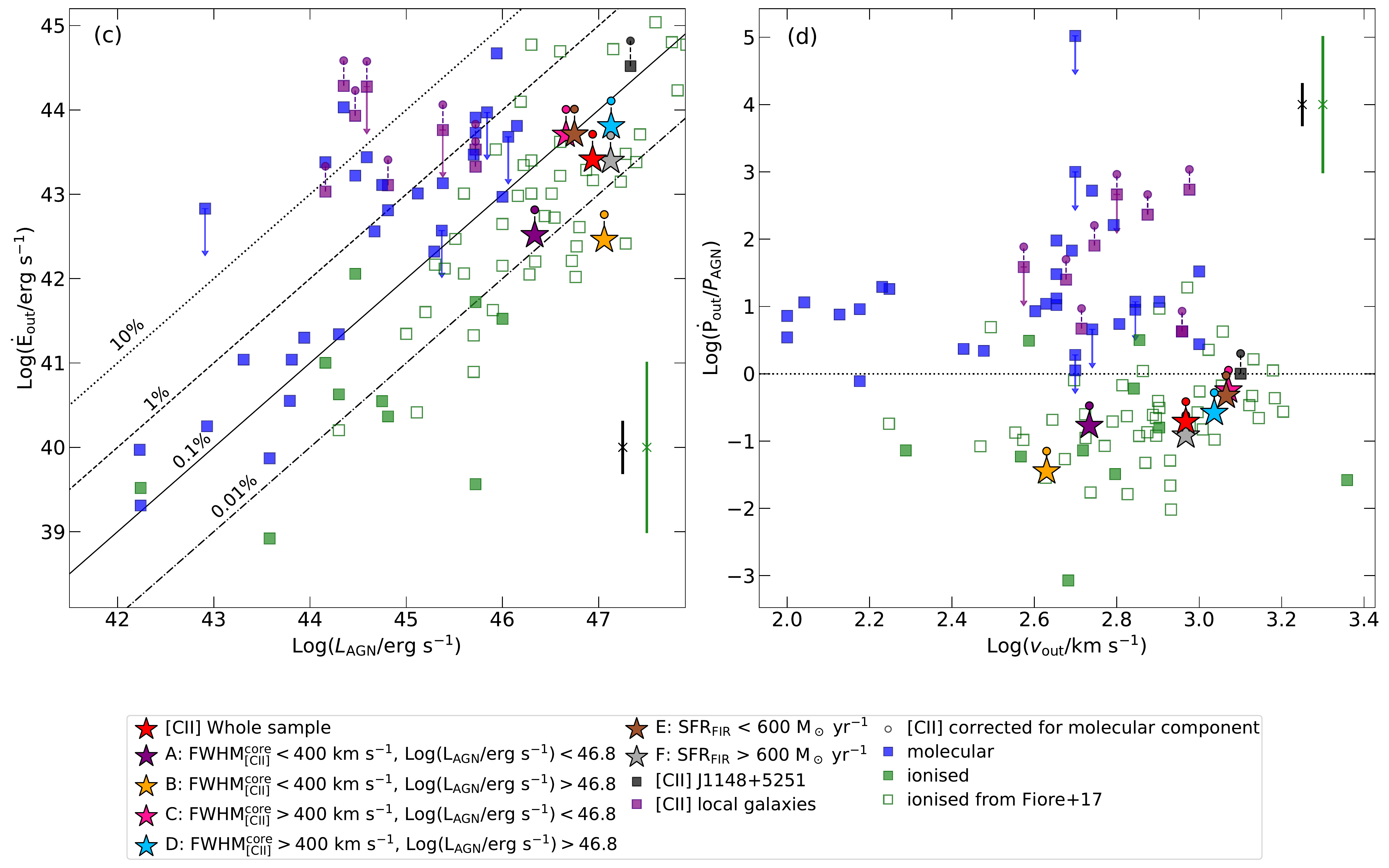}
	\caption{[CII] outflows parameters.
		\textit{(a):} mass outflow rate as a function of \lbol\ for the different stacked spectra (stars, see
		legend for details), compared to the sample of 30 low-redshift AGN from \cite{Flutsch18} for which spatially resolved molecular (blue) and, in one third of the sample, ionised
		(green) outflows have been observed. We also included the compilation of ionised outflows (hollow green
		squares) with spatial information  in $z\sim0.1-3$ AGN from \cite{Fiore17}, recomputed according to Eqs.
		\ref{eq:moutrate}-\ref{eq:pout}. Purple squares are local systems for which the outflow has
		been traced in \cii\ through observations with the \textit{Herschel Space Observatory} \citep{Janssen16,Flutsch18}.
		By applying the atomic-to-molecular outflowing gas mass correction by \cite{Flutsch18}, the molecular+atomic mass outflow rates are shown with circles. The typical $\sim0.3$ dex uncertainty on $\dot{M}_{\rm out}$ for the [CII] outflows found in our $z\sim6$ QSOs (similar to that of outflows in the atomic neutral and molecular phase in low-$z$ AGN) is shown by the black solid line, while the uncertainty on $\dot{M}_{\rm out}$ for the ionised outflows is shown by the green line.
		\textit{(b):} Outflow velocity as a function of \lbol .
		\textit{(c):} Kinetic power as a function of \lbol. The dotted, dashed, solid and dot-dashed curves
		indicate kinetic powers that are 10\%, 1\%, 0.1\% and 0.01\% of the AGN luminosity.
		\textit{(d):} Momentum load factor as a function of the outflow velocity. The horizontal line corresponds to $\dot{\rm P}_{\rm out}=P_{\rm AGN}$.}
	\label{fig:mout-lbol}
\end{figure*}
\begin{table*}
	\centering
	\caption{Outflow parameters associated with the different stacked integrated spectra. Values for the whole sample stack are listed in boldface. Columns give the following information: (1) stacked sample, (2) outflow velocity (3) atomic gas mass associated with the broad \cii\ wings, (4) mass outflow rate, computed following \cite{Flutsch18}, (5) kinetic power and (6) momentum load factor of the outflow.}
	\makebox[1\linewidth]{
		\setlength{\tabcolsep}{3 pt}
		\begin{tabular}{lccccccc}
			\toprule
			Stack & \multicolumn{2}{c}{} & $v_{\rm out}$ & $M_{\rm out}$ & $\dot{M}_{\rm out}$ & $\dot{E}_{\rm out}$ & $\dot{P}_{\rm out}$/$P_{\rm AGN}$ \\
			& & & [\kms]&[10$^8$ M$_\odot$] &  [M$_\odot$ yr$^{-1}$] &  [10$^{43}$ \ergs] & 	\\
						(1) &  &  & (2) & (3) & (4) & (5) & (6)\\
			\midrule 
			\textbf{Whole sample} & \multicolumn{2}{c}{} & \textbf{960 $\pm$ 120}  & \textbf{3.7 $\pm$ 0.7}  & \textbf{100 $\pm$ 20} & \textbf{2.6 $\pm$ 0.7} & \textbf{0.20$\pm$ 0.05} \\
			A 		  & \multicolumn{2}{c}{\fwhm$<$ 400 \kms, \lbol$<10^{46.8}$ \ergs}  & 550 $\pm$ 110 & 2.4 $\pm$ 0.9  & 35 $\pm$ 15  & 0.30 $\pm$ 0.15 &   0.17$\pm$ 0.07 \\
			B 		  & \multicolumn{2}{c}{\fwhm$<$ 400 \kms, \lbol$>10^{46.8}$ \ergs} & 440 $\pm$ 90 & 4.6 $\pm$ 1.5  & 55 $\pm$ 20  & 0.33 $\pm$ 0.15 & 0.04 $\pm$ 0.02 \\
			C 		  & \multicolumn{2}{c}{\fwhm$>$ 400 \kms, \lbol$<10^{46.8}$ \ergs} & 1180 $\pm$ 380 & 3.2 $\pm$ 1.0  & 115 $\pm$ 50 & 5.0 $\pm$ 2.3 & 0.58 $\pm$ 0.24 \\
			D 		  &  \multicolumn{2}{c}{\fwhm$>$ 400 \kms, \lbol$>10^{46.8}$ \ergs} & 1100 $\pm$ 140 &6.2 $\pm$ 1.2  & 185 $\pm$ 35 & 7.4 $\pm$ 2.0& 0.28 $\pm$ 0.07 \\
			E 		  &  \multicolumn{2}{c}{SFR$_{\rm FIR}$ $<$ 600 \msunyr} & 1210 $\pm$ 230 & 3.9 $\pm$ 1.1  & 135 $\pm$ 40 & 3.0 $\pm$ 0.7 & 0.50 $\pm$ 0.12 \\
			F 	      & \multicolumn{2}{c}{SFR$_{\rm FIR}$ $>$ 600 \msunyr} & 930 $\pm$ 140 & 3.6 $\pm$ 0.8  & 95 $\pm$ 30 & 2.5 $\pm$ 0.8& 0.12 $\pm$ 0.03\\
			\bottomrule
		\end{tabular}
	}
	\label{tab:outrate}	
\end{table*}

The resulting outflow parameters for the whole sample stack and the different subsamples considered (see Sect. \ref{sect:subsamples}) are listed in Table \ref{tab:outrate}.
We derive a mass outflow rate of $\dot{M}_{\rm out} = 100\pm20$ \msunyr\ for the stack of the whole sample, while
for the large FWHM, high-\lbol\ subgroup (stack $D$) we find $\dot{M}_{\rm out}\sim200$ \msunyr. 
These outflow rates only refer to the atomic neutral component.
\cite{Flutsch18} obtained that, for AGN-driven outflows, the molecular mass outflow rates are of the same order as the atomic neutral outflow rates, while the contribution from the ionised gas is negligible,
at least in the luminosity range probed by them. They find that the molecular-to-ionised outflow rate increases with luminosity, in contrast with what found by \cite{Fiore17}; the discrepancy may
originate from the fact that the latter study investigate disjoint samples, or may originate from the different luminosity ranges sampled. If we assume that the relations found by \cite{Flutsch18} also
apply to these distant luminous QSOs, then the implied total outflow rate is twice the value inferred from \cii.

Fig. \ref{fig:mout-lbol}a shows the mass outflow rate as a function of the AGN bolometric luminosity. Stars show the atomic neutral outflow rate inferred from the [CII] broad wings for the various stacked spectra, as indicated
in the legend. The circles, connected to the star through a dashed line,
indicate the inferred outflow rate by accounting also for the molecular
gas content in the outflow assuming the relation given by \cite{Flutsch18}. Blue, green and purple squares show the molecular, ionised and atomic neutral outflow rates measured by \cite{Flutsch18} in local AGN. In the latter case the neutral component is obtained through \cii\ observations of local galaxies performed by the \textit{Herschel} infrared space telescope \citep{Janssen16} and purple circles show the effect of correcting the atomic outflow rate as discussed
above. Hollow green squares show the ionised outflow rates inferred from \cite{Fiore17}; these are from a disjoint sample (they do not have measurements for the molecular and atomic phase) and may
be subject to different selection effects, but they have the advantage to extend to much higher luminosities than the sample in \cite{Flutsch18}.
Fig. \ref{fig:mout-lbol}a illustrates the well known phenomenon that the outflow rate increases with the AGN luminosity and that generally the outflow rate is dominated by the neutral phases (atomic and neutral).
However, at the very high luminosities probed by our stacked spectra of the most distant QSOs the outflow rates associated with the neutral phase appear to deviate from the trend observed locally, and the outflow rates seem similar to those observed in the ionised phase.

For completeness, Fig. \ref{fig:mout-lbol}b shows the outflow velocity as a function of the AGN luminosity,
illustrating that the velocity of the outflow observed in the stacked spectra is consistent with the
trend observed in other AGN and in other phases, further confirming that these outflows are QSO-driven.

Fig. \ref{fig:mout-lbol}c shows the kinetic power as a function of the AGN luminosity with the same symbols
as in Fig.\ref{fig:mout-lbol}a. For our stacked spectra, the kinetic power is between 0.01\% and 0.5\%
of \lbol, i.e. much lower than what expected from AGN "energy-driven" outflow models \citep[$\dot{E}_{\rm out}\sim0.05\times L_{\rm AGN}$, e.g.][]{DiMatteo05,ZubovasKing2012} which
ascribe outflows to
the nuclear winds that expands in an energy-conserving way.

Fig. \ref{fig:mout-lbol}d shows the outflow
momentum load factor, i.e. the outflow momentum rate relative to $ \dot{P}_{\rm AGN}$, as a function of the outflow velocity. For our stacked spectra $\dot{P}_{\rm out}/ \dot{P}_{\rm AGN}\lesssim1$, while "energy-driven" outflow models would expect momentum load factors of $\sim20$. These results suggest that the outflows in these powerful quasars are either energy-driven but with poor coupling with the ISM of the host galaxy, or are driven by direct radiation pressure onto the dusty clouds \citep[e.g.][]{Ishibashi18}. In either cases the outflow unlikely is in its "ejective" mode, i.e. very
effective in removing gas from the entire galaxy, hence in completely
suppressing star formation \citep{Costa15,Costa17,Bourne14,Bourne15,Roos15,Gabor14},
although such ejective mode can be effective in clearing of the gas content and quenching star formation in the central regions. Moreover,
the outflow can be effective in heating the circumgalactic medium and therefore preventing further accretion of fresh gas onto the galaxy, hence resulting in a delayed quenching of the galaxy by
"starvation" \citep{Peng&Maiolino15}.

It could also be, in contrast with what observed in the low-luminosity local AGN, that in these very luminous, distant QSOs the bulk of the outflow is highly ionised. The observation that in other very luminous QSOs the ionised outflow rate, kinetic power and momentum rate is similar to the same quantities locally observed in the molecular phase (Fig.\ref{fig:mout-lbol}), does suggest that the balance between the various phases is different in these systems \citep{Bischetti17,Fiore17}. However, as illustrated
in Fig. \ref{fig:mout-lbol}, even the ionised phase does not seem to be massive and powerful enough to
match the requirements of the
energy-driven scenario with high coupling.

In alternative, the interferometric data used in our stack of the \cii\ emission may miss extended, diffuse
emission associated with outflows. Indeed, a large fraction of the data have angular resolution higher
than 0.7$''$, which may prevent them to detect emission on scales larger than $\sim3-4''$.
The lack of sensitivity to extended, diffuse emission may indeed be a major problem in very distant
systems, due to the rapid cosmological dimming of the surface brightness, decreasing as $\rm \sim (1+z)^4$.
This scenario may also
explain why the \cii\ outflow rate and kinetic power in the stacked data of distant QSOs do not seem to increase significantly with respect to the local, lower-luminosity AGN (purple square symbols in Fig.\ref{fig:mout-lbol}) whose \cii\
broad wings were observed with Herschel.

Within this context it is interesting to note that in the QSO J1148$+$5251 at z$=$6.4 \cite{Maiolino12} and \cite{Cicone15} did detect a very extended outflow on scales of $\sim 6''$,
by exploiting low angular resolution observations. J1148$+$5251 (black square in Fig. \ref{fig:mout-lbol})
is indeed characterised by a larger outflow rate and higher kinetic power with respect to the stacked measurements. However, even
for J1148$+$5251 the kinetic power and momentum rate appear to be significantly lower than what expected by the simple scenario of energy-driven outflows with high coupling with the ISM.

\section{Conclusions}
In this work we have presented the stacking analysis of a sample of 48 QSOs at $4.5<z<7.1$ detected in \cii\ by
ALMA, equivalent to an observation of $\sim34$ hours on-source,
aimed at investigating the presence and the properties of broad \cii\ wings tracing cold outflows.
The stack allows us to reach an improvement in sensitivity by a factor of $\sim14$ with respect to the previous observation of a
massive \cii\ outflow in J1148$+$5251 at $z\sim6.4$ \citep{Maiolino12,Cicone15}. 

\begin{itemize}
\item From the stacked integrated spectra, we clearly detect broad \cii\ wings, tracing cold outflows associated with $z\sim6$ QSOs and whose velocities exceed 1000 \kms.
This weak, broad component has not been
previously detected in single observations (except for the case of J1148$+$5251) because of insufficient sensitivity. The same
limitation applies to the stack recently performed by \cite{Decarli18} on the sample of 23 $z\sim6$ QSOs
with ALMA \cii\ detection, which were mostly observed with very short (few minutes) exposures. In fact similarly to
\cite{Decarli18}, we find no significant broad \cii\ wings in the stacked spectrum of their sources alone.

\item The redshifted [CII] wing is fainter than the blueshifted [CII] wing. This may be associated with the
asymmetric distribution of the spectral coverage of the spectra used in the stacked spectrum. However, if confirmed
with additional data, this asymmetry would suggest that in these systems
  the dusty gas in the host galaxy has a column density high enough to obscure the receding component of the
  outflows, with respect to our line of sight. High dust column densities capable of absorbing even at far-IR and sub-mm
  wavelengths have been observed in local ULIRGs.
  
\item By splitting the sample in AGN luminosity and SFR bins, we observe that the strength of the stacked broad component correlates with the AGN luminosity, but does not depend on the SFR. This indicates that the QSOs are the primary driving mechanism of the \cii\ outflows in these systems.
Moreover, we find that the broad component is very blushifted in the stack with high SFR and nearly symmetric in the stack with low SFR. Since the SFR correlates with the gas and dust content in the galaxy,
 this finding corroborates the interpretation that the blueshift of the [CII] broad component might be associated with heavy dust absorption.

\item By stacking the ALMA data cubes, we investigate the morphology of the \cii\ outflows in our sample
and find that the high-velocity \cii\ emission extends
up to $R_{\rm out}\sim3.5$ kpc.
However, we cannot exclude that additional, more extended emission is present but missed by the
interferometric data used for the stacking. Moreover, averaging outflows with different orientations and
clumpiness may result into dilution effects affecting the observed intensity and extension of the
\cii\ broad wings in the stacked cube.

\item From the stacked cube we infer an average atomic mass outflow rate $\dot{M}_{\rm out} \sim100$ \msunyr,
which doubles for the stack of the most luminous sources.
By correcting for the atomic-to-molecular gas ratio found by \cite{Flutsch18}, the former value translates into
a total mass outflow rate of about 200 \msunyr.
The associated kinetic powers are consistent with 0.1\% of \lbol\ for most stacks, while momentum load factors span the range $0.1-1$; these $\dot{M}_{\rm out}$ are lower than what observed in cold outflows
associated with local, lower luminosity AGN, and are lower than the expectations of standard energy-driven outflow models (hence
indicating either a low coupling with the ISM and/or a different driving mechanism, such as direct radiation pressure
on the dusty clouds). 

As a consequence,
QSO-driven outflows in the early universe may have not been very effective in clearing the galaxy from their gas content, although
they may have been effective in clearing and quenching their central regions, and also heating the galaxy halo hence resulting into a delayed star formation quenching as a consequence of starvation.

\end{itemize}

Future deep ALMA follow-up observations will allow us to confirm the presence of \cii\ outflows in individual high-$z$ QSOs. Furthermore, the increasing number of available sources on the ALMA archive will increase the statistics, enabling us to reduce the uncertainties on the cold outflows parameters in the early Universe.

\begin{acknowledgements}
We are grateful to the anonymous referee for valuable feedback that helped us to improve the paper. This paper makes use of the following ALMA data: ADS/JAO.ALMA\#2011.0.00243.S, 2012.1.00604.S,
2012.1.00676.S, 2012.1.00882.S, 2013.1.01153.S, 2015.1.01115.S and
2016.1.01515.S. ALMA is a partnership of ESO (representing its member
states), NSF (USA) and NINS (Japan), together with NRC (Canada), MOST and ASIAA (Taiwan), and KASI (Republic of
Korea), in cooperation with the Republic of Chile. The Joint ALMA Observatory is operated by ESO, AUI/NRAO and
NAOJ. MB and EP acknowledge financial support from  ASI and INAF under the contract 2017-14-H.0 ASI-INAF. RM and AF acknowledge ERC Advanced Grant 695671 "QUENCH" and support by the Science and Technology Facilities Council (STFC).  FF and EP acknowledge financial support from INAF under the contract PRIN INAF 2016 FORECAST.
\end{acknowledgements}

\bibpunct{(}{)}{;}{a}{}{,} 
\bibliographystyle{aa} 
\bibliography{biblioCII} 

\newpage
\onecolumn
\section*{Appendix A: spectral fitting}

\begin{figure*}[h]
	\centering

	\includegraphics[width=0.4\textwidth]{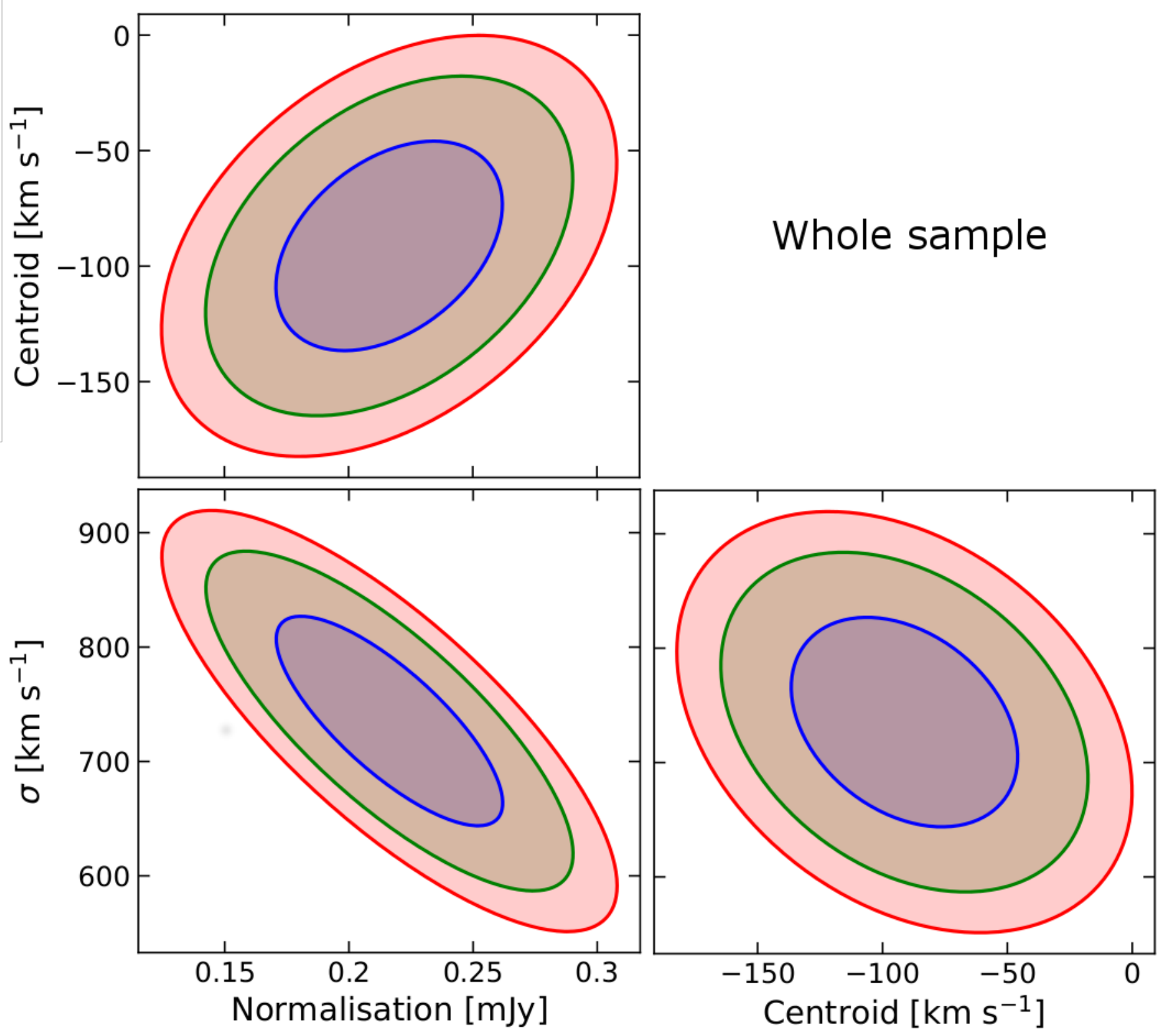}
	\includegraphics[width=0.4\textwidth]{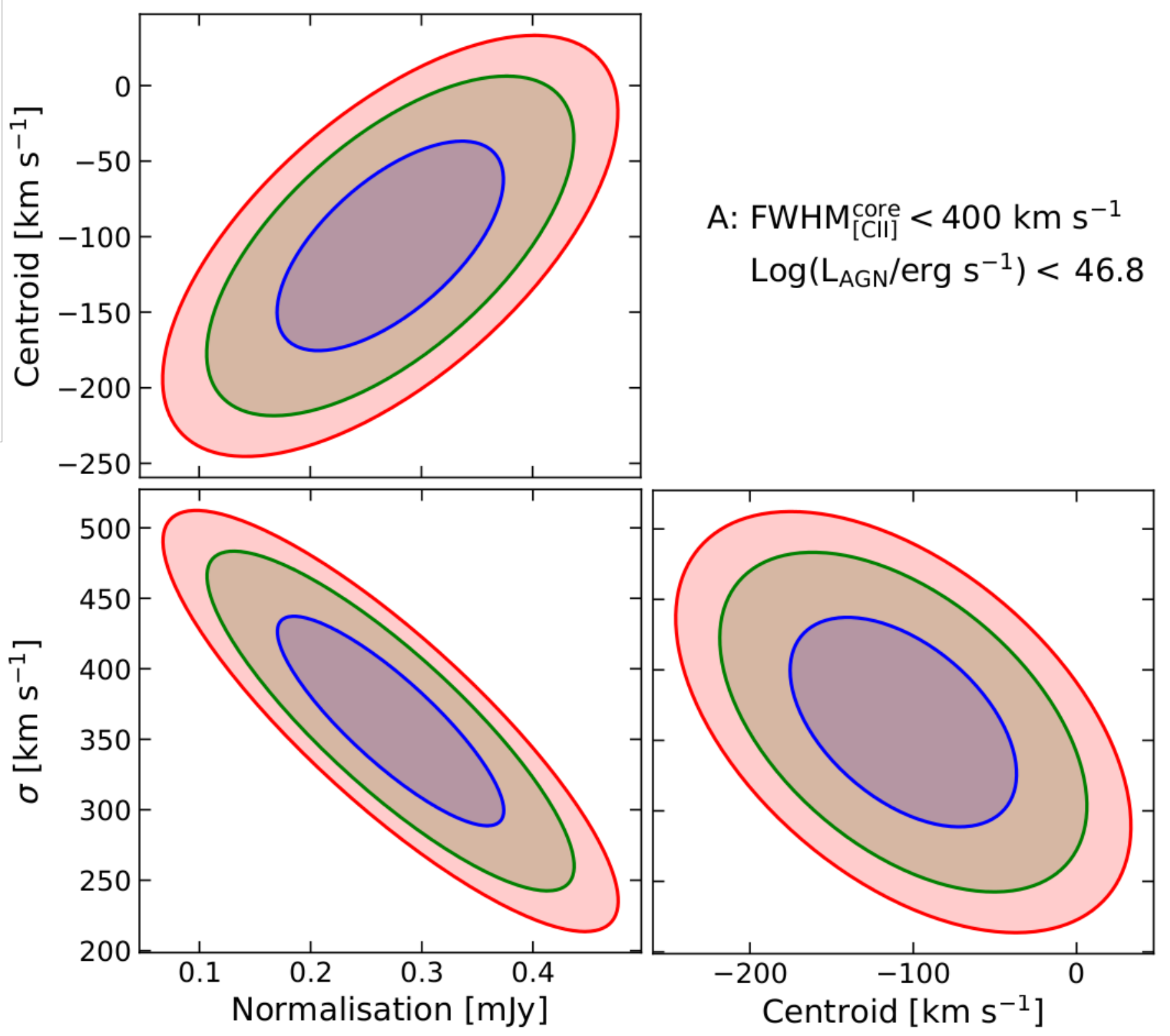}
	\includegraphics[width=0.4\textwidth]{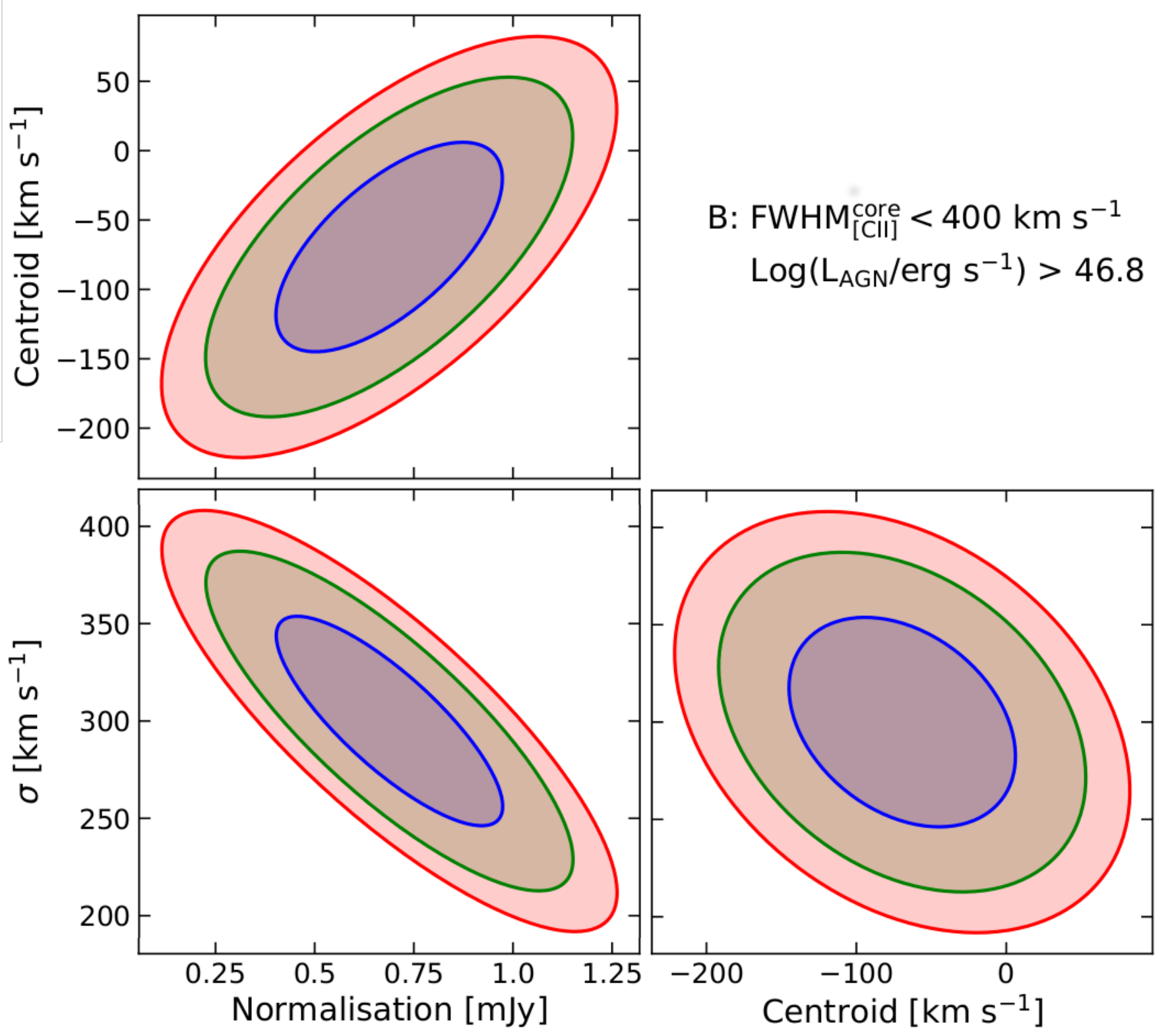}
	\includegraphics[width=0.4\textwidth]{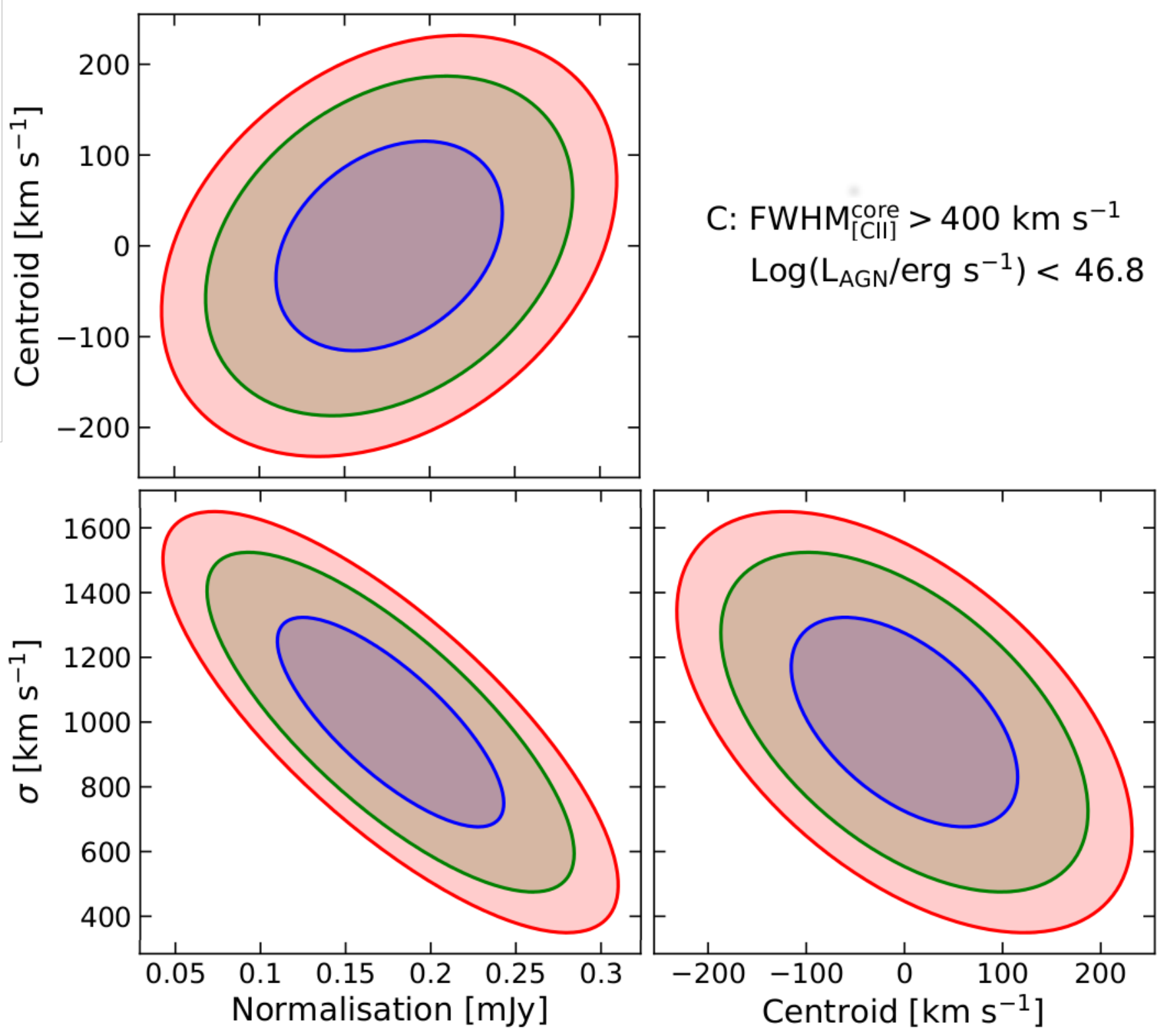}
	\caption{Confidence ellipses for the parameters of the broad Gaussian component modelling the \cii\ wings. For each stacked spectrum (indicated by the top label) the $1\sigma, 2\sigma$ and $3\sigma$ confidence intervals are indicated by the blue, green and red ellipses, respectively.}
	\label{fig:contours_1}

\end{figure*}

In order to derive the parameters of the \cii\ core and broad \cii\ wings emission, we performed spectral fitting of the stacked spectra through $\chi^2$ minimisation, by using the python \textit{scipy optimize} package \textit{curve\_fit}. We used a fitting model that includes one or two Gaussian components to reproduce the \cii\ core plus an additional broad Gaussian component to account for the high-velocity \cii\ wings. All parameters were left free to vary in the fits with no constrains. 
In Fig. \ref{fig:contours_1}, we show the confidence ellipses associated with the parameters of the broad Gaussian component modelling the \cii\ wings. Despite the correlation between the line width and the normalisation of the wings, the free outflow parameters remain well constrained.

In addition to the significance estimated through the spectral fitting analysis, i.e. from the errors on the best-fit parameters, we provided additional significance estimates by considering the pure statistical uncertainties of the stacked spectra. After removing the modelled \cii\ core emission from the stacked spectra, we calculated the integrated flux associated with the broad component alone. We therefore computed the statistical uncertainty of the ALMA map integrated over the same spectral interval. The statistical significance can be therefore derived as the ratio between the integrated flux and the statistical uncertainty. This estimate results in a higher confidence than that estimated from the fit, as it does not take into account the uncertainties in modelling the \cii\ core. 
Moreover, to estimate the significance of the high-velocity channels alone, we repeated the same procedure but excluding the velocity range covered by the \cii\ core. 

In Fig. \ref{fig:fitting-1G}, we also show the best fit 1-Gaussian component model for each stacked spectrum. In almost all panels, although with different significance levels, emission in excess of a single Gaussian component is present.

\renewcommand{\thefigure}{\arabic{figure} (Continued)}
\addtocounter{figure}{-1}
\begin{figure*}[h]
	\centering

	\includegraphics[width=0.4\textwidth]{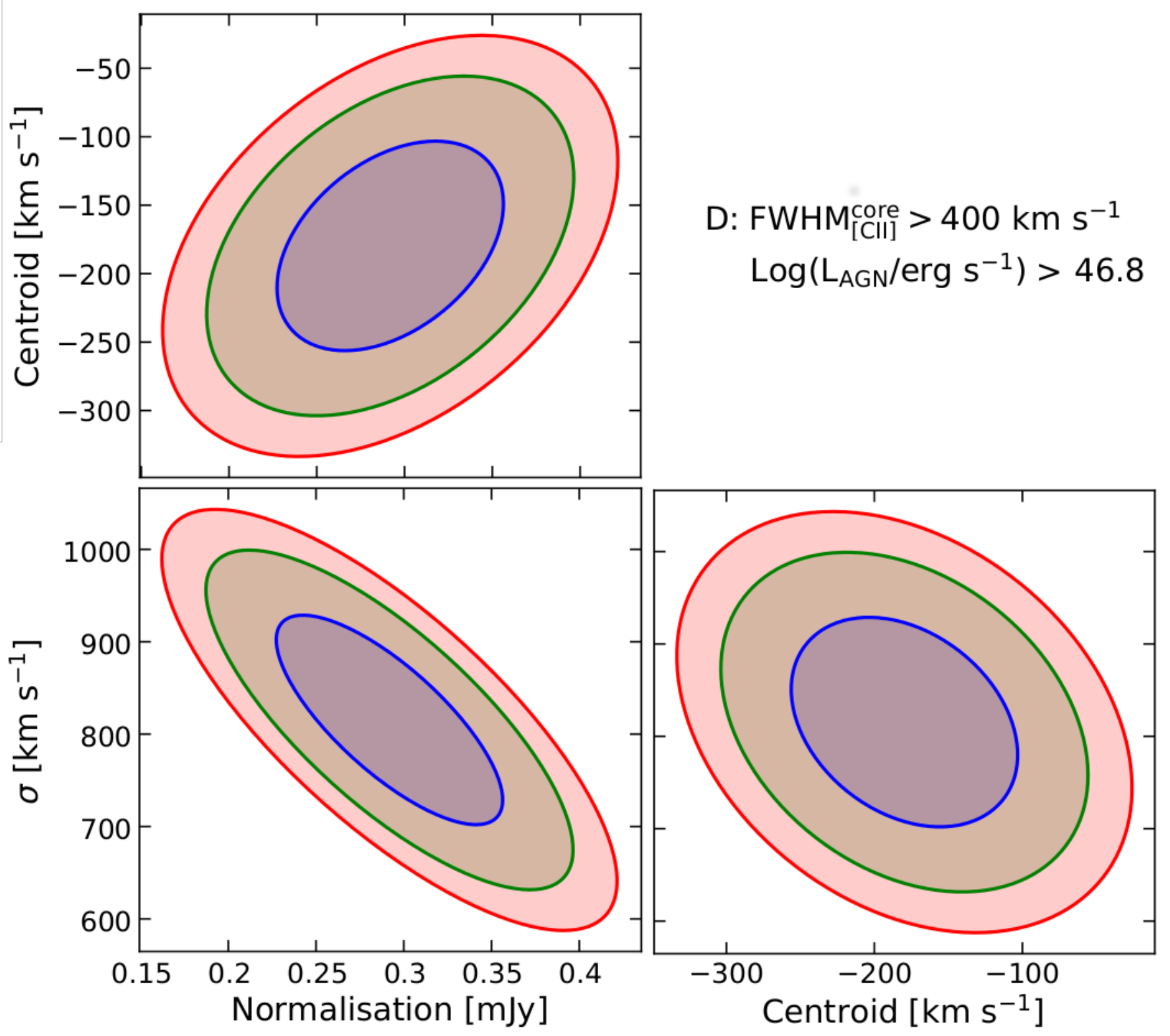}
	\includegraphics[width=0.4\textwidth]{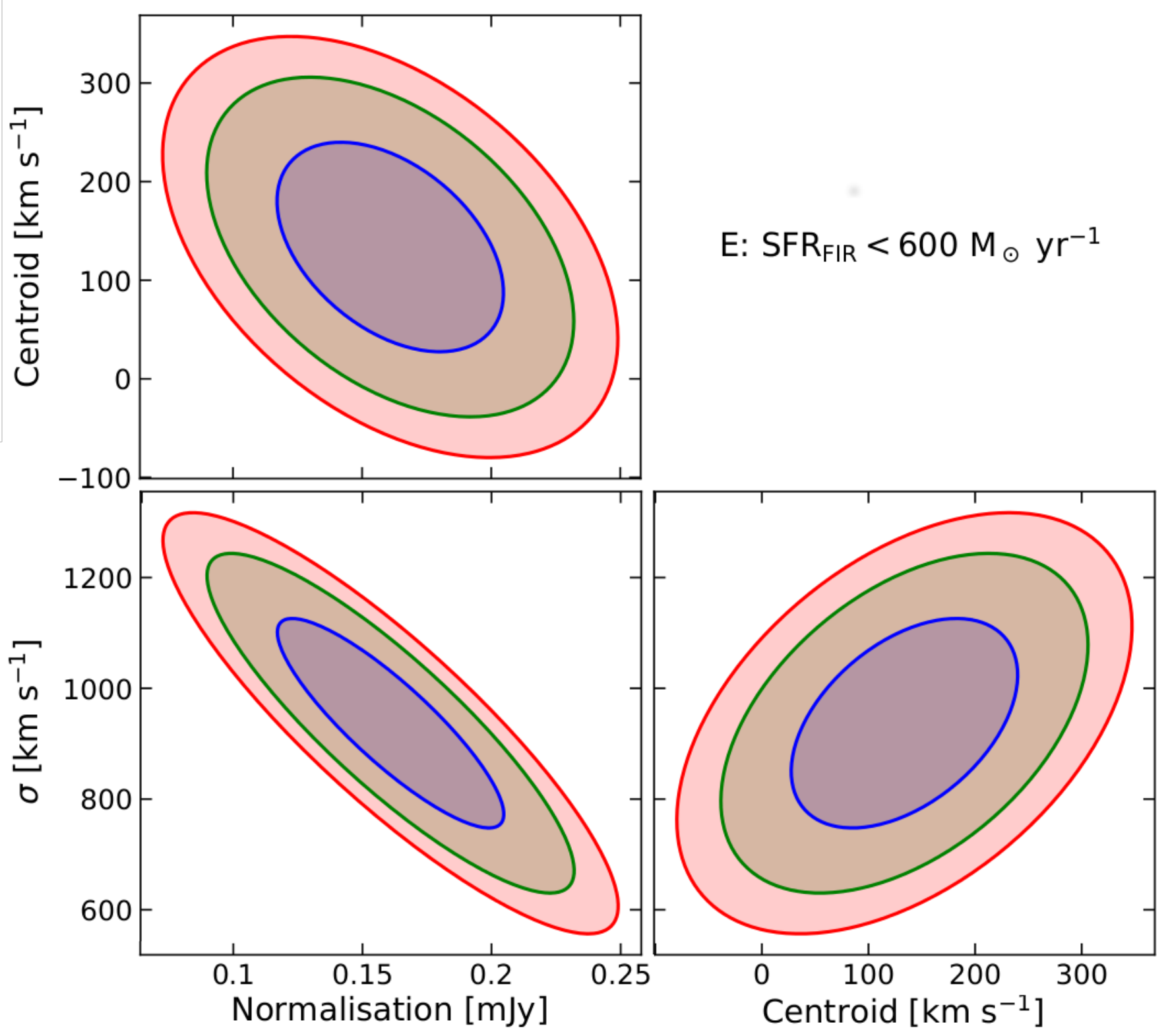}
	\includegraphics[width=0.4\textwidth]{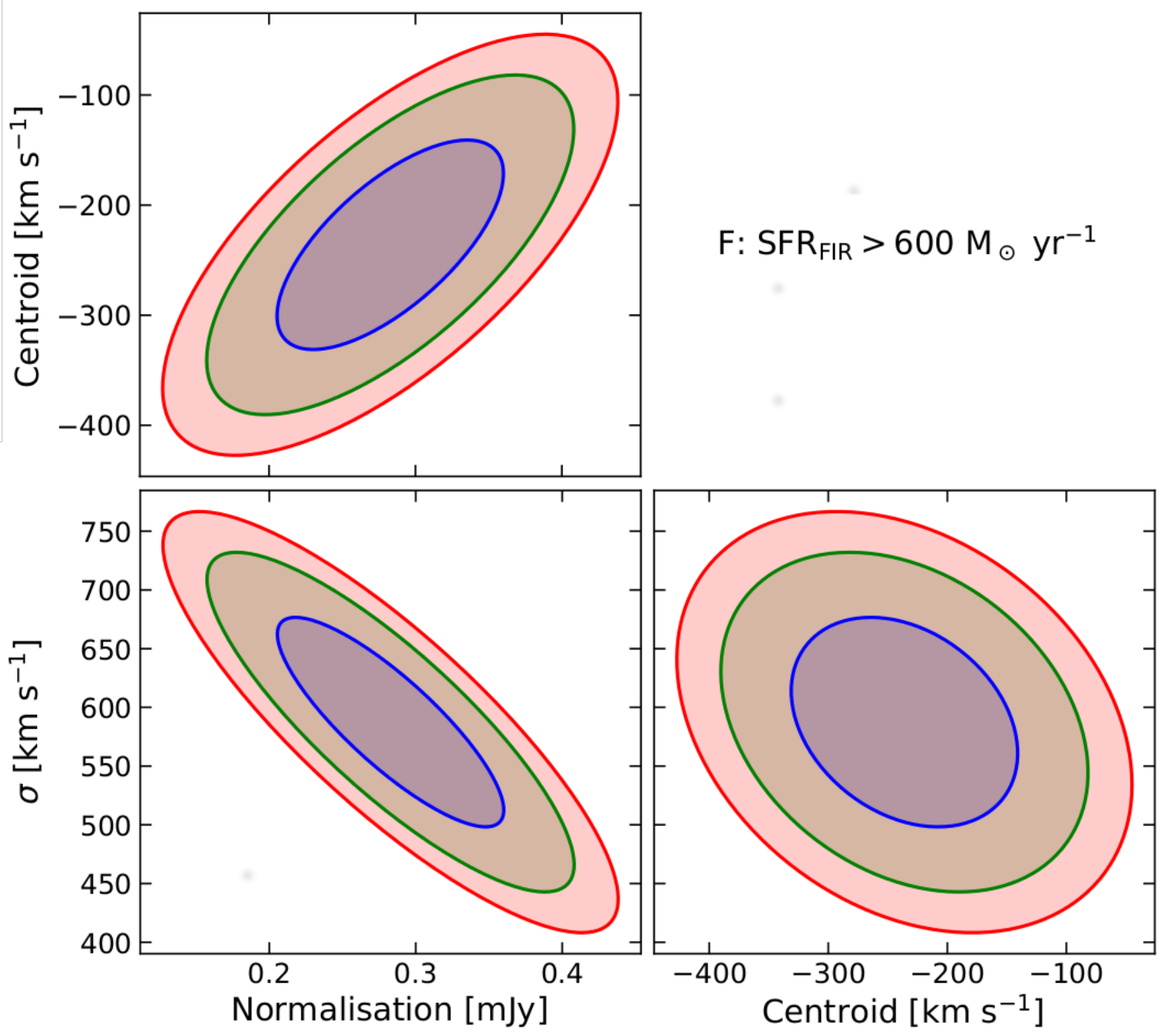}\hspace{0.405\textwidth}
	\caption{}
	\label{fig:contours_2}

\end{figure*}
\renewcommand{\thefigure}{\arabic{figure}}

\begin{figure*}[h]
	\centering
	\vspace{0.5cm}
	\includegraphics[width=0.365\textwidth]{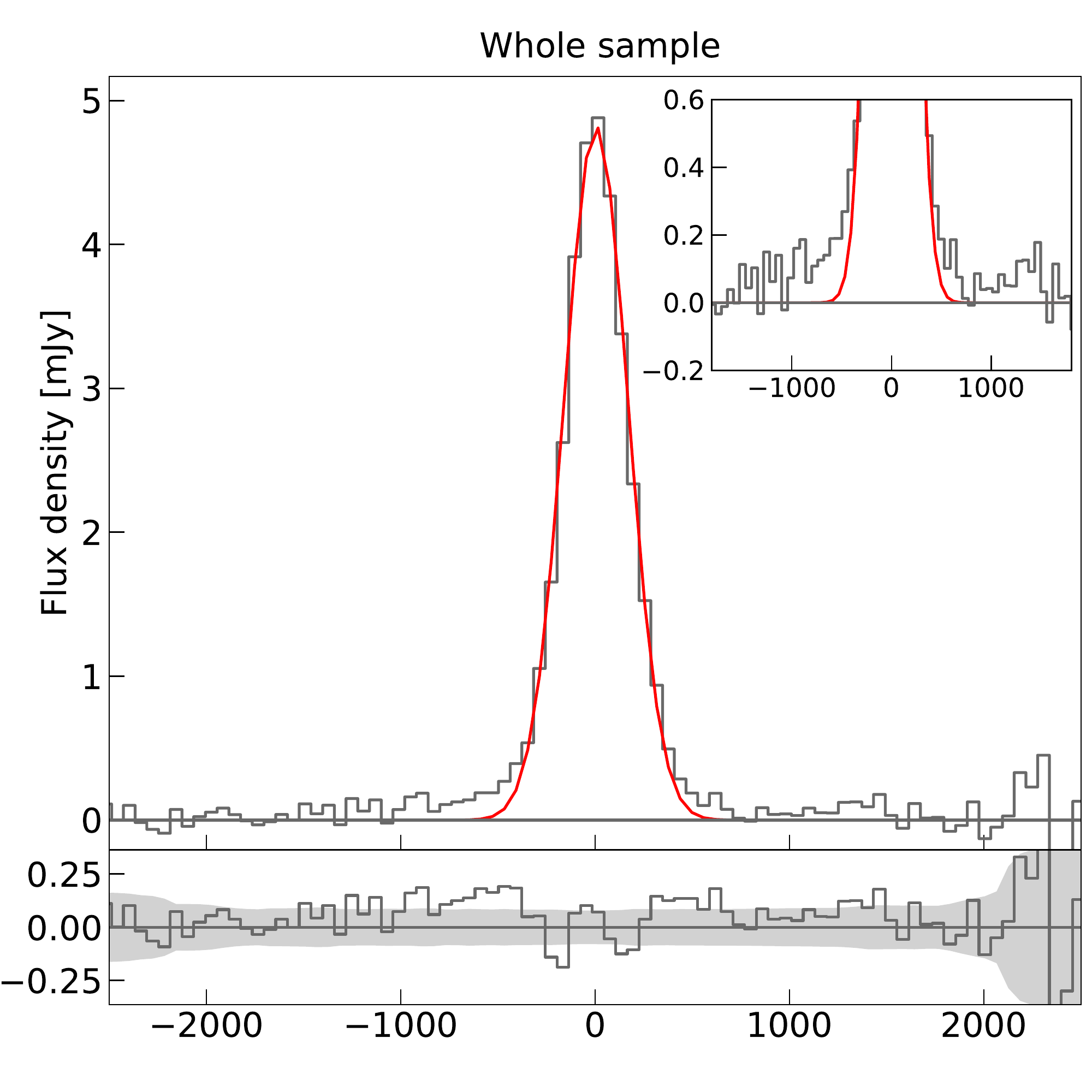}
	\includegraphics[width=0.365\textwidth]{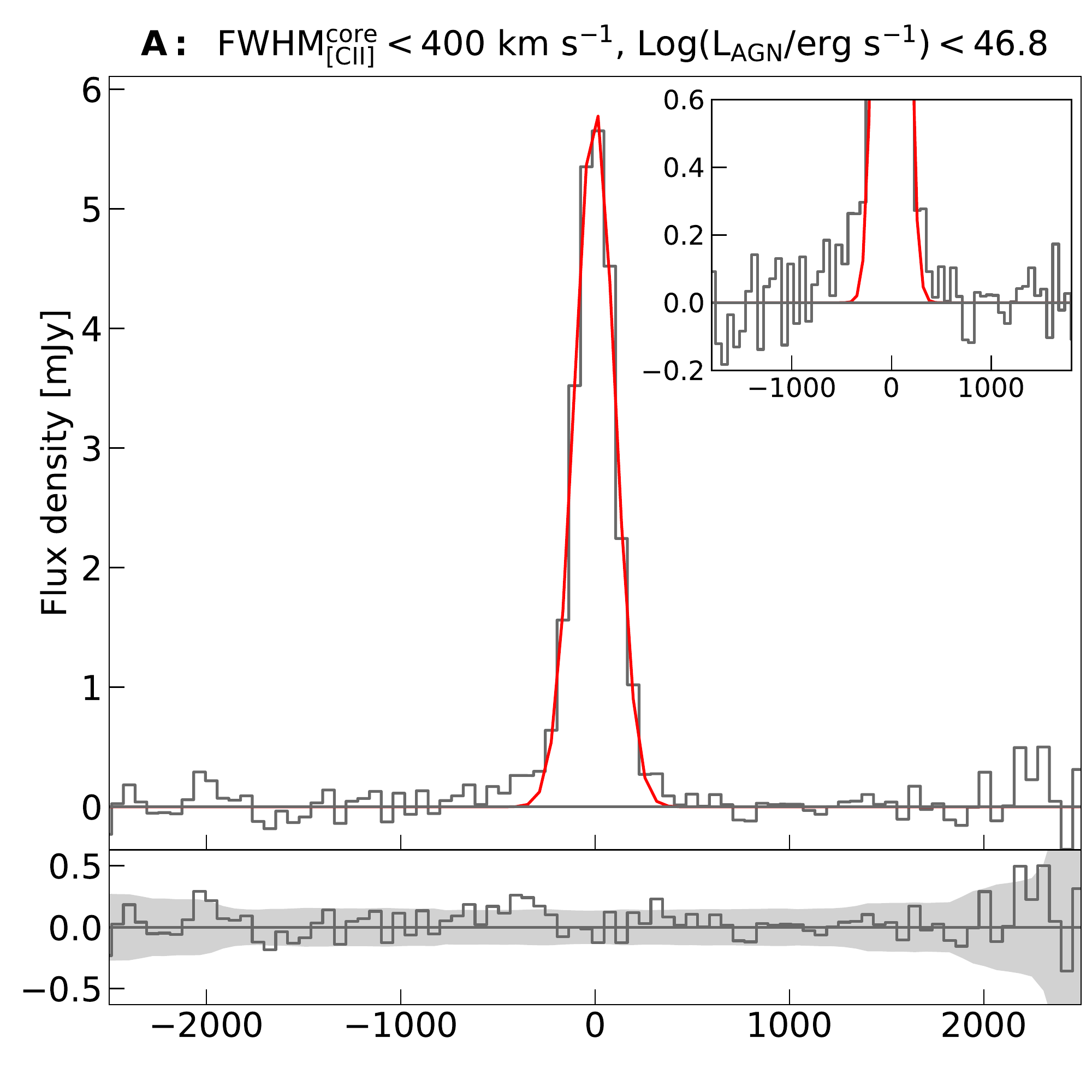}

	\caption{Stacked integrated spectra for the whole sample and all the QSOs subgroups presented in Sect. \ref{sect:stacked-cube} and Sect. \ref{sect:subsamples}. For each plot, the top panel shows the \cii\ flux density as a function of velocity, in bins of 60 \kms. The red curve represents the best-fit 1 Gaussian component model. The inset 
	zooms on the $v\in[-1500,1500]$ \kms\ region, while the bottom panel shows the residuals at different velocities. The 1$\sigma$ rms of the spectrum is also indicated by the shaded region.}
	\label{fig:fitting-1G}

\end{figure*}

\renewcommand{\thefigure}{\arabic{figure} (Continued)}
\addtocounter{figure}{-1}
\begin{figure*}[]
	\centering
	\includegraphics[width=0.365\textwidth]{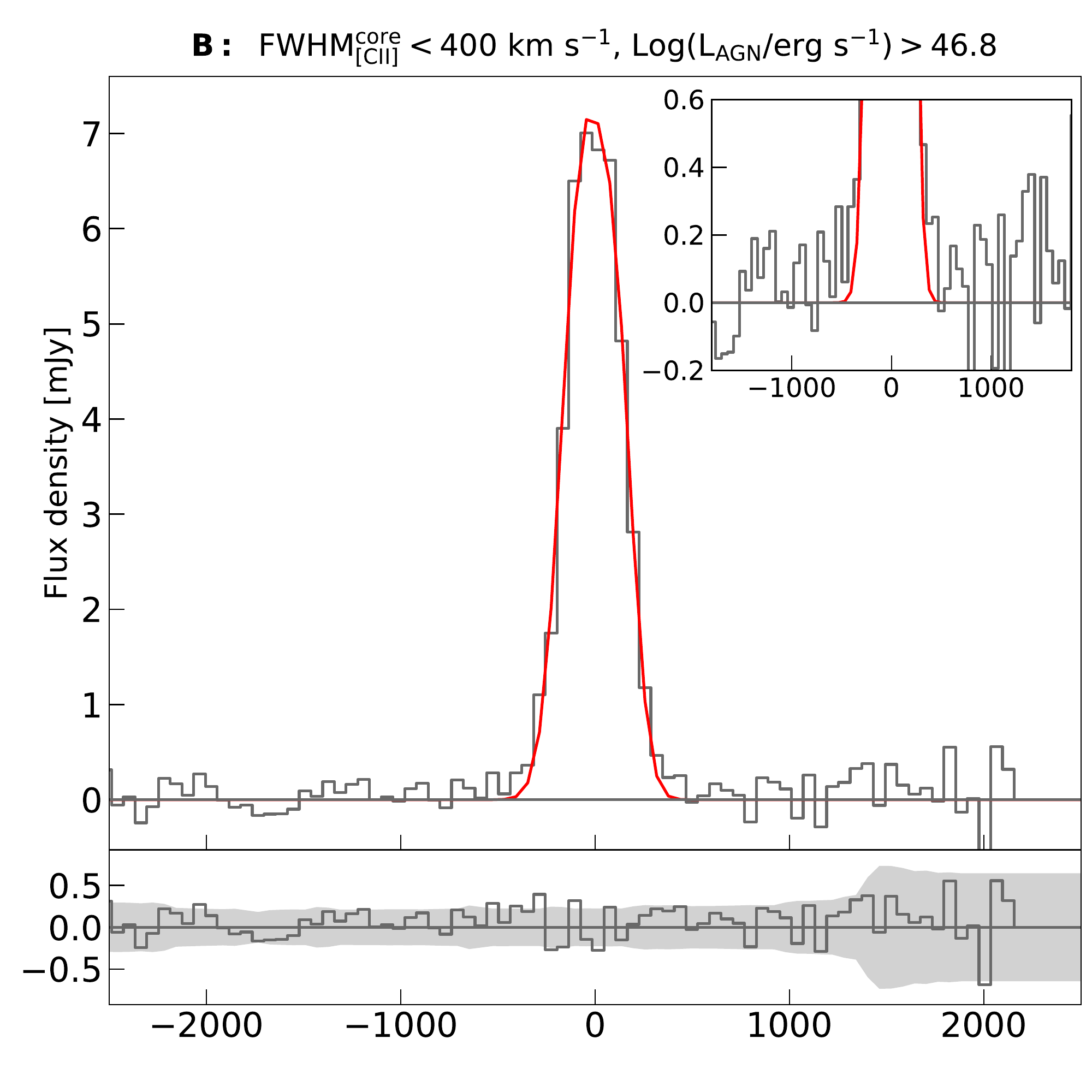}
	\includegraphics[width=0.365\textwidth]{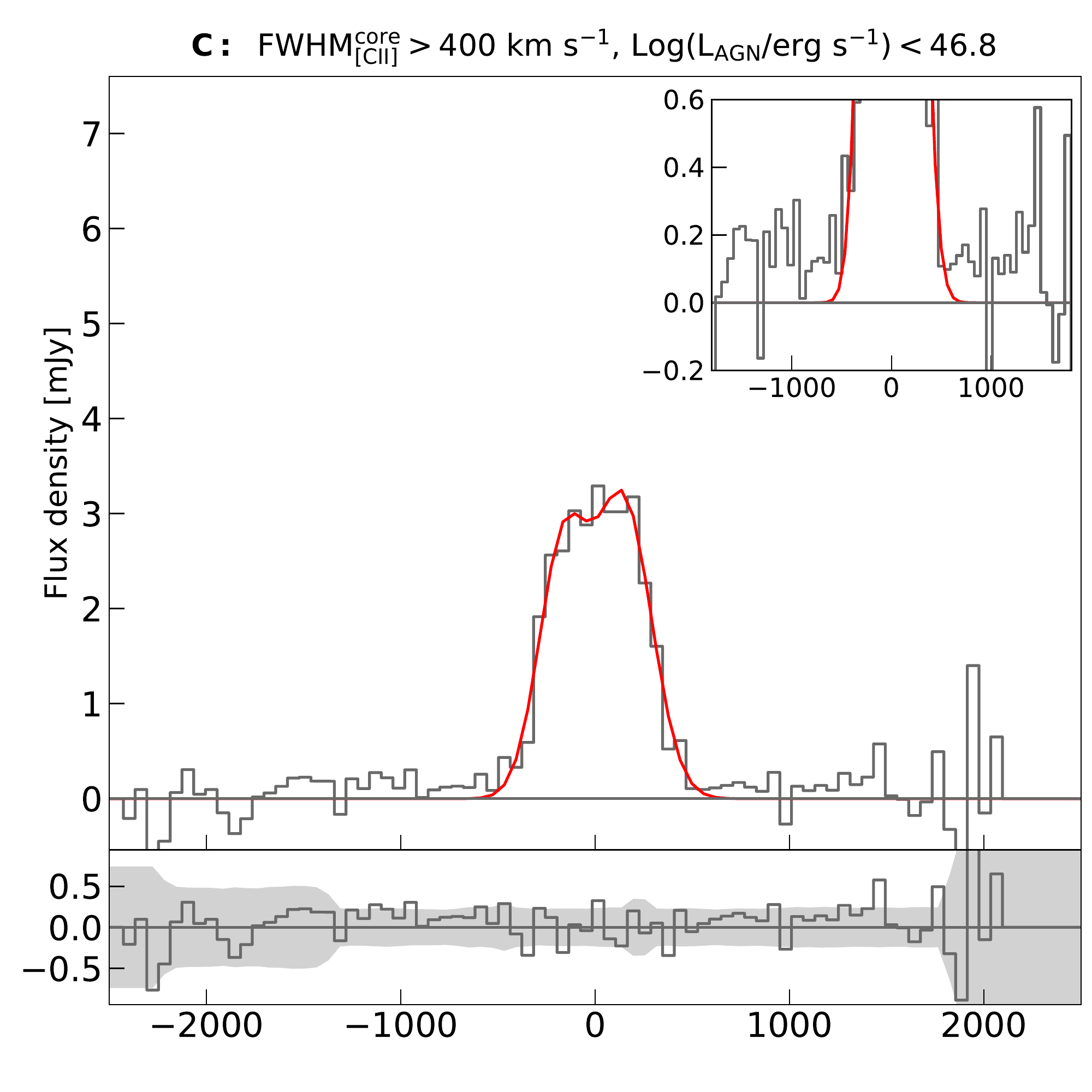}
	\includegraphics[width=0.365\textwidth]{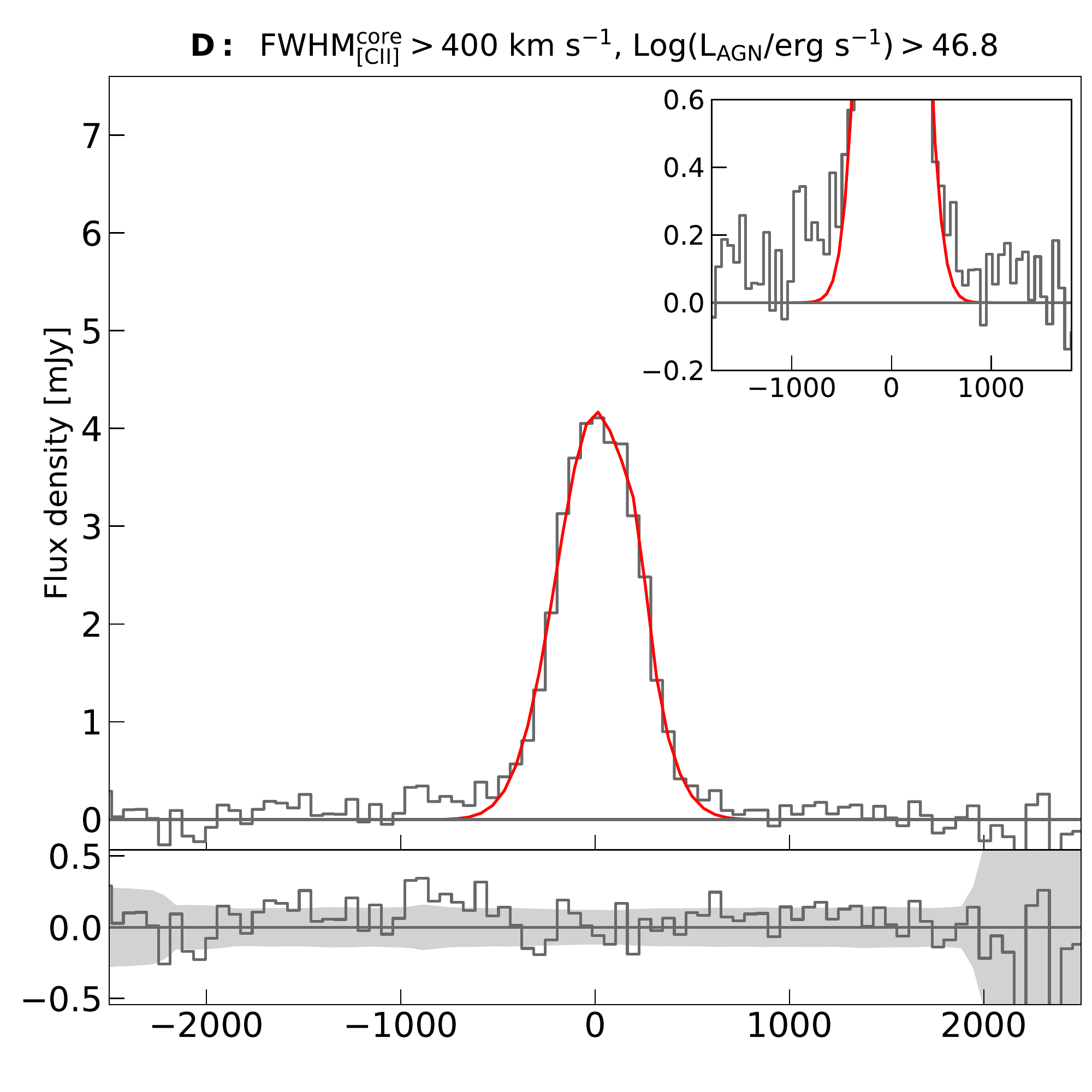}
	\includegraphics[width=0.365\textwidth]{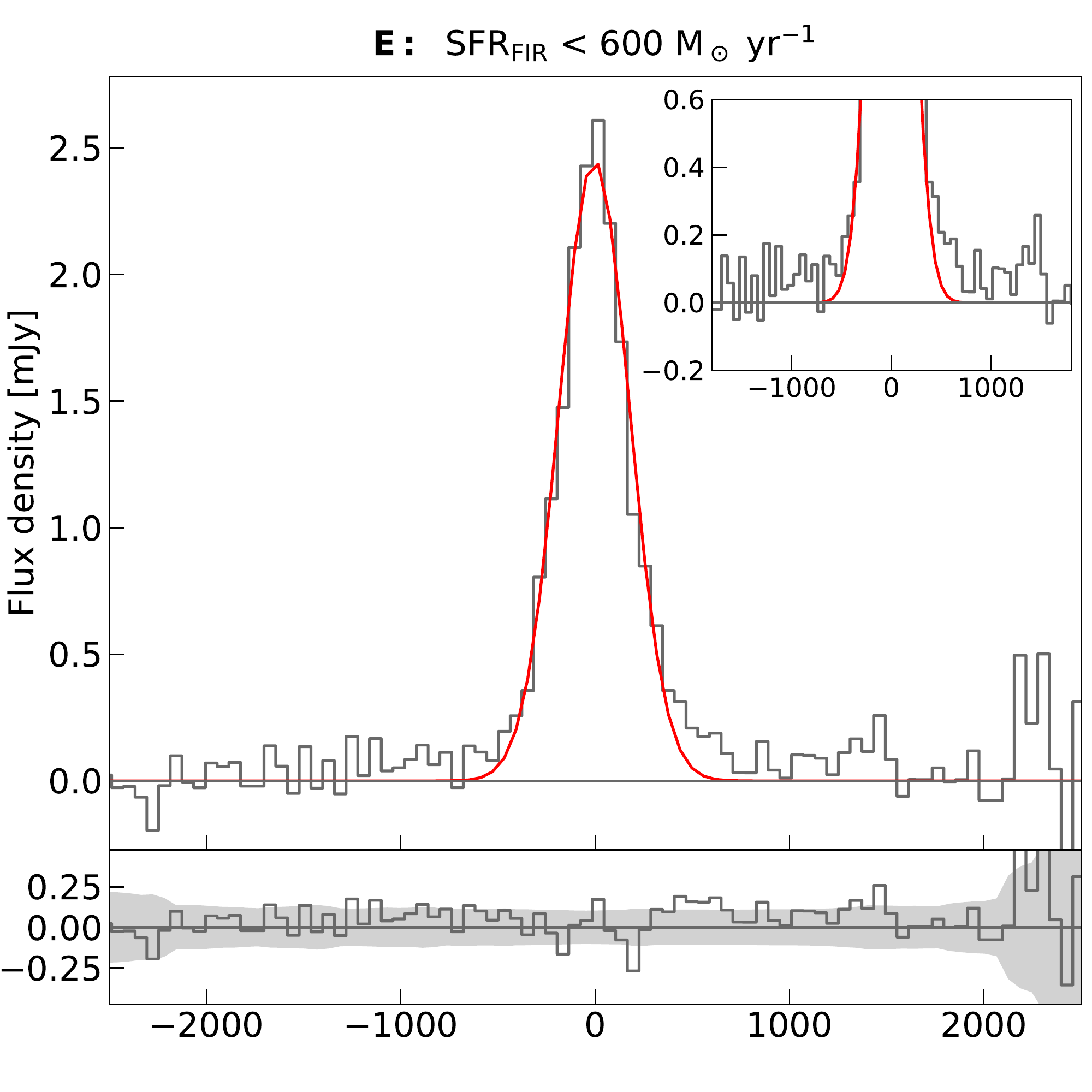}
	\includegraphics[width=0.365\textwidth]{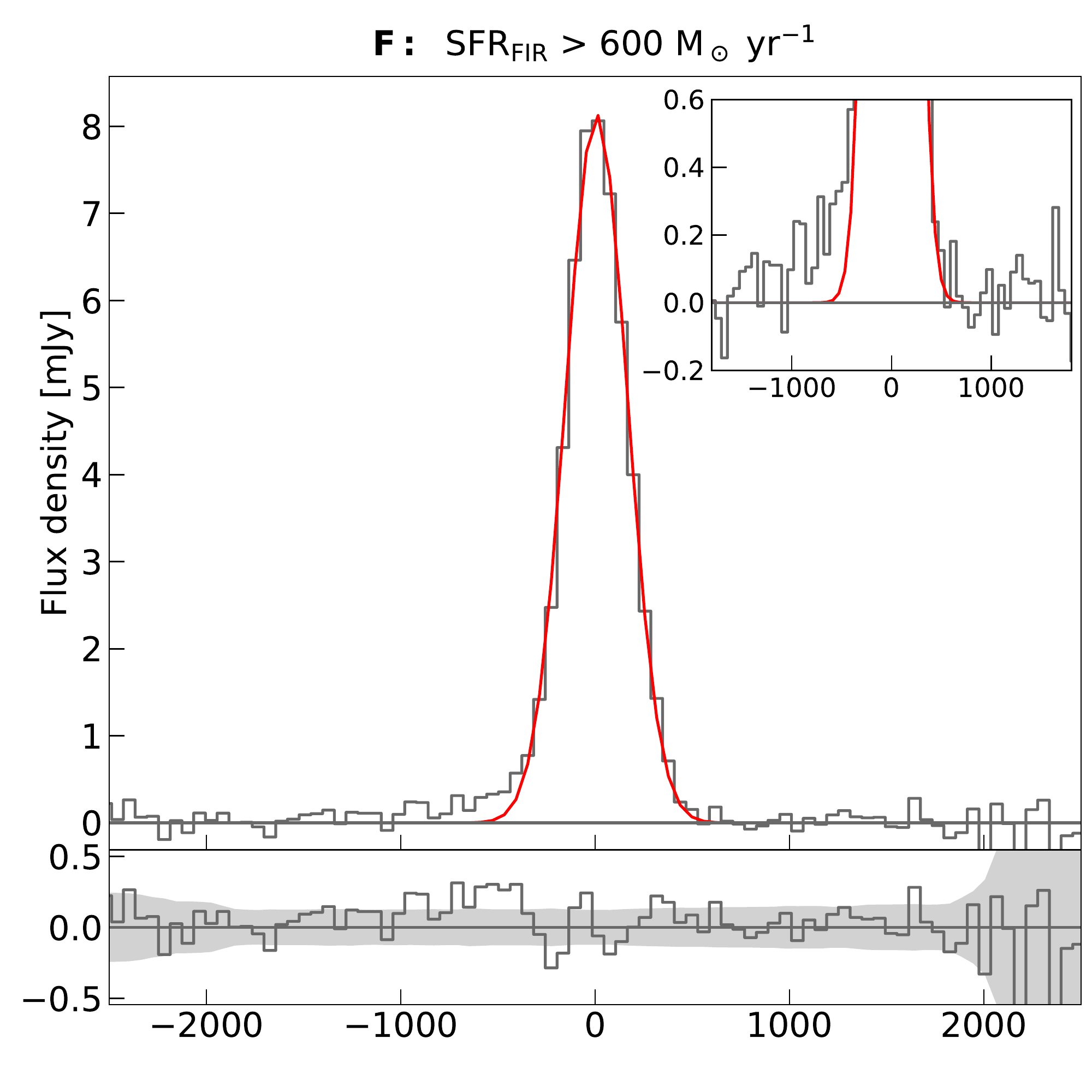}
	\caption{}
\end{figure*}
\renewcommand{\thefigure}{\arabic{figure}}
	
\clearpage

\section*{Appendix B: Stack in luminosity density units}
\begin{figure}[h]
\floatbox[{\capbeside\thisfloatsetup{capbesideposition={left,center},capbesidewidth=8cm}}]{figure}[\FBwidth]
{\caption{Whole sample integrated spectrum, stacked in luminosity density units. The first panel from top shows the \cii\ flux density as a function of velocity, in bins of 60 \kms. The red curve represents the best-fit 2 Gaussian components model; the two individual components are shown with blue and green curves. Labels indicate the number of stacked sources and the luminosity of the broad \cii\ wings. The inset shows a zoom on the broad component. \textit{Second panel from top:} residuals from the subtraction of the core emission (blue line in first panel). The green curve shows the best fit broad component. \textit{Third panel:} residuals from the two Gaussian components fitting. The 1$\sigma$ rms of the spectrum is also indicated by the shaded region.}\label{fig:lum-dens-stack}}
{\includegraphics[width=0.365\textwidth]{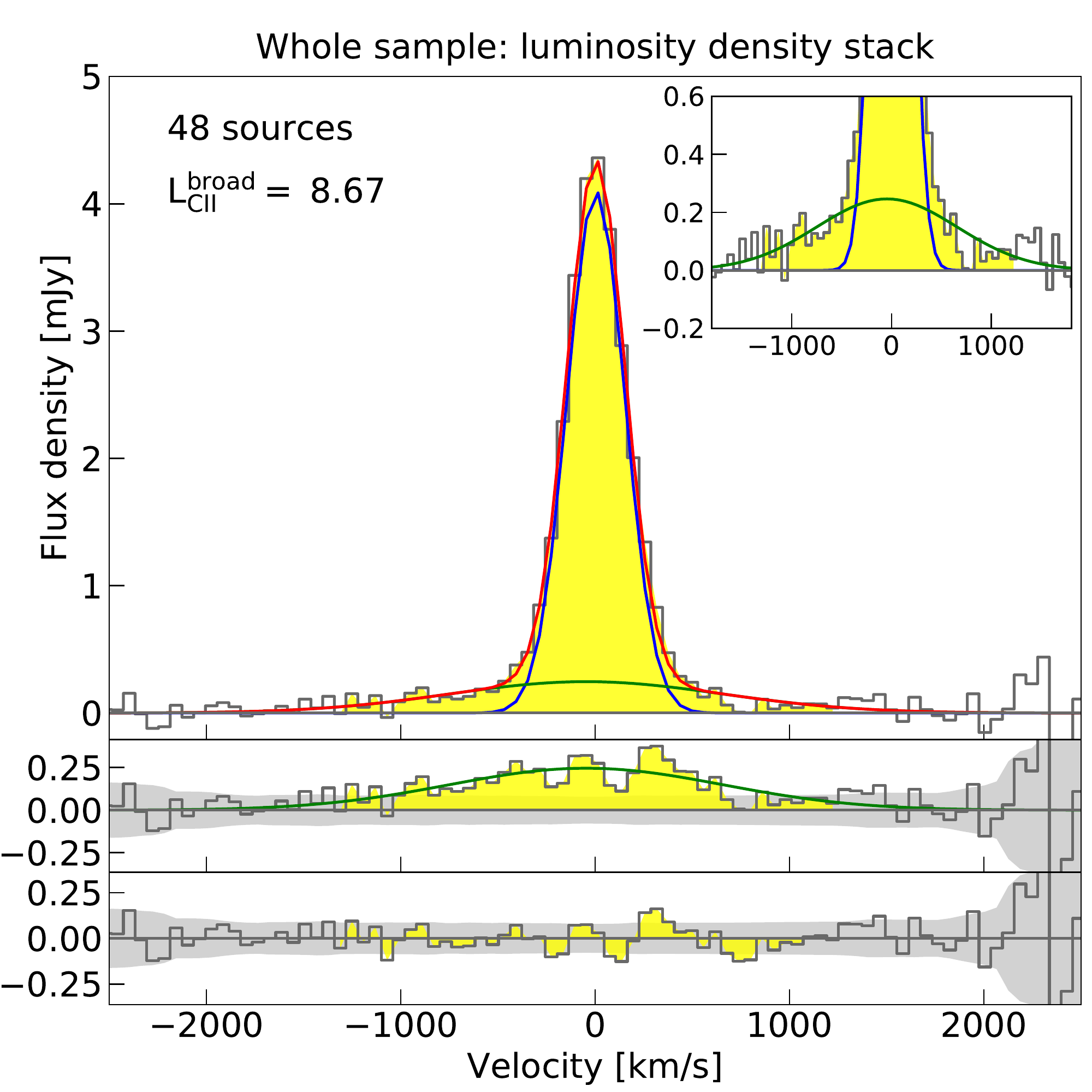}}
\vspace{-0.5cm}

\end{figure}

The QSOs in our sample span the redshift interval $4.5\lesssim z \lesssim 7$ and, therefore, a factor of 1.7 in luminosity distance. Accordingly, we verified that our results are still valid if performing a stack in luminosity density units.
We used the \zcii-based (see Table \ref{tab:sample}) luminosity distance to convert each spectrum in luminosity units and then performed the stack of the whole sample by using Eq. \ref{eq:stack_1} and Eq. \ref{eq:stack_2}. By applying the same fitting procedure presented in Sect. \ref{sect:totstack}, we derived a luminosity of the broad \cii\ wings of \lbroad$\sim4.7\times10^8$ \lsun\ (see Fig. \ref{fig:lum-dens-stack}), similar to the value estimated from the original stack in flux density units (reported in Table \ref{tab:outflow}).

\section*{Appendix C: Stack over same physical scales}

The ALMA observations of the QSOs in our sample are characterised by different angular resolutions, with sizes of the ALMA beam in the range $\sim0.4-1.2$ arcsec. This implies that emission from different spatial scales may contribute to the stacked spectrum of the total sample (Sect. \ref{sect:totstack}). To ensure that broad \cii\ wings persist when combining spectra from the same physical scales around the QSO, we extracted spectra of all QSOs in our sample from a circular aperture with radius equal to 8 kpc, i.e. the largest scale probed by our observations (Table \ref{tab:sample}). We then stacked them by following the procedure presented in Sect. \ref{sect:methods}. Fig. \ref{fig:samesize} shows that, although the increased noise fluctuations in the stacked spectrum, associated with the chosen large extraction aperture size, the \cii\ wings are still present and characterised by a luminosity \lbroad $\sim4\times10^8$ \lsun, close to the value estimated from the original stack in Sect. \ref{sect:totstack}.

\floatbox[{\capbeside\thisfloatsetup{capbesideposition={left,center},capbesidewidth=8cm}}]{figure}[\FBwidth]
{\caption{Whole sample integrated spectrum, obtained by stacking spectra of the inner 8 kpc around the QSOs in our sample. Panels are as in Fig. \ref{fig:lum-dens-stack}.}\label{fig:samesize}}{\includegraphics[width=0.365\textwidth]{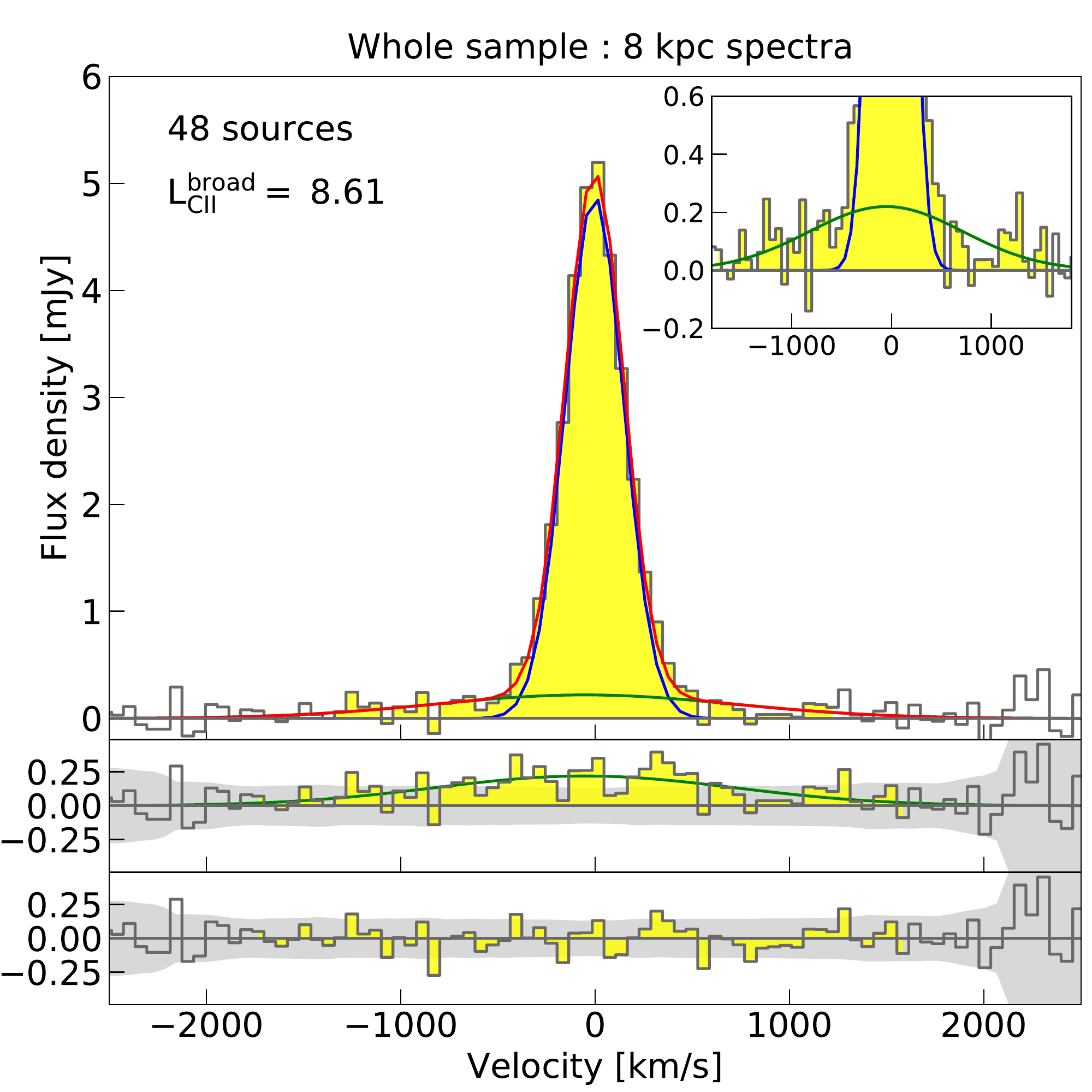}}

\end{document}